\documentclass[11pt, onecolumn]{IEEEtran}
\usepackage[letterpaper, left=1in, right=1in, bottom=1in, top=1in]{geometry}
\usepackage[english]{babel}
\usepackage{enumitem}
\usepackage{graphicx}
\usepackage{color}
\usepackage[dvipsnames]{xcolor}
\usepackage[T1]{fontenc}
\usepackage{subcaption}
\usepackage{dsfont}
\usepackage{amsfonts}
\usepackage{graphicx,wrapfig}
\usepackage[T1]{fontenc}
\usepackage{amsmath}
\usepackage{mathtools, cuted}
\usepackage{amsthm}
\usepackage{amstext}
\usepackage{amssymb}
\usepackage{mathrsfs}
\usepackage[square,numbers, sort&compress]{natbib}
\usepackage{mathtools}
\usepackage{tikz}
\usepackage{marginnote}
\usepackage{tkz-tab}
\usetikzlibrary{topaths,calc}
\usepackage{pgfplots}
\usepackage{stackengine}
\usepackage{algorithmic}
\usepackage[ruled]{algorithm} 
\usepackage[colorinlistoftodos]{todonotes}

\usetikzlibrary{arrows,positioning, shapes}
\usepackage{float}

\usepackage{setspace}
\usepackage{hyperref}
\hypersetup{colorlinks,
   colorlinks,
    citecolor=red,
    filecolor=green,
  linkcolor=blue,
    linktoc=page,
   urlcolor=blue
}

\def\R{\mathbb{R}}

\def\argmin{\mathop{\rm arg\, min}}
\def\eps{\varepsilon}

\def\D{{\mathcal D}}
\def\E{{\mathbb E}}

\def\P{{\mathcal P}}
\def\Q{{\mathcal Q}}
\def\T{{\mathcal T}}
\def\X{{\mathcal X}}

\def\M{{\mathcal M}}
\def\W{{\mathcal W}}

\def\N{{\mathcal N}}
\def\S{{\mathcal S}}
\def\L{{\mathcal L}}

\def\Z{{\mathcal Z}}

\def\sBer{{\mathsf{Bernoulli}}}

\def\sf{{\mathsf f}}

\def\sQ{{\mathsf Q}}
\def\sK{{\mathsf K}}

\def\sEs{{\mathsf E_{e^{\varepsilon}}}}
\def\sE{{\mathsf E}}
\def\sF{{\mathsf F}}

\def\tv{{\mathsf {TV}}}
\def\kl{{\mathsf {KL}}}

\usepackage[font=small,labelsep=space]{caption}
\DeclareCaptionLabelSeparator{dot}{.~}
\newcommand{\eq}[1]{\begin{equation*}
#1
\end{equation*}}
\newcommand{\eqn}[2]{\begin{equation}
\label{#1}
#2
\end{equation}}
\newcommand{\al}[1]{\begin{align*}
#1
\end{align*}}
\newcommand{\aln}[1]{\begin{align}
#1
\end{align}}

\newcounter{example}
\newenvironment{example}[1][]{\refstepcounter{example}\par\medskip
   \noindent \textit{Example~\theexample. #1} \rmfamily}{\medskip}
\newtheorem{definition}{Definition}
\newtheorem{theorem}{Theorem}
\newtheorem{corollary}{Corollary}

\newtheorem{proposition}{Proposition}
\newtheorem{lemma}{Lemma}
\newtheorem{remark}{Remark}

\usepgfplotslibrary{fillbetween}
\usepackage{times}
\usepackage{tikz}
\usepackage{amsmath, bm}
\usepackage{verbatim}
\usetikzlibrary{arrows,shapes}
\tikzstyle{RectObject}=[rectangle,fill=white,draw,line width=0.2mm]
\tikzstyle{line}=[draw]
\tikzstyle{arrow}=[draw, -latex]
\usetikzlibrary{decorations.pathmorphing}
\usetikzlibrary{calc,shapes, positioning}
\usepackage{graphicx}
\usetikzlibrary{shapes.geometric}

\DeclareFontFamily{U}{BOONDOX-calo}{\skewchar\font=45 }
\DeclareFontShape{U}{BOONDOX-calo}{m}{n}{
	<-> s*[1.05] BOONDOX-r-calo}{}
\DeclareFontShape{U}{BOONDOX-calo}{b}{n}{
	<-> s*[1.05] BOONDOX-b-calo}{}
\DeclareMathAlphabet{\mathcalboondox}{U}{BOONDOX-calo}{m}{n}
\SetMathAlphabet{\mathcalboondox}{bold}{U}{BOONDOX-calo}{b}{n}
\DeclareMathAlphabet{\mathbcalboondox}{U}{BOONDOX-calo}{b}{n}

\definecolor{DukeBlue}{HTML}{001A57}
\definecolor{DarkRed}{rgb}{0.75, 0.0, 0.0}
\definecolor{DarkGreen}{rgb}{0.0, 0.5, 0.0}

\usepackage[normalem]{ulem}
\allowdisplaybreaks
\usepackage{setspace}
\author{}
\date{}

\begin{document}
	
	\title{\vspace{5.5mm}Contraction of $\sE_\gamma$-Divergence and Its Applications to Privacy\thanks{This work was supported in part by NSF under grants CIF 1900750 and CIF CAREER 1845852. Parts of the results in this paper were presented at the International Symposium on Information Theory 2020 and 2021 \cite{Asoodeh_Contraction_ISIT,Asoodeh_LDP_Isit21}.}}
	\author{%
		Shahab Asoodeh, Mario Diaz, and Flavio P. Calmon\thanks{S. Asoodeh and F. P. Calmon are with School of Engineering and Applied Science, Harvard University (e-mails: \{shahab, flavio\}@seas.harvard.edu). M.~Diaz is with the Instituto de Investigaciones en Matem\'{a}ticas Aplicadas y en Sistemas (IIMAS), Universidad Nacional Aut\'{o}noma de M\'{e}xico, Mexico City 04510, Mexico (e-mail: mario.diaz@sigma.iimas.unam.mx).}
	}
	
	\maketitle
	\begin{abstract}
		We investigate the contraction coefficients derived from strong data processing inequalities for the $\sE_\gamma$-divergence. By generalizing the celebrated Dobrushin's coefficient from total variation distance to $\sE_\gamma$-divergence, we derive a closed-form expression for the contraction of $\sE_\gamma$-divergence. This result has fundamental consequences in two privacy settings. First, it implies that local differential privacy can be equivalently expressed in terms of the contraction of $\sE_\gamma$-divergence. This equivalent formula can be used to precisely quantify the impact of local privacy in (Bayesian and minimax) estimation and hypothesis testing problems in terms of the reduction of effective sample size. Second, it leads to a new information-theoretic technique for analyzing privacy guarantees of online algorithms. In this technique, we view such algorithms as a composition of amplitude-constrained Gaussian channels and then relate their contraction coefficients under $\sE_\gamma$-divergence to the overall differential privacy guarantees.  As an example, we apply our technique to derive the differential privacy parameters of gradient descent. Moreover, we also show that this  framework can be tailored to  batch learning algorithms that
		can be implemented with one pass over the training dataset. %In particular, we consider the framework of privacy amplification by iteration, recently proposed by Feldman et al., and derive significantly tighter bounds on the differential privacy parameters of the projected noisy stochastic gradient descent algorithm with hidden intermediate updates.
	\end{abstract}
	
	\section{Introduction}
Consider the  Markov chain
\begin{equation}\label{MarkovChain}
    X_1\to Y_1\to X_2\to \dots\to X_n\to Y_n\to X_{n+1},
    %X_1\to X_2\to\dots\to  Y_n,
\end{equation}
where the $d$-dimensional random variable $X_1$ is the input of the first Markov kernel (or channel) $P_{Y_1|X_1}$ and is assumed to satisfy $\|X_1\|^2\leq d A$ almost surely (a.s.) for some $A>0$. Each kernel $P_{X_{t+1}|X_t}$ is composed of the concatenation of two channels:
\begin{enumerate}
\item $P_{Y_{t}|X_t}$ is a vector-Gaussian kernel of dimension $d$, i.e.,
\begin{equation}\label{Eq:GaussianChannel}
P_{Y_t|X_t=x} = \mathcal{N}(x, \sigma^2_t\mathbf{I}_d),
\end{equation}
\item  $P_{X_{t+1}|Y_t}$ is an arbitrary kernel that ensures $X_{t+1}$ satisfies the same amplitude constraint as $X_1$. Thus, we have for all $t\in [n+1]\coloneqq \{1, \dots, n+1\}$
\begin{equation}\label{Eq:AplitudeAWGN}
    \|X_t\|^2\leq d A, \qquad \text{a.s.}
\end{equation}
For example, $P_{X_{t+1}|Y_t}$ can be the projection of $Y_t$ onto the $\ell_2$-ball of radius $(dA)^2$.
\end{enumerate}
One specific instantiation of this Markov chain will be given in Section~\ref{Sec:Privacy_Iterative} (see Fig.~\ref{fig:SDPI}) for characterizing the privacy guarantee of the online gradient descent algorithm, whose each iterations is modelled by the composition of a Gaussian kernel and a projection operator.

Let $\mu_{n+1}$ and $\mu'_{n+1}$ be the distributions of the output $X_{n+1}$ of chain \eqref{MarkovChain} when $X_1$ has distribution $\mu_1$ and $\mu_1',$ respectively. 
The main goal of this paper is to characterize the divergence between $\mu_{n+1}$ and $\mu'_{n+1}$  in terms of  $\mu_1$ and $\mu_1'$. 
%
% It is worth mentioning that such channel can be viewed as the random mapping $x\mapsto \Pi\left(x+\sigma_t Z_t\right)$ where $\{Z_t\}$ is sampled i.i.d.\ from $\N(0, \mathbf{I}_d)$ and $\Pi(\cdot)$ denotes the projection operator onto the ball in $\R^d$ with diameter $D$. 
%
More specifically, we derive \textit{strong data processing inequalities}  \cite{ahlswede1976} for  Gaussian channels satisfying the amplitude constraint \eqref{Eq:AplitudeAWGN}. 

When measuring the distance between $\mu_{n+1}$ and $\mu'_{n+1}$ via $f$-divergences, this framework has been extensively studied in information theory literature \cite{ahlswede1976, Dobrushin, Dobrushin_Maxim, Makur_SDPIjournal, Raginsky_SDPI,Yury_Dissipation, Polyanskiy_SDPI_network, Flavio_Polyankiy}.  Given a convex function $f:(0,\infty)\to\mathbb{R}$ such that $f(1)=0$, the $f$-divergence between two probability measures $\mu$ and $\nu$ with $\mu\ll\nu$ is defined in \cite{Ali1966AGC,Csiszar67} as
\begin{equation*}
    D_f(\mu\|\nu)\coloneqq \E_{\nu}\left[f\left(\frac{\text{d}\mu}{\text{d}\nu}\right)\right].
\end{equation*}
Specific instances of $f$-divergences include KL-divergence, where $f(t) = t\log(t)$,   and total variation, in which case  $f(t) = \frac{1}{2}|t-1|$. 
Let $\D$ and $\W$ be subsets of (potentially different) Euclidean spaces and let $\sK:\D\to\P(\W)$ be a Markov kernel (i.e., channel), where $\P(\W)$ denotes the set of all probability measures on $\W$. Following the information-theoretic approach described in Ahlswede and G\'acs \cite{ahlswede1976}, we define the \textit{contraction coefficient} (or strong data processing coefficient) of $\sK$ under $D_f(\cdot\|\cdot)$ as
% \begin{equation}\label{Eq:SDPI_f_Div}
%     \eta_f(\sK)\coloneqq \sup_{\mu, \nu: \atop D_f(\mu\|\nu)\neq 0}\frac{D_f(\mu \sK\|\nu\sK)}{D_f(\mu\|\nu)}.
% \end{equation}
\begin{equation}\label{Eq:SDPI_f_Div}
    \eta_f(\sK)\coloneqq \sup_{\substack{\mu,\nu\in\P(\D):\\ D_f(\mu\|\nu)\neq 0}}\frac{D_f(\mu \sK\|\nu\sK)}{D_f(\mu\|\nu)},
\end{equation}
where $\mu\sK$ and $\nu\sK$ denote the distribution on $\W$ induced by the push-forward of $\mu$ and $\nu$, respectively.   
This quantity has been extensively studied for general channels in \cite{ahlswede1976, Dobrushin, Dobrushin_Maxim, Makur_SDPIjournal, Raginsky_SDPI} and  for Gaussian channels with power and amplitude constraints (see e.g., \cite{Yury_Dissipation, Polyanskiy_SDPI_network, Flavio_Polyankiy}).   
%This quantity has been long studied for several $f$-divergences, e.g., $\mathsf{KL}$-divergence for which $f(t) = t\log(t)$, $\chi^2$-divergence for which $f(t) = (t-1)^2$, and total variation distance for which $f(t) = \frac{1}{2}|t-1|$. 
Most notably, Dobrushin \cite{Dobrushin} showed that $\eta_{\mathsf{TV}}$---the contraction coefficient under total variation distance---has a remarkably simple expression: 
%$$ \eta_{\mathsf {TV}}(\sK)\coloneqq \sup_{\substack{\mu, \nu:\\ \mathsf{TV}(\mu,\nu)\neq 0}}\frac{\mathsf{TV}(\mu \sK,\nu\sK)}{\mathsf{TV}(\mu,\nu)},$$
\begin{equation}
\label{eq:Dobrushin}
    \eta_\mathsf{TV}(\sK) = \sup_{x_1,x_2\in\D} \mathsf{TV}(\sK(\cdot \vert x_1), \sK(\cdot \vert x_2)).
\end{equation}
This formula has found several applications in the study of ergodicity of Markov processes as well as Gibbs measures, e.g., see  \cite{Dobrushin, Dobrushin_Maxim, Yury_Dissipation, Raginsky_SDPI}. 

Throughout this work, we focus on an instantiation of $f$-divergence named $\sE_\gamma$-divergence (also know as hockey-stick divergence) \cite{polyanskiy2010channel, hockey_stick, Csiszar_Sheilds}. Given $\gamma\geq 0$, the $\sE_\gamma$-divergence between two probability measures $\mu$ and $\nu$ in $\P(\D)$ is defined as
\small
\begin{align}
	\sE_\gamma(\mu\|\nu) &\coloneqq  \int \text{d}(\mu-\gamma\nu)^+ - (1-\gamma)^+ \label{eq:DefEgamma}\\
	\label{Defi_HS_Divergence} &= \sup_{A\subset\D} \left[\mu(A) - \gamma\nu(A)\right]- (1-\gamma)^+\\
	%	&=& \frac{1}{2}\int |\text{d}P-\gamma \text{d}Q| - \frac{1}{2}(\gamma -1)\\
	&=  \mu\left(\imath_{\mu\|\nu}>\log\gamma\right)-\gamma\nu\left(\imath_{\mu\|\nu}>\log\gamma\right)- (1-\gamma)^+, \label{eq:EgammaPDif}
\end{align}
\normalsize 
where $(\mu-\gamma\nu)^+$ is the positive part of the signed measure $\mu-\gamma\nu$ and $\imath_{\mu\|\nu}(t) \coloneqq \log\frac{\text{d}\mu}{\text{d}\nu}(t)$ denotes the \textit{information density} between $\mu$ and $\nu$. It can be directly verified that $\sE_\gamma$-divergence is in fact the $f$-divergence associated with $$f(t) = (t-\gamma)^+-(1-\gamma)^+,$$ where $(a)^+\coloneqq \max(0, a)$, and also that $$\sE_1(\mu\|\nu) = \mathsf{TV}(\mu,\nu).$$ We adopt $\sE_\gamma$-divergence for two main reasons:
\begin{itemize}
    \item Since it is a generalization of total variation distance, its contraction coefficient can potentially broaden the applicability of Dobrushin's result \eqref{eq:Dobrushin},
    \item  As we will see in Sections~\ref{Sec:LDP} and  \ref{Sec:Privacy_Iterative}, $\sE_\gamma$-divergence has a close connection with both local and central differential privacy (see Theorems~\ref{thm:LDP_Contraction} and Definition~\ref{Def:DP_Online}). A characterization of the contraction coefficient of $\sE_\gamma$-divergence leads to a simple and precise privacy  analysis of several applications  in statistics and machine learning.  
\end{itemize}
%(i) Since it is a generalization of total variation distance, its contraction coefficient can potentially broaden the applicability of Dobrushin's result \eqref{eq:Dobrushin}, and (ii) As we will see in Sections~\ref{Sec:LDP} and  \ref{Sec:Privacy_Iterative}, $\sE_\gamma$-divergence has a close connection with both local and central differential privacy and its contraction coefficient leads to significantly tighter privacy analysis in several applications in statistics and machine learning.  

Our main result (i.e., Theorem~\ref{Thm:Contraction_EGamma}) is a  formula for the contraction coefficient of channels under $\sE_\gamma$-divergence which  extends Dobrushin's result \eqref{eq:Dobrushin} from total variation distance to $\sE_\gamma$-divergence for any $\gamma\geq 0$. This result has several direct consequences. First, it allows us to write an equivalent expression for local differential privacy, which lends itself well to studying the (minimax and Bayesian) estimation and testing problems under local differential privacy. Following this path, we quantify the impact of local privacy in such problems in terms of the reduction of effective sample size.  Second, it enables us to derive a sharp upper bound for the deviation of $\mu_{n+1}$ from $\mu'_{n+1}$, i.e., $\sE_\gamma(\mu_{n+1}\|\mu'_{n+1})$, in terms of the amplitude constraint $A$ and each channel's noise variance --- a result which turns out instrumental in the privacy analysis of iterative algorithms in Section~\ref{Sec:Privacy_Iterative}.
% Throughout this work, we measure the distance between two distributions using a certain $f$-divergence called $\sE_\gamma$-divergence. Given $\gamma\geq 0$, the $\sE_\gamma$-divergence between two probability distribution $\mu$ and $\nu$ is defined as $\sE_\gamma(\mu\|\nu)\coloneqq \int \text{d}(\mu-\gamma \nu)^+-(1-\gamma)^+$ where $(a)^+\coloneqq \max(0, a)$. We discuss this divergence and its connection with privacy in Section~\ref{Sec:E_gamma}.

\begin{theorem}\label{Thm:Informal_SDPI}
Let $X_1\sim \mu_1$ and $X'_1\sim\mu'_1$ be two inputs of the Markov chain \eqref{MarkovChain} where channels $\{P_{X_{t+1}|X_t}\}_{t=1}^n$ satisfy \eqref{Eq:GaussianChannel} and \eqref{Eq:AplitudeAWGN}. Let $X_{n+1}\sim \mu_{n+1}$ and $X'_{n+1}\sim \mu'_{n+1}$ be the output of the Markov chain when the input is $X_1$ and $X'_1$, respectively. Then,  we have 
$$\sE_\gamma(\mu_{n+1}\|\mu'_{n+1})\leq \sE_\gamma(\mu_1\|\mu'_1)\prod_{t=1}^n\theta_{\gamma\vee\frac{1}{\gamma}}\Big(\frac{2 d A}{\sigma_t}\Big),$$
where $\theta_\gamma(r)\coloneqq \sQ\left(\frac{\log\gamma}{r} 
- \frac{r}{2}\right) - \gamma \sQ\left(\frac{\log\gamma}{r} + \frac{r}{2}\right)$ and $\sQ(a) \coloneqq \frac{1}{\sqrt{2\pi}} \int_a^\infty e^{-u^2/2}\textnormal{d}u$.
\end{theorem}
When $\gamma = 1$ and $\sigma_t =\sigma$, this theorem reduces to the estimate
$$\tv(\mu_{n+1}, \mu'_{n+1})\leq \tv(\mu_1, \mu'_1) (1-2\sQ(\sqrt{dA}))^n,$$ that was derived in \cite{Yury_Dissipation} by a direct application of Dobrushin's result \eqref{eq:Dobrushin}. However, unlike \cite{Yury_Dissipation}, we are interested in settings where $\gamma$ need not be equal to $1$.  In particular, we make use of Theorem~\ref{Thm:Informal_SDPI} to study the cost of differential privacy in online algorithms such as online gradient descent in which $\gamma$ is set to be $e^\eps$, where $\eps\geq 0$ is the differential privacy guarantee.   

\subsection{Main Contributions}
\noindent \textbf{Contraction coefficient under $\sE_\gamma$-divergence:} We prove that, similar to $\eta_\tv$, the contraction coefficient of a Markov kernel $\sK$ under $\sE_\gamma$-divergence, denoted by $\eta_\gamma(\sK)$, enjoys a remarkably simple two-point characterization. More precisely, we show in Theorem~\ref{Thm:Contraction_EGamma} that  
    $$\eta_\gamma(\sK) = \sup_{x_1, x_2\in \D}\E_\gamma(\sK(\cdot \vert x_1)\Vert\sK(\cdot \vert x_2)),$$ for all $\gamma\geq 1$, thus generalizing \eqref{eq:Dobrushin} from total variation distance (i.e., $\gamma =1$) to $\sE_\gamma$-divergence for any $\gamma>1$. We then use basic properties of $\sE_\gamma$-divergence to show $\eta_\gamma(\sK) = \eta_{1/\gamma}(\sK)$ for $\gamma<1$.  
    
We apply this result to the integral representation of $f$-divergences in terms of $\sE_\gamma$-divergence \cite[Corollary 3.7]{cohen1998comparisons} to obtain the estimate 
 \begin{equation}
     D_f(\mu\sK\|\nu\sK)\leq \int_0^{\infty}\eta_\gamma(\sK)f''(\gamma)\sE_\gamma(\mu\|\nu)\text{d}\gamma,
\end{equation} for all $\mu, \nu\in \P(\D)$ and all $f$-divergences with twice-differentiable $f$. This result is provably tighter than the bound \cite[Proposition II.4.10]{cohen1998comparisons} %$$\eta_f(\sK)\leq \eta_\tv(\sK)$$  or, equivalently,
$$D_f(\mu\sK\|\nu\sK)\leq \eta_\tv(\sK)D_f(\mu\|\nu).$$
%for any $\mu, \nu\in \P(\D)$ and any choice of $f$-divergence.
While this upper bound is known to be typically strict, it has been used extensively in the information theory literature \cite{Gaussian_Inf_loss, Yury_Dissipation,Raginsky_ISIT_converses}. We numerically compare these two upper bounds in  Examples~\ref{example_UB_Df1} and \ref{example_UB_Df2} for $\chi^2$-divergence. 
 
\noindent \textbf{Local differential privacy:} As another application, we use  Theorem~\ref{Thm:Contraction_EGamma} to  show that  local differential privacy (LDP) can be equivalently expressed in terms of the contraction coefficient under $\sE_\gamma$-divergence (see \ Theorem~\ref{thm:LDP_Contraction}): A randomized mechanism $\sK$ is $(\eps, \delta)$-LDP if and only if $$\eta_\gamma(\sK)\leq \delta \mbox{ for } \gamma = e^\eps.$$ Using the relationship between $\sE_\gamma$-divergence and general $f$-divergences, this result implies $$\eta_f(\sK)\leq 1-(1-\delta)e^{-\eps}$$ for any $(\eps, \delta)$-LDP mechanism $\sK$. This estimate of the output $f$-divergence of an LDP mechanism can be directly used to quantify the impact of local privacy in Bayesian  and minimax estimation problems. First, we introduce a new variant of  Le Cam's converse technique \cite{lecam1973} to derive a lower bound for the minimax estimation risk under LDP constraints. In contrast to existing results on minimax risk under LDP \cite{Duchi_LDP_MinimaxRates, Duchi_Federatedprotection, Duchi_SDPI, Geometrizing_LDP,Acharya_MinimaxDis,Acharya_infConstraint},  our result holds for any values of $\eps\geq 0$ and $\delta\in [0,1]$. 
%It must be noted that our technique is similar in nature to that of Duchi and Rogers \cite{Duchi_SDPI} where the private minimax risk was studied via the strong data processing inequality for KL divergence. 

Second, we develop a framework for characterizing the Bayesian estimation risk under LDP. To do so, we first derive a lower bound for the non-private Bayesian risk in terms of  $\sE_\gamma$-information, defined as the $\sE_\gamma$-divergence between the joint and product distributions of two random variables (i.e., the $\sE_\gamma$ counterpart of mutual information). Combining this  bound with Theorem~\ref{Thm:Contraction_EGamma}, we derive a lower bound for private Bayesian risk in Theorem~\ref{Thm:Bayesian_LB}.  Our results indicate that the cost of $(\eps, \delta)$-LDP in one-dimensional minimax and Bayesian estimation problems is to reduce the effective sample size from $n$ to $n(1-(1-\delta)e^{-\eps})$.  %Moreover, since $\sE_\gamma$-divergence is a fundamental  

\noindent \textbf{Central differential privacy:} 
As a final application, we apply Theorem~\ref{Thm:Contraction_EGamma} (more specifically, Theorem~\ref{Thm:Informal_SDPI}) to investigate the privacy guarantees of online iterative algorithms. In general, an online learning algorithm proceeds as follows:  a learner first selects a random point $W_1$ from a convex set $\W\subset  \R^d$. After committing to $W_1$, the cost function $\sf_1$ is revealed to her by nature, specifying the cost $\sf_1(W_1)$ incurred by the choice $W_1$. Upon observing the cost function $\sf_t$ at time $t$, the learner constructs $W_{t+1}\in\W$ at time $t+1$ according to some update rules based \textit{only} on $\sf_t$ (as opposed to the entire $\{\sf_1, \dots, \sf_t\}$). Denoting this update rule at time $t$ by $\Psi_{\sf_t}:\W\to \R^d$, the iterative algorithm can be expressed by 
$$W_{t+1} =\Pi_\W(\Psi_{\sf_t}(W_t)),$$ 
where $\Pi_\W(\cdot)$ is the projection operator onto $\W$ (see Section~\ref{Sec:Privacy_Iterative} for examples of this algorithms). To ensure privacy of the learner, one standard way is to add calibrated noise to $\Psi_{\sf_t}$ at each iteration \cite{DP_OnlineLearning, Online_DP_Smith, Online_DP_Agarwal, Online_Cumming}. Thus, the private version of such algorithm can typically be expressed as  
\begin{equation}\label{Iterative1}
    W_{t+1} = \Pi_{\mathcal{W}}\left(\Psi_{\sf_t}(W_t) + \sigma_t Z_{t}\right),
\end{equation}
where $\{Z_t\}$ is the collection of independent and identically distributed (i.i.d.) noise variables sampled from a known density with covariance matrix $\mathbf{I}_d$ and $\sigma_t$ specifies the magnitude of noise at time $t$. 
%Considering $\{\sf_1, \dots, \sf_n\}$ as the input of online algorithms, DP ensures that the output  remains nearly the same if a single function $\sf_i$, for some $i\in [n]\coloneqq \{1, \dots, n\}$, changes to a different $\sf'_i$. 
%Interestingly, this iterative process can be viewed as an instantiation of Markov chain \eqref{MarkovChain} with $W_t$ in place of $X_t$ and $P_{X_{t+1}|X_t}$ given by the composition of Gaussian kernel and projection operator. 
In Theorem~\ref{Thm:General_Random_DP}, we characterize the central differential privacy (DP) guarantee of such algorithm when the learner discloses $W_{n+1}$ and $\{Z_t\}$ are sampled from the Gaussian distribution. Viewing each iteration of this algorithm as a composition of a Gaussian kernel and a projection operator (see \eqref{Eq:GaussianChannel} and \eqref{Eq:AplitudeAWGN}), we precisely model this iterative algorithm by the Markov chain \eqref{MarkovChain}. Invoking Theorem~\ref{Thm:Informal_SDPI}, we can thus obtain an upper bound on the $\sE_\gamma$-divergence between the distributions of the output of the process \eqref{Iterative1} after $n$ iterations when cost function $\sf_j$ changes to $\sf_j'$ for some $j\in [n]$. The relationship between $\sE_\gamma$-divergence and DP enables us to directly translate this bound to a bound on the DP parameters $\eps$ and $\delta$. We instantiate Theorem~\ref{Thm:General_Random_DP} to characterize privacy guarantees in \textit{one-pass} stochastic gradient descent (SGD) in Corollary~\ref{cor:PNSGD} and the online gradient descent in Proposition~\ref{Prop:Utility}.    

\subsection{Additional Related Work}
Strong data processing inequalities (SDPI) for KL divergence and total variation distance are ubiquitous in information theory and statistics. They appear, for example, in the study of  ergodicity of Markov processes \cite{Dobrushin}, the uniqueness of Gibbs measures \cite{Dobrushin}, contraction of mutual information (and generalized mutual information) in Markov chains \cite{Anantharam_SDPI, Yury_Dissipation, Dobrushin_Maxim, Flavio_Polyankiy, Anurak_SDPI} and in Bayesian networks \cite{Polyankiy_SDPI_Networks}, comparison of channels \cite{Comparison_Channel_Polyankiy}, distributed estimation \cite{Maxim_SDPI_ISIT}, communication complexity of statistical estimation \cite{Communication_Complexity_SDPI},  distributed function computation \cite{Raginsky_SDPI_Function_Computation}, and private estimation problems \cite{Duchi_SDPI, LDP_Gaussian}. 
%Incidentally, the contraction coefficients of general Markov kernels under total variation distance and KL divergence have simple expressions, as outlined in \cite{Dobrushin} and \cite{ahlswede1976, Anantharam_SDPI}, respectively. In particular, Dobrushin \cite{Dobrushin} derived a remarkably simple expression for the case of total variation distance. 
A formula for the contraction coefficients under general $f$-divergences is derived in \cite[Theorem 5.2]{Raginsky_SDPI} for differentiable $f$. This formula, however, is not applicable for the specific case of $\sE_\gamma$-divergence as $f(t) = (t-\gamma)^+$ is not differentiable.   
More recently, Kamalaruban \cite[Theorem 3.10]{Phd_Thesis_SDPI_ANU} derived a closed-form expression for the contraction coefficient under another $f$-divergence, namely  \textit{DeGroot's statistical information} \cite{degroot1962}. 
%To the best of our knowledge, the contraction coefficient of general Markov kernels under $\sE_\gamma$-divergence and demonstrate that it has a simple closed-form expression (Theorem~\ref{Thm:Contraction_EGamma}) of which Dobrushin's characterization is a special case.

The study of statistical efficiency under \textit{pure} LDP constraints was initiated by Duchi et al. \cite{Duchi_LDP_MinimaxRates} in the minimax setting and has since gained considerable attention, e.g., \cite{Shiva_subsampling, Duchi_SDPI, LDP_gaboardi19a, Duchi_Federatedprotection, kairouz2014extremal_JMLR,LDP_Fisher, LDP_Acharya1, Acharya_MinimaxDis, LDP_DistributionEstimation, LDP_LinearRegression, Geometrizing_LDP, evfimievski2003limiting, LDP_interactiveFaster,Acharya_infConstraint}.  While the original bounds on the private minimax risk in \cite{Duchi_LDP_MinimaxRates} were meaningful only in the high privacy regime (i.e., small $\eps$), the order optimal bounds were recently given by Duchi and Rogers in \cite{Duchi_SDPI} for the general privacy regime. Interestingly, their technique relies on the decay rate of mutual information over a Markov chain, which is known to be equivalent to the SDPI for KL divergence \cite{Anantharam_SDPI}. Unlike their technique, ours is based on the SDPI for $\sE_\gamma$-divergence and allows us to study minimax risk under \textit{approximate} LDP (i.e., $\delta>0$).

%Almost all of these works consider $\eps$-LDP (i.e., $\delta =0$) and provide meaningful bounds
% only for sufficiently small values of $\eps$ (i.e., the high privacy regime). % and their technique fail to apply to the approximate LDP setting where $\delta>0$.
% For instance, Duchi et al.\ \cite{Duchi_LDP_MinimaxRates} studied minimax estimation problems under $\eps$-LDP constraints and showed that for $\eps\leq 1$, the price of privacy is to reduce the effective sample size from $n$ to $\eps^2 n$. A slightly improved version of this result appeared in \cite{kairouz2014extremal_JMLR}.  %Given the promise of privacy amplification for LDP,  it is essential to study statistical problems under $(\eps, \delta)$-LDP for any $\eps\geq 0$ and $\delta>0$.  
% More recently, Duchi and Rogers \cite{Duchi_SDPI}
% developed a framework based on 
%Our results differ from these existing results in that they hold for $(\eps, \delta)$-LDP for any possible values of $\eps$ and $\delta$. 

Online learning \cite{Hazan_Online, Shalev_Online_LectureNote} is a framework where a sequence of predictions are made given the knowledge
of past actions.
%of the outputs or consequences of past actions. 
This framework is particularly well-suited for dynamic and adversarial environments where learning from data must be done in real-time. Online learning is ubiquitous in practical applications such as 
recommender systems \cite{Path_online}, spam detection \cite {Online_Spam}, portfolio optimization \cite{Online_Finance1,Online_Finance2,Online_Finance3}, and convex optimization \cite{Online_Convex_Optimization, Online_Convex}, to name a few.  The problem of analyzing DP guarantees of online algorithms was introduced in \cite{Continual_Observation} for the simple setting where $\W$ is a probability simplex and the cost functions are  linear with binary coefficients. Inspired by this work, \cite{DP_OnlineLearning, Online_DP_Smith, Online_DP_Agarwal} developed techniques for making a large class of online learning algorithms differentially private. The settings studied in these works differs from ours in that they violate two implicit assumptions made in our setting: 
\begin{enumerate}
    \item Each update $W_{t+1}$ depends only on $\sf_t$, conditioned on $W_t$. However, the aforementioned works rely on the tree-based aggregation \cite{Continual_Observation} which selects $W_{t+1}$ based on the entire $\sf_1, \dots, \sf_{t}$. 
    \item The output of each iteration is kept private. Thus the privacy guarantee must be quantified against $W_{n+1}$ and not the entire $W_1, \dots, W_{n+1}$.
\end{enumerate}
The Markovity condition in Assumption 1 is equivalent to a memory constraint on the  online algorithm.  Moreover, it is trivially satisfied by several popular online algorithms such as online (stochastic) gradient descent \cite{Online_SGD}, online mirror descent \cite{Hazan_Online}, implicit gradient descent \cite{Online_ConvexProblem}, the passive-aggressive algorithm \cite{PAssive_Aggressive}, composite mirror descent \cite{Duchi_Composite_Mirror}, and  the Frank-Wolfe algorithm \cite{Frank_Wolf_Algorithm}. Nevertheless, this assumption rules out  algorithms such as the follow-the-leader (FTL) algorithm and its variants. We do note, however, that a particular popular variant of the FTL algorithm, namely Regularized FTL \cite{RegularizedFTL1, RegularizedFTL2} is equivalent to online mirror descent (see \cite[Lemma 1]{Hazan08extractingcertainty} and  \cite{RFTL_Equivalence_Mirror_Mcmahan}), rendering our framework applicable in this specific case.

Assumption 2 is also present in  several recent works on last-iterate convergence of learning algorithms \cite{Last_Iterate_SGD_Rakhlin, Last_Iterate_SGD_Shamir, Last_Iterate_abernethy2019lastiterate, Last_Iterate_lei2020iterate, last-iterate_SGD},  the \textit{privacy amplification by iteration} framework \cite{Feldman2018PrivacyAB}, its variants in \cite{Feldman_LinearTime, Balle2019mixing}, and privacy-preserving generative model for data inspection \cite{DP_FL_generative_model}. Another practical scenario where Assumption 2 naturally arises is as follows: A learner intends to fit a model privately  and publicly releases the model parameters  after a target level of accuracy is met (for instance,  after a certain number of iterations).  Hiding intermediate updates not only leads to privacy amplification \cite{Feldman2018PrivacyAB, Balle2019mixing}, it may also drive the final parameters to be sparse \cite{last-iterate_SGD} --- a property which is often crucial for many applications  \cite{last-iterate_SGD}.

The iterative process \eqref{Iterative1} can also correspond to batch (i.e., offline) algorithms where a dataset is fixed and is available to the learner in advance. For instance, consider the empirical risk minimization (ERM) problem: Given a dataset $\{x_1, \dots, x_n\}\in \X^n$, a learner %[?learner?] 
seeks to solve $\min_{w\in \W}\frac{1}{n}\sum_{t=1}^n\ell(w, x_t)$ for some convex loss function $\ell:\W\times \X\to \R_+$. 
The problem of ERM with DP constraints has been widely-studied with known asymptotically tight upper and lower bounds for the excess loss (i.e., the difference between achieved loss and the true minimum), e.g., \cite{SGD_Exponential_VS_Gaussian, chaudhuri2011differentially, Chaudhuri_Subsampling, Bassily_ERM, BAssily_NIPS19, Chaudhuri_logistic, Thakurta_LASSO,Sarwate_SGD_Update,duchi2013local, Smith_interaction,Talwar_Privacte_LASSO,Private_ERM_General}.  This problem can be viewed as an instance of the online learning problem with cost functions $\sf_t(w) = \ell(w, x_t)$ \cite{Kakade_Off_Online}. This observation enables us to translate the privacy analysis of the process \eqref{Iterative1} to the privacy guarantee of batch algorithms. The caveat here is that each data point $x_i$, $i\in [n]$ must be involved in the training process exactly once. Examples of such algorithms include SGD with sub-sampling \textit{without} replacement. 
%A closely related idea of randomly \textit{partitioning} the dataset has already proved to be efficient for privacy in \cite{PATHE1, PATHE2}. 

The privacy properties of SGD algorithms with Gaussian perturbations under similar assumptions as ours was initiated in \cite{Feldman2018PrivacyAB} where the privacy guarantee was given in terms of a variant of DP, namely \textit{R\'enyi differential privacy} (RDP) \cite{RenyiDP}. The RDP guarantee can easily be converted to the DP guarantee. Recently, an \textit{optimal} conversion formula between RDP and DP was presented in \cite{asoodeh2020variants}. In Section~\ref{Sec:Privacy_Amplification}, we revisit the model studied in \cite{Feldman2018PrivacyAB}, and analytically and numerically demonstrate that our result  can be substantially tighter than what would be obtained by converting their results from RDP to DP via the formula given in \cite{asoodeh2020variants}.

\subsection{Paper Organization}
The rest of the paper is organized as follows. Section~\ref{Section:ContractionEgamma} presents our main result on  the contraction coefficient $\eta_\gamma$,  its multi-letter generalizations, and an explicit formula for contraction in amplitude-constrained additive Gaussian kernels. Section~\ref{Section:ContractionEgamma} shows how to equivalently express LDP constraints in terms of $\eta_\gamma$. This section also presents three applications of $\sE_\gamma$-contraction to quantify the impact of LDP in estimation in terms of minimax risk (Section~\ref{Sec:ApplicationsMinimax}), Bayesian risk  (Section~\ref{Sec:ApplicationsBayesian}), and binary hypothesis testing (Section \ref{Sec:ApplicationsHypothesis}). Section~\ref{Sec:Privacy_Iterative} demonstrates how typical machine-learning iterative algorithms can be viewed as a composition of Markov kernels and how their contraction coefficients can be used to obtain tighter privacy guarantees for projected noisy stochastic gradient descent in Section~\ref{Sec:Privacy_Amplification} and online gradient descent in Section~\ref{Sec:OnlinePrivacy}.

\subsection{Notation}
We use upper-care letters (e.g., $X$) to denote random variables and  calligraphic letters to represent their alphabets (e.g., $\X$). The set of all distributions on $\X$ is denoted by $\P(\X)$. 
For a signed measure $\phi$ over $\X$, its total variation is defined as
\begin{equation*}
    \lVert\phi\rVert := \phi^{+}(\X) + \phi^{-}(\X),
\end{equation*}
where $(\phi^{+},\phi^{-})$ is the Hahn-Jordan decomposition of $\phi$ (see, e.g., \cite[Page 421]{Billingsly}). Observe that if $\mu$ and $\nu$ are probability measures,
\begin{equation*}
    \lVert\mu-\nu\rVert = 2 \sup_{A\subset\X} \big[\mu(A) - \nu(A)\big].
\end{equation*}
The total variation distance between two distributions $\mu$ and $\nu$ is 
\begin{equation*}
    \mathsf{TV}(\mu,\nu) := \frac{1}{2} \lVert\mu-\nu\rVert.
\end{equation*}
A Markov kernel (or channel) $\sK:\D\to \P(\W)$ is specified by a collection of distributions $\{\sK(x)\in \P(\W):x\in \S\}$. Given a Markov kernel $\sK:\D\to \P(\W)$ and $\mu\in \P(\D)$, we denote by $\mu\sK$ the push-forward of $\mu$ under $\sK$, i.e., the output distribution of $\sK$ when the input is distributed according to $\mu$, and is given by $\mu\sK\coloneqq \int\mu(\text{d}x)\sK(x)$. We use $\E_\mu[\cdot]$ to write the expectation with respect to $\mu$ and $[n]$ for an integer $n\geq 1$ to denote $\{1, \dots, n\}$.  For $a\in [0,1]$, we define $\bar{a} = 1-a$. For a set $\mathcal{D}\subset\mathbb{R}^d$, we let $\textnormal{dia}(\mathcal{D})$ be its diameter, i.e.,
\begin{equation*}
	\textnormal{dia}(\mathcal{D}) = \sup_{x_1,x_2\in\mathcal{D}} \lVert x_{2}-x_{1}\rVert.
\end{equation*}

	\section{Contraction Coefficient of Markov Kernels under $\sE_\gamma$-Divergence}
\label{Section:ContractionEgamma}

In this section, we establish a closed-form expression for the contraction coefficient of Markov kernels under $\sE_\gamma$-divergence which generalizes Dobrushin's formula. Relying on this expression, we compute the contraction coefficient of the amplitude-constrained additive Gaussian kernel. Finally, we improve classical estimates for the output $f$-divergence based on our generalization of Dobrushin's formula.

Similar to Dobrushin's formula \eqref{eq:Dobrushin}, the next theorem reduces the computation of the contraction coefficient of a Markov kernel under $\sE_{\gamma}$-divergence to maximizing the output $\sE_{\gamma}$-divergence for atomic inputs. In the sequel, we always assume that $\mathcal{D}$ and $\mathcal{W}$ are subsets of (potentially different) Euclidean spaces. Recall that $\eta_\gamma(\sK)$ denotes the contraction coefficient of kernel $\sK$ under $\sE_\gamma$-divergence. 

\begin{theorem}
	\label{Thm:Contraction_EGamma}
	Let $\sK:\D\to\P(\W)$ be a Markov kernel. For any $\gamma\geq1$, we have
	\begin{equation}
		\label{eq:DobrushinEgamma}
		\eta_\gamma(\sK)= \sup_{x_1,x_2\in\D} \sE_{\gamma}(\sK(\cdot \vert x_1)\|\sK(\cdot \vert x_2)).
	\end{equation}
	Furthermore, for any $\gamma<1$, we have
	\begin{equation}
		\label{eq:DobrushinEgamma2}
		\eta_\gamma(\sK)= \eta_{1/\gamma}(\sK).
	\end{equation}
\end{theorem}

The proof of this theorem is given in Appendix~\ref{Appendix:Proof_THM_Contraction}. In view of the reciprocity relation \eqref{eq:DobrushinEgamma2}, in the sequel we focus on the case $\gamma\geq1$.

%It is known that $\sE_\gamma$-divergence can be used to describe a large family of $f$-divergences (see, e.g., \cite{Yury_Dissipation}). The following lemma establishes an upper bound for $\eta_f$ in terms of $\eta_\gamma$. 

Liu et al.\ \cite[Proposition 4]{E_gamma} showed that, for $\gamma\geq 1$ and any probability measures $\mu$ and $\nu$,
\begin{equation*}
	\tv(\mu, \nu)\leq 1-\frac{1-\sE_\gamma(\mu\|\nu)}{\gamma}.
\end{equation*}
Applying Theorem~\ref{Thm:Contraction_EGamma} to this inequality leads to the following 
general relation between the contraction coefficient of a Markov kernel under total variation and the corresponding one under $\sE_{\gamma}$-divergence.

\begin{lemma}\label{Lemma:UB_f_Div_general}
	% If $\gamma\geq1$, then, for every $\mu,\nu$,
	% \begin{equation*}
		%     \sE_{\gamma}(\mu \Vert \nu) \leq \tv(\mu,\nu) \leq 1 - \frac{1-\sE_{\gamma}(\mu \Vert \nu)}{\gamma}.
		% \end{equation*}
	Let $\sK:\D\to\P(\W)$ be a Markov kernel. For any $\gamma\geq1$, we have
	$$\eta_{\tv}(\sK)\leq 1-\frac{1-\eta_\gamma(\sK)}{\gamma}.$$
\end{lemma}

%The previous lemma could be readily extended for $\gamma<1$ via the reciprocity relation \eqref{eq:DobrushinEgamma2}. % We already said this, sounds very repetitive.

Recall that, for any convex function $f$ satisfying $f(1)=0$ and any Markov kernel $\sK$,
\begin{equation}\label{eq:General_UB_TV2}
	\eta_f(\sK)\leq \eta_{\mathsf{TV}}(\sK).
\end{equation}
This result, originally proved in \cite{cohen1998comparisons} for the discrete case and subsequently extended to the general case in \cite{Del_Moral_Contraction}, and our previous lemma produce the estimate
\begin{equation*}
	\eta_{f}(\sK)\leq 1-\frac{1-\eta_\gamma(\sK)}{\gamma},
\end{equation*}
or, equivalently, for any probability measures $\mu$ and $\nu$,
\begin{equation}
	\label{eq:ContractionDfEtaGamma}
	D_f(\mu\sK\|\nu\sK)\leq \left(1-\frac{1-\eta_\gamma(\sK)}{\gamma}\right) D_f(\mu \| \nu).
\end{equation}
In Section~\ref{Sec:LDP} we apply the latter bound to quantify the statistical cost of local differential privacy.

Next we extend Lemma~\ref{Lemma:UB_f_Div_general} to the tensor product of Markov kernels. Given Markov kernels $\sK_1, \dots, \sK_n$, we let $\sK^{\otimes n}$ be their $n$-fold tensor product, i.e., the memoryless channel defined by
\begin{equation*}
	\sK^{\otimes n} = \sK_1 \otimes \cdots \otimes \sK_n.
\end{equation*}
Recall that the contraction coefficient $\eta_{\tv}$ satisfies (see, e.g., \cite[Corollary 9]{Polyankiy_SDPI_Networks}, \cite[Lemma 3]{Raginsky_ISIT_converses} and \cite[Eq. (62)]{Makur_SDPIjournal}),
\begin{equation*}
	\eta_\tv(\sK^{\otimes n}) \leq \max_{i\in [n]}\left[1-(1-\eta_{\tv}(\sK_i))^n\right].
\end{equation*}
The next lemma is an immediate consequence of this inequality and Lemma~\ref{Lemma:UB_f_Div_general}.

\begin{lemma}\label{lemma:UB_Eta_Kn_general}
	Let $\sK_{1},\dots,\sK_n$ be Markov kernels. For any $\gamma\geq1$, we have
	$$\eta_{\tv}(\sK^{\otimes n})\leq \max_{i\in [n]}\left[1-\left(\frac{1-\eta_\gamma(\sK_{i})}{\gamma}\right)^n\right].$$
\end{lemma}

As before, we can combine the previous lemma with \eqref{eq:General_UB_TV2} to obtain an upper bound for $\eta_{f}(\sK^{\otimes n})$ with $f$ a convex function satisfying $f(1)=0$.
%%%%%%%

\subsection{Contraction Coefficient of the Amplitude-Constrained Additive Gaussian Kernel}

Next we compute the contraction coefficient of the amplitude-constrained additive Gaussian kernel under $\sE_{\gamma}$-divergence. To this end, we start by recalling the following lemma.

% \begin{lemma}
	% \label{Lemma:EgammaLaplace}
	% For $m\in\mathbb{R}$ and $v>0$, let $\mathcal{L}(m,v)$ denote the Laplace distribution with mean $m$ and variance $2v^2$. If $m_1,m_2\in\mathbb{R}$ and $v>0$, then
	% \begin{equation}
		% \label{Exa_Contraction_Laplacian}
		%     \sE_\gamma(\mathcal{L}(m_1,v)\|\mathcal{L}(m_2,v)) =\left[1- e^{\frac{v\log(\gamma) -|m_1-m_2|}{2v}}\right]^+.
		% \end{equation}
	% \end{lemma}
% \begin{proof}
	% For ease of notation, let $\mathcal{L}_i=\mathcal{L}(m_i, v)$ for $i=1, 2$. It follows from the definition that 
	% \al{\sE_\gamma(\L_1\|\L_2)&= \frac{1}{2v}\int_{-\infty}^\infty \left[e^{-\frac{|t-m_1|}{v}} - e^{\frac{v\log\gamma-|t-m_2|}{v}}\right]^+\text{d}t\\
		% & = \frac{1}{2v}\int_{-\infty}^\infty \left[e^{-\frac{|t-\tilde m|}{v}} - e^{\frac{v\log\gamma-|t|}{v}}\right]^+\text{d}t
		% }
	% where $\tilde m = m_2-m_1$ (assuming $m_2\geq m_1$). Clearly, the above integral is non-zero only if $\tilde m\geq v\log\gamma$. With this in mind, we can write 
	% \al{\sE_\gamma(\L_1\|\L_2)&= \frac{1}{2v}\int_{\frac{\tilde m +v\log\gamma}{2}}^\infty \left[e^{-\frac{|x-\tilde m|}{v}} - e^{\frac{v\log\gamma-|x|}{v}}\right]\text{d}x = 1- e^{\frac{v\log\gamma -\tilde m}{2v}},}
	% and hence $$\sE_\gamma(\L_1\|\L_2) =\left[1- e^{\frac{v\log\gamma -\tilde m}{2v}}\right]^+.$$
	% The case $m_1\geq m_2$ is similar. In general, we can write
	% $$\sE_\gamma(\L_1\|\L_2) =\left[1- e^{\frac{v\log\gamma -|m_1-m_2|}{2v}}\right]^+.$$
	% \end{proof}

\begin{lemma}[{\cite[Lemma~6]{balle2018improving}}]
	\label{Lemma:EgammaGaussian}
	For $m\in\mathbb{R}^d$ and $\sigma>0$, let $\mathcal{N}(m,\sigma^2\mathbf{I}_d)$ denote the multivariate Gaussian distribution with mean $m$ and covariance matrix $\sigma^2\mathbf{I}_d$. If $m_1,m_2\in\mathbb{R}^d$ and $\sigma>0$, then
	\begin{align*}
		&\sE_{\gamma}(\mathcal{N}(m_1,\sigma^2\mathbf{I}_d)\| \mathcal{N}(m_2,\sigma^2\mathbf{I}_d)) =\sQ\left(\frac{\log\gamma}{\beta} - \frac{\beta}{2}\right) - \gamma \sQ\left(\frac{\log\gamma}{\beta} + \frac{\beta}{2}\right),
	\end{align*}
	with $\displaystyle \beta = \frac{\|m_2-m_1\|}{\sigma}$ and $\displaystyle \sQ(t) = \frac{1}{\sqrt{2\pi}} \int_t^\infty e^{-u^2/2}\text{d}u$.
\end{lemma}

This lemma motivates the following definition.

\begin{definition}\label{Def:theta}
	For $\gamma\geq1$, we let $\theta_\gamma:[0,\infty)\to [0,1]$ be the function defined by
	\begin{align*}
		\theta_\gamma(r) &\coloneqq \sE_\gamma\left(\mathcal{N}(ru,\mathbf{I}_d) \| \mathcal{N}(0,\mathbf{I}_d)\right)\\
		&= \sQ\left(\frac{\log\gamma}{r} - \frac{r}{2}\right) - \gamma \sQ\left(\frac{\log\gamma}{r} + \frac{r}{2}\right),
	\end{align*}
	where $u\in\mathbb{R}^d$ is any vector of unit norm.
\end{definition}

It is straightforward to verify that $r \mapsto \theta_{\gamma}(r)$ is a non-decreasing mapping for every $\gamma\geq1$. This  intuitive property is useful to compute the contraction coefficient of the amplitude-constrained additive Gaussian kernel.

Given $\mathcal{D}\subset\mathbb{R}^{d}$ and $\sigma>0$, the $\mathcal{D}$-constrained additive Gaussian kernel $\sK:\mathcal{D}\to\P(\mathbb{R}^d)$ is defined by
\begin{equation*}
	\sK(\cdot \vert x) = \mathcal{N}(x,\sigma^2 \mathbf{I}_d).
\end{equation*}
Note that $\mathcal{D} = \{x\in\mathbb{R}^{d} : \lVert x \rVert^2 \leq d A\}$ recovers the usual amplitude-constrained additive Gaussian kernel in \eqref{Eq:AplitudeAWGN}. The following proposition is obtained by combining the previous lemma and our generalization of Dobrushin's formula in Theorem~\ref{Thm:Contraction_EGamma}, as shown in Appendix~\ref{Appendix:Prop_etaPAGK}. Recall that
\begin{equation*}
	\textnormal{dia}(\mathcal{D}) = \sup_{x_{1},x_{2}\in\mathcal{D}} \lVert x_{2} - x_{1} \rVert.
\end{equation*}

\begin{proposition}
	\label{Proposition:etaPAGK}
	Let $\gamma\geq1$, $\mathcal{D}\subset\mathbb{R}^d$, and $\sigma>0$. If $\sK$ is the $\mathcal{D}$-constrained additive Gaussian kernel, then
	\begin{equation*}
		\eta_\gamma(\sK) = \theta_\gamma\left(\frac{\textnormal{dia}(\mathcal{D})}{\sigma}\right).
	\end{equation*}
\end{proposition}

The contraction coefficient of the additive Gaussian kernel under $\sE_\gamma$-divergence is trivial\footnote{Recall that $\eta_{\mathsf{TV}}(\sK) = 1$ for any Gaussian channel $\sK$ without input constraints \cite{Yury_Dissipation} and, in addition, $\eta_{\mathsf{TV}}(\sK)=1$ if and only if $\eta_{f}(\sK)=1$ for any non-linear function $f$ \cite[Prop.~II.4.12]{cohen1998comparisons}.}, i.e., $\eta_\gamma(\sK) = 1$. Hence, Proposition~\ref{Proposition:etaPAGK} shows that constraints to the channel input must be imposed in order for the  contraction coefficient to be non-trivial. Given this proposition, Theorem~\ref{Thm:Informal_SDPI} is a simple consequence of the data processing inequality for $f$-divergences. In Section~\ref{Sec:Privacy_Iterative} we provide differential privacy guarantees for different noisy iterative algorithms based on these results.

\begin{remark}
	Note that the $\sE_\gamma$-divergence between Gaussian distributions $\N(m_1, \sigma^2\mathbf{I}_d)$ and $\N(m_2, \sigma^2\mathbf{I}_d)$ can be concisely expressed as
	\begin{equation}
		\label{theta_Gaussian}
		\sE_{\gamma}(\N(m_1, \sigma^2\mathbf{I}_d)\| \N(m_2, \sigma^2\mathbf{I}_d)) = \theta_\gamma\left(\frac{\lVert m_2 - m_1\rVert}{\sigma}\right).
	\end{equation}
	The mapping $r\mapsto \theta_\gamma(r)$ is closely related to 
	the functions $\theta(r)$ by Polyanskiy and Wu \cite[Sec.~2.2]{Yury_Dissipation}, $h(\eta)$ by Balle and Wang in \cite[Lemma~7]{balle2018improving}, and $R_\alpha$ by Feldman et al.\ \cite[Def.~10]{Feldman2018PrivacyAB}.
	%Here, $\theta_\gamma$ has a similar role as: (i) the function $R_\alpha$ introduced by Feldman et al. in \cite{Feldman2018PrivacyAB} and (ii) the function $\theta(r)\coloneqq \mathsf{TV}(P_Z, P_{Z+ru})$, with $u$ being a vector of unit norm and $Z$ being a noise variable with density $P_Z$, introduced in \cite{Yury_Dissipation}.
\end{remark}
%%%%%%%

\subsection{A General Upper Bound for Output $f$-divergence}

The fundamental inequality in \eqref{eq:General_UB_TV2} implies that, for any probability measures $\mu$ and $\nu$,
\begin{equation}
	\label{eq:DfEtaTVDf}
	D_f(\mu\sK\|\nu\sK)\leq \eta_{\tv}(\sK) D_f(\mu \| \nu),
\end{equation}
where $f$ is a convex function with $f(1)=0$. We end this section by proposing an improvement to this inequality which is very effective in combination with Theorem~\ref{Thm:Contraction_EGamma}.

As established in \cite[Proposition 3]{Verdu:f_divergence}, if $f$ is twice differentiable and convex, then $D_f(\mu\|\nu)$ can be expressed in terms of the $\sE_\gamma$-divergence as
\begin{align}
	\nonumber D_f(\mu \Vert \nu) &= \int_{0}^\infty f''(\gamma) \sE_\gamma(\mu \Vert \nu) \text{d}\gamma\nonumber\\
	&= \int_{1}^{\infty} \big[f''(\gamma) \sE_\gamma(\mu \Vert \nu) + \gamma^{-3} f''(\gamma^{-1}) \sE_{\gamma}(\nu \Vert \mu) \big] \text{d}\gamma.\label{Eq:D_f_integral}
\end{align}
This expression and the definition of the contraction coefficient $\eta_{\gamma}(\sK)$ directly yield the following corollary.

\begin{corollary}\label{Cor:UB_Df_general}
	Let $f:(0, \infty)\to \R$ be a twice differentiable convex function with $f(1)=0$. If $\sK:\D\to \P(\W)$ is a Markov kernel, then, for any probability measures $\mu, \nu\in \P(\D)$,
	\begin{align*}
		%D_f(\mu\sK\|\nu\sK) &\leq \int_{0}^\infty \eta_\gamma(\sK)f''(\gamma)\sE_\gamma(\mu\|\nu)\textnormal{d}\gamma,\\
		D_f(\mu\sK\|\nu\sK) &\leq \int_{1}^{\infty} \eta_{\gamma}(\sK) \big[f''(\gamma) \sE_\gamma(\mu \Vert \nu)
		\label{Eq:D_f_integral}  + \gamma^{-3} f''(\gamma^{-1}) \sE_{\gamma}(\nu \Vert \mu) \big] \textnormal{d}\gamma.
	\end{align*}
\end{corollary}

Observe that, by \eqref{eq:General_UB_TV2}, $\eta_{\gamma}(K) \leq \eta_{\tv}(\sK)$ for all $\gamma>0$. Hence, the integral representation \eqref{Eq:D_f_integral} shows that the bound in the previous corollary improves over \eqref{eq:DfEtaTVDf}. To illustrate the magnitude of this improvement, we consider the following two examples. Recall that $\chi^2$-divergence is the $f$-divergence with $f(t) = (t-1)^2$.

\begin{example}\label{example_UB_Df1}
	Let $X\sim \N(0,1)$ and $Y\sim \N(2,1)$. In addition, let $X_\D$ and $Y_\D$ be the projections of $X$ and $Y$ onto the set
	\begin{equation*}
		\D = \left\{x\in \R: |x|\leq 1/2\right\},
	\end{equation*}
	respectively, and denote by $\mu$ and $\nu$ their corresponding distributions.
	
	Let $\sK$ be the $\D$-constrained additive Gaussian kernel. By Proposition~\ref{Proposition:etaPAGK}, $\eta_\tv(\sK) = \theta_1(1)$ and $\eta_\gamma(\sK) = \theta_\gamma(1)$. Thus, \eqref{eq:DfEtaTVDf} implies that
	\begin{align*}
		\chi^2(\mu\sK\|\nu\sK)&\leq \eta_\tv(\sK)\chi^2(\mu\|\nu)\\
		%\leq 2\eta_\tv(\sK)\int_{1}^{\infty} \left[\sE_\gamma(\mu\|\nu) + \gamma^{-3}\sE_\gamma(\nu\|\mu)\right]\textnormal{d}\gamma\\
		& =  2\theta_1\left(1\right)\int_{1}^{\infty} \sE_\gamma(\mu\|\nu) + \gamma^{-3} \sE_\gamma(\nu\|\mu) \textnormal{d}\gamma,
	\end{align*}
	where the equality follows from \eqref{Eq:D_f_integral}. By the data processing inequality,
	\begin{align*}
		\sE_\gamma(\mu\|\nu) &\leq \sE_{\gamma}(\N(0,1) \Vert \N(2,1)) = \theta_{\gamma}(2),\\
		\sE_\gamma(\nu\|\mu) &\leq \sE_{\gamma}(\N(2,1) \Vert \N(0,1)) = \theta_{\gamma}(2).
	\end{align*}
	Therefore, Lemma~\ref{Lemma:EgammaGaussian} implies that
	\begin{align*}
		\chi^2(\mu\sK\|\nu\sK) %&\leq 2\theta_1\left(1\right)\int_{0}^{\infty} \sE_\gamma(\N(0, 1)\|\N(2, 1)) \textnormal{d}\gamma\\
		& \leq 2\theta_1\left(1\right) \int_{1}^{\infty} (1+\gamma^{-3}) \theta_\gamma\left(2\right)\textnormal{d}\gamma\\
		& = 0.49.
	\end{align*}
	Note, however, that Corollary~\ref{Cor:UB_Df_general} implies that
	\begin{align*}
		\chi^2(\mu\sK\|\nu\sK)
		%&\leq 2\int_{1}^{\infty} \eta_\gamma(\sK)\left[\sE_\gamma(\mu\|\nu) + \gamma^{-3}\sE_\gamma(\nu\|\mu)\right]\textnormal{d}\gamma\\
		& \leq  2\int_{1}^{\infty} \theta_\gamma\left(1\right) \big[\sE_\gamma(\mu\|\nu) + \gamma^{-3} \sE_\gamma(\nu\|\mu)\big] \textnormal{d}\gamma.
	\end{align*}
	As before, the data processing inequality and Lemma~\ref{Lemma:EgammaGaussian} lead to
	\begin{align*}
		\chi^2(\mu\sK\|\nu\sK)
		%&\leq 2\int_{1}^{\infty} \eta_\gamma(\sK)\left[\sE_\gamma(\mu\|\nu) + \gamma^{-3}\sE_\gamma(\nu\|\mu)\right]\textnormal{d}\gamma\\
		%&\leq 2\int_{1}^{\infty} \theta_\gamma\left(1\right)\sE_\gamma(\N(0, 1)\|\N(2, 1)) \textnormal{d}\gamma\\
		&\leq 2\int_{1}^{\infty} (1+\gamma^{-3})\theta_\gamma\left(1\right)\theta_\gamma\left(2\right) \textnormal{d}\gamma\\
		&= 0.26.
	\end{align*}
	In conclusion, the bound for $\chi^2(\mu\sK\|\nu\sK)$ obtained from \eqref{eq:DfEtaTVDf} is almost twice the corresponding bound obtained from Corollary~\ref{Cor:UB_Df_general}.
\end{example}

\begin{example}\label{example_UB_Df2}
	Let $\sK$ be the binary-input binary-output channel with crossover probabilities $a, b\in [0,\frac{1}{2}]$, i.e.,
	$$\sK= \begin{bmatrix}
		\bar a& a\\
		b& \bar b
	\end{bmatrix}.$$
	Theorem~\ref{Thm:Contraction_EGamma} implies that, for all $\gamma\geq1$,
	\begin{equation*}
		\eta_\gamma(\sK) = \max\{(\bar a-\gamma b)^+, (\bar b-\gamma a)^+\},
	\end{equation*}
	and, in particular,
	\begin{equation*}
		\eta_\tv(\sK) = 1-a-b.
	\end{equation*}
	Thus, \eqref{eq:DfEtaTVDf} implies that, for any binary probability measures $\mu$ and $\nu$,
	\begin{align*}
		\chi^2(\mu\sK\|\nu\sK)&\leq   
		(1-a-b) \chi^2(\mu\|\nu)\\
		& = 2(1-a-b)\int_{1}^{\infty} \big[\sE_\gamma(\mu\|\nu) + \gamma^{-3} \sE_{\gamma}(\nu\Vert\mu)\big]\textnormal{d}\gamma,
	\end{align*}
	where the equality comes from the integral representation \eqref{Eq:D_f_integral}. In particular, for $a = 0.1$, $b=0.4$, $\mu = \sBer(0.1)$, and $\nu = \sBer(0.4)$, we obtain
	\eq{\chi^2(\mu\sK\|\nu\sK)\leq 0.19.}
	% \begin{align}D_f(\mu\sK_{\mathsf{BSC}}^{\tau}\|\nu\sK_{\mathsf{BSC}}^{\tau})&\leq (1-2\tau)D_f(\mu\|\nu)\nonumber\\
		% &= (1-2\tau)\int_{1}^{\infty} \left[f''(\gamma)\sE_\gamma(\mu\|\nu) + \gamma^{-3}f''(\gamma^{-1})\sE_\gamma(\nu\|\mu)\right]\textnormal{d}\gamma, \label{SDPI_General1}
		% \end{align}
	% where the equality comes from \eqref{Eq:D_f_integral}. 
	On the other hand, Corollary~\ref{Cor:UB_Df_general} renders \begin{align*}
		\chi^2(\mu\sK\|\nu\sK)&\leq  2\int_{1}^{\frac{\bar b}{a} \vee \frac{\bar a}{b}} \left[(\bar{a}-\gamma b)^+ \vee (\bar{b}-\gamma a)^+\right] \big[\sE_\gamma(\mu\|\nu) + \gamma^{-3} \sE_\gamma(\nu\|\mu)\big] \textnormal{d}\gamma.
		%&\leq  2(1-a-b)\int_{1}^{\max\{\frac{\bar b}{a}, \frac{\bar a}{b}\}} \left[\sE_\gamma(\mu\|\nu) + \gamma^{-3}\sE_\gamma(\nu\|\mu)\right]\textnormal{d}\gamma
	\end{align*}
	For our particular choice of $a$, $b$, $\mu$, and $\nu$, we obtain
	\eq{\chi^2(\mu\sK\|\nu\sK)\leq 0.17.}
	Hence, as in the previous example, Corollary~\ref{Cor:UB_Df_general} produces a tighter estimate for $\chi^2(\mu\sK\|\nu\sK)$ than \eqref{eq:DfEtaTVDf}.
\end{example}
%%%%%%%

	\section{LDP as the Contraction of $\sE_\gamma$-Divergence}
\label{Sec:LDP}

In this section we establish the equivalence between the notions of local differential privacy (LDP) and contraction under $\sE_{\gamma}$-divergence. We apply this equivalence in a plug-and-play manner to obtain estimates for the cost of privacy in the minimax risk, Bayesian risk, and binary hypothesis testing settings. Motivated by the equivalence between LDP and contraction under $\sE_{\gamma}$-divergence, we obtain these estimates by bounding the cost of privacy directly in terms of $\sE_{\gamma}$-divergence.

A privacy mechanism is a (potentially random) mapping from a given set $\D$ into another set $\W$. As a result, it is customary to represent privacy mechanisms as Markov kernels $\sK:\D\to\P(\W)$. In this context, LDP can be expressed as follows.

\begin{definition}[\cite{evfimievski2003limiting,Shiva_subsampling}]\label{Def:LDP}
Let $\eps\geq 0$ and $\delta\in [0,1]$. A privacy mechanism $\sK:\D\to\P(\W)$ is $(\eps,\delta)$-LDP if 
\begin{equation*}
    \sup_{x_{1},x_{2}\in\D} \sup_{A\subset\W} \left[\sK(A \vert x_{1})-e^{\eps}\sK(A \vert x_{2})\right] \leq \delta.
\end{equation*}
We let $\Q_{\eps,\delta}$ be the family of all Markov kernels that are $(\eps,\delta)$-LDP. An $(\eps,0)$-LDP mechanism is called $\eps$-LDP. \end{definition}

Observe that, by \eqref{Defi_HS_Divergence}, a Markov kernel $\sK$ is $(\eps,\delta)$-LDP if and only if
\begin{equation*}
    \sup_{x_{1},x_{2}\in\D} \sE_{\gamma}(\sK(\cdot \vert x_{1}) \Vert \sK(\cdot \vert x_{2})) \leq \delta.
\end{equation*}
Thus, an immediate application of Theorem~\ref{Thm:Contraction_EGamma} shows that $(\eps, \delta)$-LDP is \textit{equivalent} to contraction under $\sE_\gamma$-divergence.

\begin{theorem}
\label{thm:LDP_Contraction}
A mechanism $\sK$ is $(\eps, \delta)$-LDP if and only if  $\eta_{e^\eps}(\sK)\leq \delta$, i.e.,
\begin{equation*}
    \sE_{e^\eps}(\mu\sK\|\nu\sK)\leq \delta \sE_{e^\eps}(\mu\|\nu), \quad \forall \mu,\nu.
\end{equation*}
\end{theorem}

We note that Duchi et al.\ \cite{Duchi_LDP_MinimaxRates} showed that if $\sK$ is $\eps$-LDP then
\begin{equation}
\label{eq:DuchiObservation}
    D_{\kl}(\mu\sK\|\nu\sK)\leq 2 (e^\eps-1)^2\tv^2(\mu,\nu).
\end{equation}
From this result, they concluded that $\eps$-LDP acts as a contraction on the space of probability measures. Theorem \ref{thm:LDP_Contraction} makes this observation precise by showing that $\eps$-LDP is in fact equivalent to contraction under $\sE_{e^{\eps}}$-divergence. Indeed, according to Theorem \ref{thm:LDP_Contraction}, a privacy mechanism $\sK$ is $\eps$-LDP if and only if $\sE_{e^{\eps}}(\mu\sK\|\nu\sK) =0$ for any distributions $\mu$ and $\nu$. An example of a Markov kernel satisfying the latter property is the randomized response mechanism, as described next.

\begin{example}\label{example:RRMechanism}
Let $\D=\W=\{0,1\}$ and $\eps\geq0$. The randomized response mechanism \cite{warner1965randomized}, denoted by $\sK^{\eps}_{\mathsf{RR}}$, is the mechanism implemented by the binary symmetric channel with crossover probability $\omega_\eps \coloneqq (1+e^\eps)^{-1}$. It is known that $\sK^{\eps}_{\mathsf{RR}}$ is $\eps$-LDP. For the sake of illustration, we verify this fact below using Theorem~\ref{thm:LDP_Contraction}.

Let $\mu = \sBer(p)$ and $\nu=\sBer(q)$ with $p,q\in [0,1]$. Observe that $\mu\sK^\eps_{\mathsf{RR}} = \sBer(p*\omega_\eps)$ and $Q\sK^\eps_{\mathsf{RR}} = \sBer(q*\omega_\eps)$ where
\begin{equation*}
    a*b \coloneqq a(1-b) + (1-a)b.
\end{equation*}
It is straightforward to verify that, for any $p,q$,
\begin{align*}
    p*\omega_\eps-e^\eps q*\omega_\eps &\leq 0,\\
    1-p*\omega_\eps-e^\eps (1-q*\omega_\eps) &\leq 0.
\end{align*}
The definition of the $\sE_{\gamma}$-divergence \eqref{eq:DefEgamma} implies that
\begin{equation*}
    \sE_{e^\eps}(\mu\sK^\eps_{\mathsf{RR}}\|\nu\sK^\eps_{\mathsf{RR}}) = 0.
\end{equation*}
Thus, by Theorem~\ref{thm:LDP_Contraction}, we conclude that $\sK^\eps_{\mathsf{RR}}$ is $\eps$-LDP.

A generalization of this mechanism for $|\D| = k \geq 2$ has been reported in the literature (see, e.g., \cite{kairouz16_LDPEstimation, kairouz2014extremal_JMLR}). Specifically, the so-called $k$-ary randomized response mechanism $\sK^{\eps}_{\mathsf{kRR}}:\D\to\P(\D)$ is defined by
\begin{equation*}
    \sK^{\eps}_{\mathsf{kRR}}(z \vert x) = \begin{cases}\frac{e^\eps}{k-1+e^\eps} & z = x,\\ \frac{1}{k-1+e^\eps} & z \neq x.\end{cases}
\end{equation*}
Using an argument similar to the one in the paragraph above, the reader can verify that $\sK^{\eps}_{\mathsf{kRR}}$ is $\eps$-LDP.
\end{example}

For each $\eps\geq0$ and $\delta\in[0,1]$, we define
\begin{equation*}
    \varphi(\eps,\delta) \coloneqq 1-(1-\delta)e^{-\eps}.
\end{equation*}
By combining \eqref{eq:ContractionDfEtaGamma} and Theorem~\ref{thm:LDP_Contraction}, we conclude that if $\sK$ is $(\eps,\delta)$-LDP, i.e., $\sK\in \Q_{\eps, \delta}$, then
\begin{equation}
\label{eq:LDPImpliesContractionDf}
    D_f(\mu\sK\|\nu\sK)\leq \varphi(\eps, \delta) D_f(\mu\|\nu) \qquad \forall \mu,\nu.
\end{equation}
This direct consequence of Theorem~\ref{thm:LDP_Contraction} reveals a remarkable structural property of LDP mechanisms: any LDP mechanism is contractive under \textit{all} $f$-divergences. 
% Consequently, an  \fc{[I would consider removing the next statement. Note that \eqref{eq:DuchiObservation} involves mixed divergences.... perhaps we can derive a similar result by apply (19) and Pinsker's? ] discussed after \eqref{eq:DuchiObservation}.} \MD{I'm fine removing it, we are emphasizing the (rather straightforward) bound of Duchi et al.\ a bit too much.}
%Recall that \eqref{eq:LDPImpliesContractionDf} holds for every $f$-divergences and every  $(\eps, \delta)$-LDP mechanism. Nonetheless, this bound can be improved if one considers a particular $f$-divergence or mechanism. For instance, it is known that $\eta_\kl(\mathsf{BSC}(\omega)) = (1-2\omega)^{2}$ (see, e.g., \cite{ahlswede1976}). Thus, for the randomized response mechanism $\sK^\eps_{\mathsf{RR}}$ introduced in Example~\ref{example:RRMechanism}, we have that
% \begin{equation*}
%     \eta_\kl(\sK^\eps_{\mathsf{RR}}) = \left(\frac{e^\eps-1}{e^\eps+1}\right)^2.
% \end{equation*}
%The latter expression improves over $\eta_\kl(\sK^\eps_{\mathsf{RR}})\leq 1-e^{-\eps}$, as obtained from \eqref{eq:LDPImpliesContractionDf}. Needless to say, $\eta_\kl$ is difficult to compute in closed form for general Markov kernels, in which case   \eqref{eq:LDPImpliesContractionDf} provides a useful alternative. 

In the sequel we focus on two multi-user settings which are common in the literature. Assume there are $n$ users, each in possession of a datapoint $X_i\in\mathcal{X}$, $i\in [n]$. Furthermore, assume that the users wish to apply a mechanism $\sK_i$ that generates a privatized version of $X_i$, denoted by $Z_i\in\mathcal{Z}_{i}$.

\subsubsection{Non-interactive setting} The collection of mechanisms $\{\sK_i\}$ is  said to be \textit{non-interactive} if the distribution of $Z_{i}$ is entirely determined by $X_i$ and independent of $(X_j,Z_j)$ for $j\neq i$. When all users apply the same mechanism $\sK$, we can view $Z^n \coloneqq (Z_1, \ldots, Z_n)$ as independent applications of $\sK$ to each $X_i$. In this case, the overall mechanism is the $n$-fold tensor power of $\sK$, denoted by $\sK^{\otimes n}$.

\subsubsection{Sequentially interactive setting}
The collection of mechanisms $\{\sK_i\}$ is  said to be \textit{sequentially interactive} \cite{Duchi_LDP_MinimaxRates} if the distribution of $Z_{i}$ depends on $X_{i}$, the datapoint in possession of user $i$, and $Z_{1},\ldots,Z_{i-1}$, the output of the $i-1$ previous mechanisms. Note that $K_{i}$ is a Markov kernel with domain $\mathcal{D} = \mathcal{X}\times\mathcal{Z}_{1}\times\cdots\times\mathcal{Z}_{i-1}$.
In this case, we denote the overall mechanism by $\sK^n$.

Next, we  extend \eqref{eq:LDPImpliesContractionDf} for non-interactive mechanisms. For each $n\in\mathbb{N}$, $\eps\geq0$, and $\delta\in[0,1]$, we define
\begin{equation*}
    \varphi_{n}(\eps,\delta) \coloneqq 1-e^{-n\eps}(1-\delta)^{n}.
\end{equation*}
Fix an $(\eps, \delta)$-LDP mechanism $\sK$ and consider the corresponding non-interactive mechanism $\sK^{\otimes n}$. By combining Lemma~\ref{lemma:UB_Eta_Kn_general} and Theorem~\ref{thm:LDP_Contraction}, we obtain that
\begin{equation*}
    \eta_f(\sK^{\otimes n})\leq \varphi_n(\eps, \delta),
\end{equation*}
where $f$ is any convex function with $f(1)=0$. From this inequality and \eqref{eq:ContractionDfEtaGamma}, we conclude that
\begin{equation}
\label{eq:LDPImpliesContractionDf_n}
    D_f(\mu\sK^{\otimes n}\|\nu\sK^{\otimes n})\leq \varphi_{n}(\eps, \delta) D_f(\mu\|\nu) \qquad \forall \mu,\nu,
\end{equation}
which generalizes \eqref{eq:LDPImpliesContractionDf} as desired. In the rest of this section we rely on the results derived so far to estimate the cost of local differential privacy in some statistical settings.
%%%%%%%

\subsection{Private Minimax Risk}
\label{Sec:ApplicationsMinimax}

Let $\P \subseteq \P(\X)$ be a family of distributions, let $\vartheta$ be a parameter space, and let $\theta:\P\to\vartheta$ be a function assigning a parameter to each distribution in $\mathcal{P}$. Also, let $X^n = (X_1,\dots, X_n)$ be independent and identically distributed (i.i.d.) samples drawn from a distribution $P$ with parameter $\theta(P)$. We assume that each user possesses a sample $X_i$ and applies a sequentially interactive privacy-preserving mechanism $\sK_i$ to obtain $Z_i$. Given the sequences $\{Z_i\}_{i=1}^n$, the goal is to  estimate $\theta(P)$ through an estimator $\Psi:\Z^n\to\vartheta$.  
The quality of such estimator is assessed by a semi-metric $\ell:\vartheta\times\vartheta\to\R_+$ and is used to define the private minimax risk
\begin{equation*}
    \mathcal R_n(\P, \ell, \eps, \delta)\coloneqq \inf_{\sK_{i}\in\Q_{\eps, \delta}}\inf_{\Psi}\sup_{P\in \P}\E[\ell(\Psi(Z^n), \theta(P))],
\end{equation*}
where the first infimum is taken over all $\sK_{1},\ldots,\sK_{n}$ which are $(\eps,\delta)$-LDP. The quantity $R_n(\P, \ell, \eps, \delta)$ uniformly characterizes the optimal rate of private statistical estimation over the family $\P$ using the best possible estimator and privacy-preserving mechanisms in $\Q_{\eps, \delta}$. In the absence of privacy constraints (i.e., $Z^n = X^n$), we denote the minimax risk by $\mathcal R_n(\P, \ell)$. 

A typical first step in deriving information-theoretic lower bounds for the minimax risk is to reduce the above estimation problem to a testing problem via Le Cam's, Fano's, or Assouad's method \cite{Yu1997, Barron_ITMinimax, Tsybakov_Book}. For ease of exposition, in the sequel we focus only on Le Cam's method, which relies on binary hypothesis testing, although a similar reasoning could be applied to multiple hypothesis testing settings (e.g., Fano's and Assouad's methods). The canonical binary hypothesis testing problem is defined as follows: Nature chooses a random variable $V$ uniformly at random from $\{0,1\}$, then, conditioned on $V = v$, the samples $X^n$ are drawn i.i.d.\ from $P_v\in\mathcal{P}$, denoted by $X^n\sim P^{\otimes n}_v$. It is well-known \cite{Yu1997, Barron_ITMinimax, Tsybakov_Book} that if $\ell(\theta(P_{0}), \theta(P_{1}))\geq 2\tau$ for some $\tau>0$, then
\begin{equation*}
    \mathcal R_n(\P, \ell)\geq \tau \mathsf{P}_{\mathsf e}(V|X^n),
\end{equation*}
where $\mathsf{P}_{\mathsf e}(V|X^n)$ denotes the probability of error in guessing $V$ given $X^n$. In its simplest form, Le Cam's method relies on the inequality
\begin{equation*}
    \mathsf{P}_{\mathsf{e}}(V|X^n) \geq \frac{1}{2} \left[1-\tv(P^{\otimes n}_0, P^{\otimes n}_1)\right],
\end{equation*}
see, e.g., \cite[Lemma 1]{Yu1997} or \cite[Theorem 2.2]{Tsybakov_Book}, which yields the following lower bound for the minimax risk
\aln{
\mathcal R_n(\P, \ell)&\geq \frac{\tau}{2} \left[1-\tv(P^{\otimes n}_0, P^{\otimes n}_1)\right]\label{LeCam_TV}\\
& \geq \frac{\tau}{2}\left[1-\sqrt{\frac{n}{2} D_\kl(P_0\| P_1)}\right], \label{LeCam_KL}}
where the second inequality follows from Pinsker's inequality\footnote{Observe that Pinsker's inequality is ineffective when $D_\kl(P_0\| P_1)$ is sufficiently large. In that case, Pinsker's inequality could be replaced by the Bretagnolle-Huber inequality \cite{bretagnolle1978estimation}
\begin{equation*}
    \tv(P_{0},P_{1}) \leq \sqrt{1 - e^{-D_{\kl}(P_{0} \Vert P_{1})}},
\end{equation*}
or Vajda's inequality \cite{vajda1970note}
\begin{equation*}
\setlength{\belowdisplayskip}{0pt}
    \log\left(\frac{1+\tv(P_{0},P_{1})}{1-\tv(P_{0},P_{1})}\right) - \frac{2\tv(P_{0},P_{1})}{1+\tv(P_{0},P_{1})} \leq D_{\kl}(P_{0} \Vert P_{1}).
\end{equation*}} and the chain rule for KL-divergence.

In the presence of privacy, the estimator $\Psi$ depends on $Z^n$ which is generated upon $X^n$ by a sequentially interactive mechanism $\sK^n$. To derive the private counterpart of \eqref{LeCam_TV}, we need to replace $P^{\otimes n}_{v}$, the marginal distribution of $X^{n}$ conditioned on $V = v$, with $P^{\otimes n}_v\sK^n$, the marginal distribution of $Z^n$ conditioned on $V=v$. A lower bound for $\mathcal R_n(\P, \ell, \eps, \delta)$ is therefore obtained  by deriving an upper bound for 
$\tv(P_0^{\otimes n}\sK^n,P_1^{\otimes n}\sK^n)$ for all $\sK^n\subset \Q_{\eps, \delta}$. This strategy is implemented in the following lemma, whose proof could be found in Appendix~\ref{Appendix:ProofLemmaLB_Minimax}.

\begin{lemma}\label{Lem:LB_Minimax}
If $P_0, P_1\in \P$ satisfy $\ell(\theta(P_0), \theta(P_1))\geq 2\tau$ for some $\tau>0$, then
\begin{equation*}
    \mathcal R_n(\P, \ell, \eps, \delta)\geq \frac{\tau}{2}\left[1-\sqrt{\frac{\varphi(\eps, \delta) n}{2}D_\kl(P_0\| P_1)}\right],
\end{equation*}
where $\varphi(\eps,\delta)\coloneqq 1-(1-\delta)e^{-\eps}$.
\end{lemma}

By comparing the previous lemma with the original non-private Le Cam's method \eqref{LeCam_KL}, we observe that the effect of $(\eps, \delta)$-LDP is to reduce the effective sample size from $n$ to $(1-(1-\delta)e^{-\eps})n$. Setting $\delta =0$ and $\eps<0.224$, this result strengthens Duchi et al. \cite[Corollary 2]{Duchi_LDP_MinimaxRates}, where the effective sample size was shown to be $4\eps^2 n$ for sufficiently small $\eps$.
%\fc{and $\epsilon>0.224$} \MD{I think the sign $>$ is the other way around, a bigger factor is better as the effective sample size is reduced less.}, 
%$4\eps^2 n$ for $\eps\leq 1$ [Should this be 1/2?].  

\begin{example}[(1-dimensional mean estimation)]
For some $k > 1$, we assume that $\P = \P_k$ is given by
$$\P_k \coloneqq \left\{P\in \P(\X):~|\E_P[X]|\leq 1, \E_P[|X|^k]\leq 1\right\}.$$
Consider the problem of estimating $\theta(P) = \E_P[X]$ when $\ell = \ell_2^2$, i.e., the loss is given by the the squared $\ell_2$ metric. This problem was first studied by Duchi et al.\ in  \cite[Prop.~1]{Duchi_LDP_MinimaxRates} where it was shown that $\mathcal R_n(\P_k, \ell_2^2, \eps, 0)\geq (\eps^2n)^{-(k-1)/k}$ for $\eps\leq 1$. Applying our framework to this example, we obtain a similar lower bound that holds for all $\eps\geq 0$ and $\delta\in [0,1]$, stated below.

\begin{corollary}\label{corollary:LB_Pk}
For the $1$-dimensional mean estimation problem, for all $k> 1$, $\eps\geq 0$, and $\delta\in (0,1)$,
\begin{align}
\mathcal R_n(\P_k, \ell_2^2, \eps, \delta) &\gtrsim \min\Big\{1,\left[\varphi^2(\eps, \delta)n\right]^{-\frac{k-1}{k}}\Big\}.\label{LB_Example}
\end{align}
\end{corollary}

It is worth instantiating this corollary, whose proof can be found in Appendix~\ref{Appendix:ProofCorollaryLB_Pk}, for some special values of $k$. Consider first the usual finite variance setting, i.e., $k=2$. In the non-private case, it is known that the sample mean has mean-squared error that scales as $n^{-1}$. According to Corollary~\ref{corollary:LB_Pk}, this rate worsens to $(\varphi(\eps, \delta)\sqrt{n})^{-1}$ in the presence of an $(\eps, \delta)$-LDP requirement. Now, consider the limiting setting in which $k\to\infty$. Observe that the moment condition $\E_p[|X|^k]\leq 1$ implies the boundedness of $X$. In this case, the non-private minimax risk scales as $n^{-1}$, while Corollary~\ref{corollary:LB_Pk} implies that its LDP counterpart scales as $(\varphi^2(\eps, \delta)n)^{-1}$. 
\end{example}

We end this section with an alternative to Lemma~\ref{Lem:LB_Minimax} in the non-interactive setting. Its proof, which can be found in Appendix~\ref{Appendix:ProofLemmaLB_Minimax_Alternative}, relies on Lemma~\ref{lemma:UB_Eta_Kn_general} and leads to tighter bounds for certain values of $n$, $\eps$, and $\delta$.

\begin{lemma}
\label{Lem:LB_Minimax_Alternative}
If $P_0, P_1\in \P$ satisfy $\ell(\theta(P_0), \theta(P_1))\geq 2\tau$ for some $\tau>0$, then
\begin{equation*}
    \mathcal R_n(\P, \ell, \eps, \delta)\geq \frac{\tau}{2}\left[1-\sqrt{\frac{\varphi^2_{n}(\eps,\delta)n}{2} D_\kl(P_0\| P_1)}\right],
\end{equation*}
where $\varphi_{n}(\eps,\delta) \coloneqq 1-e^{-n\eps}(1-\delta)^{n}$.
\end{lemma}
%%%%%%%

\subsection{Private Bayesian Risk}
\label{Sec:ApplicationsBayesian}

In the minimax setting, the worst-case parameter is considered which usually leads to over-pessimistic bounds. In practice, the parameter that incurs a worst-case risk may appear with very small probability. To capture this prior knowledge, it is reasonable to assume that the true parameter is sampled from an underlying prior distribution. In this case, we are interested in the \textit{Bayesian risk} of the problem as described next. Let $\vartheta$ be a parameter space endowed with a prior distribution $P_{\Theta}$ and let $\P=\{P_{X|\Theta}(\cdot|\theta):\theta\in \vartheta\}$ be a collection of parametric probability distributions over $\X$. Given an i.i.d.\ sequence $X^n$ drawn from $P_{X|\Theta}$, the goal is to estimate $\Theta$ from a privatized sequence $Z^n$ via an estimator $\Psi:\Z^n\to\vartheta$. Throughout this section, we focus on the non-interactive setting. In this case, the private Bayesian risk is defined as
\eqn{}{R_n^{\mathsf{Bayes}}(P_\Theta, \ell, \eps, \delta) \coloneqq \inf_{\sK\in \Q_{\eps, \delta}}\inf_{\Psi}\E[\ell(\Theta, \Psi(Z^n))],} 
where the expectation is taken with respect to the randomness of both $\Theta \sim P_{\Theta}$ and $Z^n$. It is evident that $R_n^{\mathsf{Bayes}}(P_\Theta, \ell, \eps, \delta)$ must depend on the prior $P_\Theta$. One way to quantify this dependence is through the so-called \emph{small ball probability} of $\Theta$ with respect $\ell$ defined as
\begin{equation*}
    \L(\zeta)\coloneqq \sup_{\theta\in \vartheta}\Pr(\ell(\Theta, \theta)\leq \zeta).
\end{equation*}

Xu and Raginsky \cite{Raginsky_ISIT_converses} showed that the non-private Bayesian risk ($Z^n = X^n$), denoted by $R_n^{\mathsf{Bayes}}(P_\Theta, \ell)$, is lower bounded as 
\begin{equation}
\label{eq:XuRaginskyRisk}
    R_n^{\mathsf{Bayes}}(P_\Theta, \ell)\geq \sup_{\zeta>0} \zeta\left[1-\frac{I(\Theta; X^n) + \log 2}{\log(1/\L(\zeta))}\right].
\end{equation}
By replacing $I(\Theta; X^n)$ with $I(\Theta; Z^n)$ in the previous inequality, an application of \eqref{eq:LDPImpliesContractionDf_n} directly leads to the following lower bound for $R_n^{\mathsf{Bayes}}(P_\Theta, \ell, \eps, \delta)$.

\begin{corollary}
\label{Corollary:Private_Raginsky}
In the non-interactive setting, the private Bayesian risk $R_n^{\mathsf{Bayes}}(P_\Theta, \ell, \eps, \delta)$ is bounded below by
\begin{equation}
\label{Eq:Private_Raginsky}
    \sup_{\zeta>0} \zeta\left[1-\frac{\varphi_n(\eps, \delta)I(\Theta; X^n) + \log 2}{\log(1/\L(\zeta))}\right],
\end{equation}
where $\varphi_{n}(\eps,\delta) \coloneqq 1-e^{-n\eps}(1-\delta)^{n}$.
\end{corollary}

The following theorem, whose proof can be found in Appendix~\ref{Appendix:ProofTheoremBayesian_LB}, provides a lower bound for the private Bayesian risk that directly involves $\sE_\gamma$-divergence, leading to a tighter bound than \eqref{Eq:Private_Raginsky}. For $\gamma\geq 0$ and any pair of random variables $(A, B)\sim P_{AB}$ with marginals $P_A$ and $P_B$, we define their $E_\gamma$-information as 
$$I_\gamma(A; B)\coloneqq \sE_\gamma(P_{AB}\|P_AP_B).$$ 

\begin{theorem}\label{Thm:Bayesian_LB}
In the non-interactive setting, the private Bayesian risk $R_n^{\mathsf{Bayes}}(P_\Theta, \ell, \eps, \delta)$ is bounded below by
\begin{equation*}
    \sup_{\zeta>0}\zeta \left[1- \varphi_n(\eps, \delta)I_{e^\eps}(\Theta; X^n)-e^\eps\L(\zeta)\right].
\end{equation*}
Furthermore, for $n=1$ we have that
\begin{equation*}
    R_1^{\mathsf{Bayes}}(P_\Theta, \ell, \eps, \delta)\geq \sup_{\zeta>0}\zeta \left[1-\delta I_{e^\eps}(\Theta; X)-e^\eps\L(\zeta)\right].
\end{equation*}
\end{theorem}

We compare the bound in Theorem~\ref{Thm:Bayesian_LB} with  those of Corollary~\ref{Corollary:Private_Raginsky} in the next example.

\begin{example}\label{Example_UniformTheta}
Suppose $\Theta$ is uniformly distributed on $[0,1]$,  $P_{X|\Theta=\theta}=\sBer(\theta)$, and $\ell(\theta, \theta')=|\theta-\theta'|$. Observe that in this case $\L(\zeta) \leq \min\{2\zeta, 1\}$. For ease of notation, let $\gamma = e^\eps$. Note that
\begin{equation*}
    I_\gamma(\Theta; X^n)= \int_{0}^1 \sE_\gamma(P_{X^n|\Theta=\theta}\|P_{X^n})\text{d}\theta.
\end{equation*}
A routine calculation shows that, for any $\theta\in [0,1]$,
\begin{align*}
    P_{X^n|\Theta=\theta}(x^n) &= \theta^{s(x^n)}(1-\theta)^{n-s(x^n)},\\
    P_{X^n}(x^n) &= \frac{s(x^n)!(n-s(x^n))!}{(n+1)!},
\end{align*}
where $s(x^n)$ is the number of 1's in $x^n$. Given these marginal and conditional distribution, one can obtain that
\begin{equation*}
    I_\gamma(\Theta; X^n) = \sum_{s=0}^n\int_{0}^1 \left[\frac{n!\theta^{s}(1-\theta)^{n-s}}{s!(n-s)!} - \frac{\gamma}{n+1} \right]_+\text{d}\theta.
\end{equation*}
Plugging this equation into Theorem~\ref{Thm:Bayesian_LB}, we arrive at a maximization problem that can be numerically solved. Similarly, we compute
\begin{equation*}
    I(\Theta; X^n) = \int_{0}^1D_\kl(P_{X^n|\Theta=\theta}\|P_{X^n})\text{d}\theta,
\end{equation*}
plug it into Corollary~\ref{Eq:Private_Raginsky}, and numerically solve the resulting optimization problem. In Figure~\ref{fig:LDP_Uniform}, we compare these two lower bounds for $\delta = 10^{-4}$ and $n= 20$. Observe the significant advantage of Theorem~\ref{Thm:Bayesian_LB} against Corollary~\ref{Eq:Private_Raginsky} for small $\eps$.

\begin{figure}
    \centering
    \includegraphics[scale=0.5]{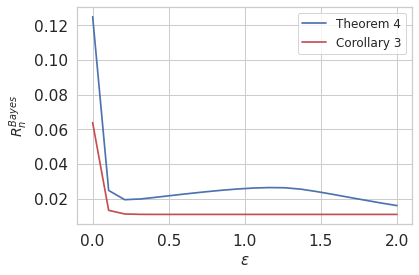}
    \caption{Comparison of the lower bounds obtained from Theorem~\ref{Thm:Bayesian_LB} and the private version of \cite[Theorem 1]{Raginsky_ISIT_converses} described in Corollary~\ref{Corollary:Private_Raginsky} for Example~\ref{Example_UniformTheta} with $\delta = 10^{-4}$ and $n= 20$.}
    \label{fig:LDP_Uniform}
\end{figure}
\end{example}

\begin{remark}
Although we are mainly interested in the private Bayesian risk, it is possible to obtain from the proof of Theorem~\ref{Thm:Bayesian_LB} that the non-private Bayesian risk $R_n^{\mathsf{Bayes}}(P_\Theta, \ell)$ is bounded below by
\eqn{Bayes_LB2}{\sup_{\gamma\geq0} \sup_{\zeta>0} \zeta \left[1-I_{\gamma}(\Theta; X^n)-\gamma\L(\zeta)-(1-\gamma)_+\right].}
In order to compare the previous bound with is non-private counterpart \eqref{eq:XuRaginskyRisk}, we consider the following example.   
Suppose $\Theta$ is uniformly distributed on $[0,1]$, $P_{X|\Theta = \theta}=\sBer(\theta)$, and $\ell(\theta, \theta') = |\theta-\theta'|$. It can be shown that $I(\Theta; X) = 0.19$ nats while 
\begin{equation*}
    I_\gamma(\Theta; X) = \begin{cases} 0.25\gamma^2 & \text{if}~\gamma\in [0,1],\\
0.25(\gamma-2)^2 & \text{if}~\gamma\in [1,2],\\
0& \text{otherwise}.
\end{cases}
\end{equation*}
As in Example~\ref{Example_UniformTheta}, it can be verified that \eqref{eq:XuRaginskyRisk} produces the lower bound $R_1^{\mathsf{Bayes}}(P_\Theta, \ell_1)\geq 0.03$, whereas our bound \eqref{Bayes_LB2} yields $R_1^{\mathsf{Bayes}}(P_\Theta, \ell_1)\geq 0.08$.
\end{remark}

\subsection{Private Binary Hypothesis Testing}
\label{Sec:ApplicationsHypothesis}

We now turn our attention to the well-known problem of binary hypothesis testing under local differential privacy constraints. 
Assume that we observe $n$ i.i.d. samples $X^n$ drawn from a distribution $Q\in \P(\X)$. We assume further that each $X_i$ is mapped to $Z_i$ via a privacy-preserving mechanism $\sK\in \Q_{\eps, \delta}$, i.e., we assume a non-interactive setting with $\sK_{i} = \sK$. Given $Z^{n}$, the goal is to distinguish between the null hypothesis $H_0: Q = P_0$ and the alternative hypothesis $H_1: Q = P_1$. Let $T$ be a binary statistic generated from a randomized decision rule $P_{T|Z^n}:\Z^n \to \mathcal P(\{0,1\})$, where $T = 1$ indicates that $H_0$ is rejected. The type I and type II error probabilities corresponding to this statistic are given by $\Pr(T = 1|H_0)$ and $\Pr(T = 0|H_1)$, respectively. To capture the optimal tradeoff between type I and type II error probabilities, it is customary to define
\begin{equation*}
    \beta^{\eps, \delta}_n(\alpha)\coloneqq \inf_{\sK\in\Q_{\eps, \delta}} \inf_{\substack{P_{T|Z^n} :\\ \Pr(T = 1|H_0)\leq \alpha}} \Pr(T = 0|H_1),
\end{equation*}

The following corollary, whose proof relies on \eqref{eq:LDPImpliesContractionDf} and can be found in Appendix~\ref{Appendix:ProofCorollaryBHT}, provides an asymptotic lower bound for $\beta_n^{\eps, \delta}(\alpha)$.

\begin{corollary}\label{Cor:BHT}
For any $\eps\geq 0$ and $\delta\in [0,1]$, we have that
\eqn{}{
\lim_{n\to\infty}\frac{1}{n}\log\beta_n^{\eps, \delta}(\alpha)\geq -\varphi(\eps, \delta)D_\kl(P_0\|P_1).}
\end{corollary}

Kairouz et al. \cite[Sec.~3]{kairouz2014extremal_JMLR} proved a similar result which is optimal for sufficiently small (albeit unspecified) $\eps$ and $\delta = 0$. Recall that the Chernoff-Stein lemma \cite[Thm.~11.8.3]{cover2012elements} establishes that $D_\kl(P_0\|P_1)$ is the asymptotic exponential decay rate of $\beta_{n}(\alpha)$ in the absence of privacy constraints. Thus, the above corollary exhibits, once again, a reduction of the effective sample size from $n$ to $\varphi(\eps, \delta)n$ in the presence of an $(\eps, \delta)$-LDP constraint.

We end this section with a remark. In the spirit of Theorem~\ref{Thm:Bayesian_LB}, it is possible to investigate $\alpha\mapsto \beta^{\eps,\delta}_n(\alpha)$ directly in terms of $\sE_{\gamma}$-divergence, for instance, by considering the dual representation of $\sE_\gamma$-divergence. This method can potentially lead to a tighter lower bound than Corollary~\ref{Cor:BHT} and is left as future work. The details of this approach for the non-private case can be found in the pioneering work of \cite{polyanskiy2010channel}.
% \SA{It is known that 
% $$\sE_\gamma(P_0\|P_1) = \alpha - \gamma\beta(\alpha),$$
% where $\gamma$ is such that $\alpha = P_0(\frac{dP_0}{dP_1}\geq \gamma)$ and $\beta(\alpha)$ denotes the type-II error prob when type-I is given by $\alpha$. See Theorem 21 in Polyanskiy's thesis. Note that he defined $\sE_\gamma$ for any $\gamma$ but it only corresponds to our $\sE_\gamma$ only for $\gamma\geq 1$.}

	\section{Differentially Private Online Learning}\label{Sec:Privacy_Iterative}

In this section, we use Theorem~\ref{Thm:Contraction_EGamma} to develop a framework for
quantifying the DP guarantees of general online learning algorithms. 
% \begin{equation}\label{Eq:Iterative_Algorithm2}
%     W_{t+1} = \Pi_{\mathcal{W}}\left(\Psi_{\sf_t}(W_t) + \sigma_t Z_{t}\right),
% \end{equation}
% where 
% \begin{itemize}
% \item[-] $\Pi_\W(\cdot)$ denotes the projection operator onto $\W\subset \R^d$,
% \item[-] $\Psi_{\sf_t}:\W\to \R^d$ denotes an update function at time $t$ depending on a \textit{convex} cost function $\sf_t$, 
% \item[-] $\{Z_t\}$ is the collection of i.i.d.\ noise variables sampled from a known density with covariance matrix $\mathbf{I}_d$,
% \item[-] $\sigma_t$ specifies the magnitude of noise at time $t$.
% \end{itemize}
% Typical example of such iterative algorithms includes \textit{online learning} in which 
% data points $x_1, \dots, x_n$ are made available to a learner \textit{one at a time}. The goal of the learner is to find a minimizer of $\sum_{t=1}^n\ell(w, x_t)$, where the minimization is taken over $w\in \W$ and $\ell(\cdot, \cdot)$ is a loss function. Applying the gradient descent algorithm to iteratively solve this minimization can be cast as \eqref{Eq:Iterative_Algorithm2} for $\sf_t(w) = \ell(w, x_t)$ and $\Psi_{\sf_t}(w) = w - \eta_t\nabla\sf_t(w)$ with $\eta_t$ being the learning rate at time $t$. 
Before delving into the technical results, we first describe the framework of online learning studied in this section and give the definition of central differential privacy. 

\subsection{Online Learning Algorithms}
Let $\W\subset\R^d$ denote a parameter space, e.g., the coefficients of a linear regression model, and let $\Pi_\W(\cdot)$ denote the projection operator onto $\W$. We describe typical \textit{online learning} algorithms in the next  definition.

\begin{definition}[Online Algorithm]
An online learning algorithm $\M$ proceeds as follows. A learner initiates the algorithm by taking a random point $W_1$ from $\W$. Once $W_1$ is chosen, a convex cost function $\sf_1:\W\to \R$ is revealed, implying that the cost associated with $W_1$ is $\sf_1(W_1)$. Upon observing $\sf_t$ and $W_t$, the learner at time $t+1$ chooses $W_{t+1}$ according to an update rule $\Psi_{\sf_{t}}:\mathcal{W}\to\mathcal{W}$, specifically,
\begin{equation*}
    W_{t+1} = \Pi_{\W}\left(\Psi_{\sf_{t}}(W_t)\right).
\end{equation*}
After $n$ iterations, the algorithm outputs $W_{n+1}$.  We say that $\M$ is \textit{randomized} if, for $\{\sf_t\}$ and $W_t$ fixed , $W_{t+1}$ is a random variable on $\W$. 
%\MD{The last sentence is not part of the definition, it is a consequence. Perhaps we could take it out of the definition environment? In fact, this is a bit out of place. I think that it is better to talk about this interpretation once we started talking about privacy (only revealing the last iteration). At this point it could be confusing, recall that the regret depends on $W_{1},\ldots,W_{n}$.}
\end{definition}
Letting $\sF$ be the collection of all possible convex cost functions, an online learning algorithm can thus be viewed as the mapping 
\eq{\M:\sF^n\to \W,}
given by 
\eq{\M(\{\sf_1, \dots, \sf_n\}) = W_{n+1}.}

For brevity, we denote $\{\mathsf{f}_1,\ldots,\mathsf{f}_n\}$ by $\{\sf_t\}$. It is worth noting that in this setting we assume that the collection of cost functions $\{\sf_1, \dots, \sf_n\}$ is fixed before the algorithm is started---often referred to as the \textit{oblivious setting} in the literature \cite[Section 5.5]{Hazan_Online}. We also assume that $\Psi_{\sf_t}$, the update function at time $t$, only depends on $\sf_t$ and not on the previous cost functions $\sf_1, \dots, \sf_{t-1}$. 
%A popular example of such algorithms is the noisy \textit{online gradient
%descent} (OGD) algorithm \cite{Online_SGD} where $W_{t+1}$ is constructed according to  $W_{t+1} = \Pi_{\W}\left(W_t - \eta_t \nabla\mathsf{f}_{t}(W_t) + Z_t\right),$
%where $\Pi_{\W}$ denotes the projection operator onto $\W$, $\eta_t >0$ is the learning rate, and $Z_t$ represents noise at time $t$.  

The goal of the learner is to minimize  the \textit{regret}, i.e., the difference between the cumulative cost of the algorithm's choices and that of the best fixed (offline) solution in hindsight. Specifically, the regret of a randomized algorithm $\M$ after $n$ iterations is defined as
\begin{equation}\label{Eq:regret}
  \E\left[\sum_{t=1}^n\mathsf{f}_t(W_t)\right] - \min_{w\in \W}\sum_{t=1}^n\mathsf{f}_t(w),
\end{equation}
where the expectation is taken over the algorithm's randomness.  
 %Two  examples of such algorithms are (1) \textit{one-pass} empirical risk minimization (ERM) where $\sf_t(w)  = \ell(w, x_t)$ with $\ell(\cdot, \cdot)$ being a  loss function and $x_t$ being the ground truth observation (e.g., stock price) at time $t$, and (2) online gradient descent where $\sf_t(w) = w-\eta_t\nabla\sf_t(w)$ with $\eta_t$ being the learning rate at time $t$. These two examples are studied in details in Sections~\ref{Sec:Privacy_Amplification} and \ref{Sec:OnlinePrivacy}, respectively.  
 %Other examples include online matrix completion and online recommendation systems \cite{Online_Recommend, Online_Recommend2, Online_Recommend3}, sequential investment \cite{Online_Finance, Online_Finance1, Online_Finance2, Online_Finance3, Online_Finance4}, and dynamic pricing \cite{Online_Pricing}.  
Typically, for online learning algorithms, a \textit{sublinear} regret is sought, as it implies that asymptotically an online algorithm performs almost as well as the optimal solution in hindsight.

%Cost functions may be chosen by the adversary as to compromise the privacy of the player which renders the need for the privacy-preserving mechanism for online algorithms very applicable. 
%Our goal is to design differentially private online algorithms which ensures that, upon observing the final parameter $W_{n+1}$, the
%amount of information an adversary learns about a particular cost function $\mathsf{f}$ is almost independent of its presence or absence in $\{\sf_t\}$. 

In many practical scenarios,  the cost functions $\{\sf_{t}\}$ might leak private information about the learner. For example, in offline setting (cf. Section~\ref{Sec:Privacy_Amplification}) $\sf_t(w)=\ell(w, x_t)$ where $x_t$ is a ground truth observation (e.g., stock price) at time $t$ and $\ell$ is loss function. If the loss function is linear, then $\sf_t(w)$ reveals significant information about the observation $x_t$. Other examples of privacy leakage in online algorithms are listed in \cite{Recomm_Online_privacy,Recomm_Online_privacy2}. As such, we focus on privacy attacks aimed to infer information about the cost function $\sf_{i}$ for some $i\in[n]$ upon observing the output of the algorithm, i.e., $\M(\{\sf_t\}) =W_{n+1}$. Thus, our goal is to design online algorithms such that $W_{n+1}$ does not reveal significant information about any single cost function in $\{\sf_t\}$. The following definition formalizes this goal. We say that two collection of cost functions $\{\sf_t\}$ and $\{\sf'_t\}$ are \textit{neighboring} at index $i\in [n]$, if $\sf_t = \sf'_t$ for all $t\in [n]\backslash{i}$ and $\sf_i\neq \sf'_i$. We denote this by $\{\sf_t\} \stackrel{i}{\sim} \{\sf'_t\}$. 

\begin{definition}[\cite{Feldman2018PrivacyAB}]\label{Def:DP_Online}
Given $\eps\geq 0$ and $\delta\in [0,1]$, a randomized online algorithm $\M$ is said to be $(\eps, \delta)$-DP at index $i\in [n]$, if 
$$\sup_{\{\sf_t\} \stackrel{i}{\sim} \{\sf'_t\}} \sE_{e^\eps}(\mu_{n+1}\|\mu'_{n+1})\leq \delta,$$
%$$\sup_{i\in [n]}\sup_{\{\sf_t\} \stackrel{i}{\sim} \{\sf'_t\}}\sup_{A\subset \W} \big[\Pr(\M(\{\sf_t\})\in A)-e^\eps \Pr(\M(\{\sf'_t\})\in A)\big]\leq  \delta.$$
where $\mu_{n+1}$ and $\mu'_{n+1}$ are the distributions of $\M(\{\sf_t\})$ and $\M(\{\sf'_t\})$, respectively. We say that $\M$ is $(\eps, \delta)$-DP if it is $(\eps, \delta)$-DP at index $i$ for all $i\in [n]$. 
\end{definition}
Notice that in light of the definition of $\sE_\gamma$-divergence in \eqref{Defi_HS_Divergence}, we can equivalently say $\M$ is $(\eps, \delta)$-DP if 
$$\Pr(\M(\{\sf_t\})\in A)-e^\eps \Pr(\M(\{\sf'_t\})\in A)\leq  \delta,$$
for any $A\subset \W$ and any pair of neighboring collections of cost functions $\{\sf_t\}$ and $\{\sf'_t\}$. While this is a more widely used definition in DP literature, we will use its equivalent form, given in Definition~\ref{Def:DP_Online}, to emphasize the connection between  $\sE_\gamma$-divergence and DP. 
%\MD{The way we did things in Section III was: we first introduce the regular definition of DP and then point out the equivalence in terms of $\sE_{\gamma}$. I suggest to do the same thing here (or the same thing there, but it does not look good...).}

\begin{remark}
The above definition of DP is specific to  online learning. Here, a pair of neighboring collections of cost functions is considered rather than the "neighboring datasets" used in  offline learning scenarios. Definition \ref{Def:DP_Online} is also the standard notion of DP for online learning algorithm considered in \cite{Online_DP_Agarwal, Online_DP_Smith, Online_Cumming, DP_OnlineLearning, Nozari_FunctionalDP}. Nevertheless, Definition~\ref{Def:DP_Online} differs from previous works on differentially private online algorithms in that  privacy is ensured with respect to \textit{only} the last parameter $W_{n+1}$ rather than the \textit{entire} set of parameters $\{W_1, \dots, W_{n+1}\}$. %\MD{Should we add a comment about when this makes sense (revealing only the last one)?}
\end{remark}

 %In order to quantify the DP parameters $\eps$ and $\delta$ of an online algorithm, we model each of its iterations as a Markov kernel (i.e., a channel) and apply the relation between the DP constraint and the \textit{contraction coefficient} of a Markov kernel. 

% For online learning setting, each iteration (say iteration $t+1$) can be thought of as a Markov kernel $\sK_{t}:\W\to \P(\W)$ with input is $W_{t}$ and output is $W_{t+1}$ and whose statistics is given by $\mathsf{f}_{t}$ and the  noise distribution.  To make this dependency clear, we  denote this Markov kernel as $\sK_{\mathsf{f}_{t}}$. Given the statistics of $\sK_{\mathsf{f}_1}$ and the distribution of the random initial point $W_1\sim \mu_1\in \P(\W)$, the distribution of $W_2$ is denoted by $W_2\sim \mu_1\sK_{\mathsf{f}_1} \coloneqq  \int_{\W}\mu_1(\text{d}w)\sK_{\mathsf{f}_1}(w)$.
% Due to its sequential nature, the resulting online algorithm can be expressed as the composition of $T$ Markov kernels, i.e., $\sK_{\mathsf{f}_{1}}\dots \sK_{\mathsf{f}_{T}}$. 

The analysis of DP mechanisms in online learning is particularly challenging: a single change in the algorithm's cost function at any time instance may have an accumulative impact on  all future parameter updates. One standard way to address this issue is to add calibrated noise to $\Psi_t$ at each iteration \cite{DP_OnlineLearning, Online_DP_Smith, Online_DP_Agarwal, Online_Cumming}.
Thus, most differentially private online learning algorithms can be expressed as 
general iterative process of the form
\begin{equation}\label{Eq:Iterative_Algorithm2}
    W_{t+1} = \Pi_{\mathcal{W}}\left(\Psi_{\sf_t}(W_t) + \sigma_t Z_{t}\right),
\end{equation}
where $\{Z_t\}$ is the collection of i.i.d.\ noise variables sampled from a known density with covariance matrix $\mathbf{I}_d$ and $\sigma_t$ specifies the magnitude of noise at time $t$.

The main goal of this section is to establish a bound for the DP parameters $\eps$ and $\delta$ achievable by adding \textit{Gaussian noise} to the update rules $\{\Psi_{\sf_t}\}$.
The subsequent analysis is based on a key observation.

\begin{figure*}
    \centering
    \includegraphics[width=1\textwidth]{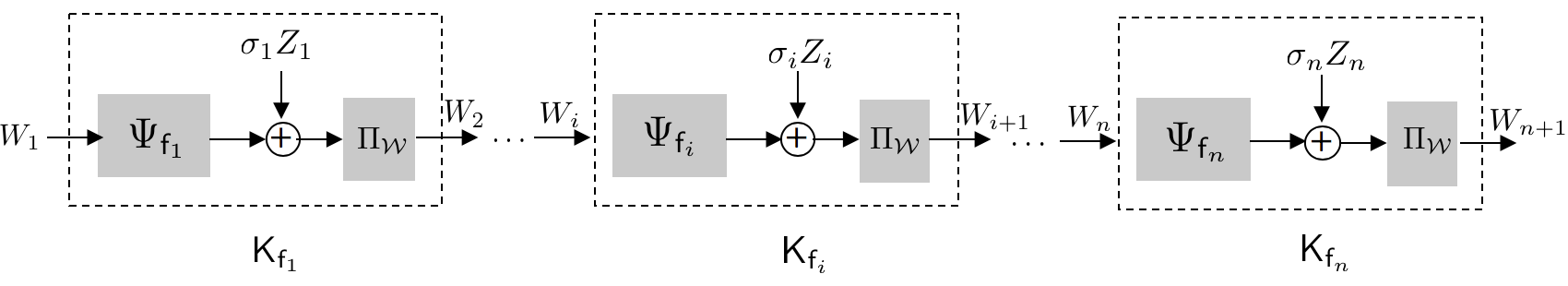}
    \caption{The schematic representation of the iterative process described in \eqref{Eq:Iterative_Algorithm2}. Each Markov kernel $\sK_{\sf_t}$, $t\in [n]$ consists of three components: mapping $w\mapsto \Psi_{\sf_t}(w)$, additive noise, and projection mapping $\Pi_\W$. The initial point $W_1\sim \mu_1$ is a random point in $\W$. Input and output of kernel $\sK_{\sf_t}$ is $W_t\sim \mu_t$ and $W_{t+1}\sim \mu_{t+1}$, respectively. }
    \label{fig:SDPI}
\end{figure*}

Assume $\{Z_t\}$ are i.i.d.\ samples drawn from a Gaussian distribution.  As illustrated in Fig.~\ref{fig:SDPI},  the iterative process \eqref{Eq:Iterative_Algorithm2} can be viewed as  a collection of Markov kernels $\{\sK_{\sf_t}\}$ for $t\in [n]$ where each $\sK_{\sf_t}$ is a concatenation of the update rule $\Psi_{\sf_t}$, an additive Gaussian kernel with noise magnitude $\sigma^2_t$, and the projection operator $\Pi_\W(\cdot)$. Consequently, we can express $\sK_{\sf_t}$ as $\Pi_{\mathcal{W}} \circ\sK_{t}\circ\Psi_{\sf_t}:\mathcal{W}\to\P(\mathcal{W})$ where $\sK_{t}$ is the $\Psi_{\sf_t}(\W)$-constrained additive Gaussian kernel with noise magnitude given by $\sigma_t^2$. Note also that $\mu_{t+1}$ the distribution of $\sK_{\sf_t}$ is given by  
$\mu_1\sK_{\sf_1}\cdots \sK_{\sf_t}$, where $\mu_1$ is the  distribution from which the initial point $W_1$ is sampled. Similarly, $\mu_{n+1}$ the distribution of
the final parameter $W_{n+1}$ is given by  $\mu_1\sK_{\sf_1}\cdots \sK_{\sf_n}$.  
If we change $\{\sf_t\}$ to $\{\sf'_t\}$ such that $\{\sf_t\}\stackrel{i}{\sim} \{\sf'_t\}$, then we denote the new distribution of the final parameter by $\mu'_{n+1}$. 
Since $\sK_{\sf_t} = \sK_{\sf'_t}$ for all $t\in [n]\backslash i$, we can write 
 \begin{align*}
     \mu_{n+1} &= \mu_{i} \sK_{\sf_i} \sK_{\sf_{i+1}} \cdots \sK_{\sf_n},\\
     \mu'_{n+1} &= \mu_{i} \sK_{\sf'_i} \sK_{\sf_{i+1}} \cdots \sK_{\sf_n}.
 \end{align*}
 A similar construction is given in \cite{Balle2019mixing} to quantify the improvement of DP parameters due to post-processing (aka privacy amplification). 
 By repeatedly invoking the definition of contraction coefficient \eqref{Eq:SDPI_f_Div}, we obtain 
 \begin{align*}
     D_f(\mu_{n+1}\|\mu'_{n+1})  %= D_f(\mu_{i} \sK_{\sf_i} \cdots \sK_{\sf_n} \| \mu_{i} \sK_{\sf'_i} \cdots \sK_{\sf_n})\\
     & \leq D_f(\mu_{i}\sK_{\sf_i}\|\mu_{i}\sK_{\sf'_i})\prod_{t=i+1}^n\eta_f(\sK_{\sf_t}),
 \end{align*}
and similarly for any $\eps\geq 0$
\begin{align}\label{Eq:EGamma_PNSGD}
   \sEs(\mu_{n+1}\|\mu'_{n+1})\leq \sE_{e^\eps}(\mu_{i}\sK_{\sf_i}\|\mu_{i}\sK_{\sf'_i})\prod_{t=i+1}^n\eta_{e^\eps}(\sK_{\sf_t}).
\end{align}
While this inequality follows from a routine application of SDPI, it provides a natural framework to study privacy guarantees of general iterative processes. In fact, inequality \eqref{Eq:EGamma_PNSGD} indicates that an upper bound for DP parameters can be easily obtained by bounding $\sEs(\mu_i\sK_{\sf_i}\|\mu_i\sK_{\sf'_i})$ and $\eta_{e^\eps}(\sK_{\sf_t})$.   
This is the linchpin for our privacy analysis in the sequel. Inequality~\eqref{Eq:EGamma_PNSGD} can also be used in applications beyond privacy. In particular, together with Theorem~\ref{Thm:Contraction_EGamma} and Proposition~\ref{Proposition:etaPAGK}, it implies Theorem~\ref{Thm:Informal_SDPI}, which is of independent interest (e.g., in relation to studying information dissipation over Markov chain \eqref{MarkovChain}). 
%Specializing this lemma to $\sE_\gamma$-divergence, we obtain that for iterative process \eqref{Eq:Iterative_Algorithm} and $\{\sf_t\}\stackrel{i}{\sim}\{\sf'_t\}$ 
% \begin{align}\label{Eq:EGamma_PNSGD}
%   \sEs(\mu_{n+1}\|\mu'_{n+1})\leq \sE_{e^\eps}(\mu_{i}\sK_{\sf_i}\|\mu_{i}\sK_{\sf'_i})\prod_{t=i+1}^n\eta_{e^\eps}(\sK_{\sf_t}).
% \end{align}
% \begin{align}\label{Eq:EGamma_PNSGD}
%   \sE_{e^\eps}(\mu_{1}\sK_{\sf_1}\cdots \sK_{\sf_n}\|\mu_{1}\sK_{\sf'_1}\cdots \sK_{\sf_n'})\leq \sE_{e^\eps}(\zeta_{x_i}\|\omega_{x_{i}})\prod_{t=i+1}^n\eta_{e^\eps}(\sK_{x_t}),\nonumber
% \end{align}
%where $\zeta_{i}\coloneqq \mu_{i}\sK_{\sf_i}$ and $\omega_{i}\coloneqq \mu_{i}\sK_{\sf'_i}$.

\begin{theorem}
\label{Thm:General_DP}
Assume that $\mathcal{W}\subset\mathbb{R}^d$ is a closed convex set and $P_Z = \N(0, \mathbf{I}_d)$.
The iterative process \eqref{Eq:Iterative_Algorithm2} is $(\eps, \delta)$-DP at index $i\in [n]$ for $\eps\geq 0$ and 
\begin{equation}\label{Theorem_general_DeltaLaplace}
    \delta = \theta_{e^\eps}\left(\frac{\psi}{\sigma_i}\right) \prod_{t=i+1}^n \theta_{e^\eps}\left(\frac{\textnormal{dia}(\Psi_{\sf_{t}}(\mathcal{W})) }{\sigma_t}\right),
\end{equation}
where $\displaystyle \psi \coloneqq \sup_{\sf_1,\sf_2\in\sF} \sup_{w\in\W} \lVert \Psi_{\sf_1}(w) - \Psi_{\sf_2}(w)\rVert$. 
\end{theorem}
The detailed proof of this theorem is given in Appendix~\ref{Appendix:Thm:General_DP}. We will use this theorem to quantify the DP guarantee of popular algorithms. Before we delve into instantiating this result, we first ``homogenize'' Theorem~\ref{Thm:General_DP} over all $i\in [n]$ and give a bound for $\delta$ independent of index $i$. To this end, we follow \cite{Feldman2018PrivacyAB} to consider the \textit{randomly-stopped} variant of process \eqref{Eq:Iterative_Algorithm2}: Instead of terminating after pre-determined $n$ iterations, the algorithm stops at a random time $T$ uniformly chosen in $[n]$. 

\begin{algorithm}
\caption{Randomly-stopped variant of process \eqref{Eq:Iterative_Algorithm2}}
\label{alg:Randomly_Stopped}
\begin{algorithmic}[1]
\REQUIRE{Collection of cost functions $\{\sf_t\}$, convex set $\W\subset \R^d$, and noise parameter $\sigma$}
\STATE{Pick a starting point $W_1 \sim \mu_1 \in \P(\W)$ sampled from some distribution $\mu_1$,  }
\STATE{Pick $T$ uniformly at random from $[n]$,}
\FOR{$t\in\{1,\ldots,T\}$}
\STATE{Play $W_t$, then obtain the cost function $\sf_t$,}
\STATE{$\displaystyle W_{t+1} =\Pi_{\W}\left(\Psi_{\sf_t}(W_t) + \sigma_t Z_t\right)$,\quad $Z_t\sim \N(0, \mathbf{I}_d)$}
\ENDFOR
\RETURN{$W_{T+1}$.} 
\end{algorithmic}
\end{algorithm}

\begin{theorem}
\label{Thm:General_Random_DP}
Assume that $\mathcal{W}\subset\mathbb{R}^d$ is a closed convex set and $P_Z = \N(0, \mathbf{I}_d)$. The randomly-stopped iterative process described in Algorithm~\ref{alg:Randomly_Stopped}, is $(\eps, \delta)$-DP for $\eps\geq 0$ and
\begin{align}
    \label{eq:PrivacyRandomlyStopper1} \delta &= \max_{i\in[n]} \left\{\frac{1}{n} \sum_{t=i}^{n} \theta_{e^\eps}\left(\frac{\psi}{\sigma_{i}}\right) \prod_{j=i+1}^{t} \theta_{e^\eps}\left(\frac{\textnormal{dia}(\Psi_{\sf_{j}}(\mathcal{W}))}{\sigma_{j}}\right)\right\}\\
    \label{eq:PrivacyRandomlyStopper2} &\leq \frac{1}{n}\theta_{e^\eps}\left(\frac{\psi}{\sigma}\right) \left[1-\theta_{e^\eps}\left(\frac{D}{\sigma}\right)\right]^{-1},
\end{align}
where $\displaystyle \psi \coloneqq \sup_{\sf_1,\sf_2\in\sF} \sup_{w\in\W} \lVert \Psi_{\sf_1}(w) - \Psi_{\sf_2}(w)\rVert$, $\displaystyle \sigma \coloneqq \min_{t\in [n]}\sigma_t$ and $\displaystyle D \coloneqq \max_{t\in [n]}\textnormal{dia}(\Psi_{\sf_{t}}(\mathcal{W}))$. 
\end{theorem}
The detailed proof of this theorem is given in Appendix~\ref{Appendox:Thm_General_RandomDP}.
%Theorem~\ref{Thm:General_Random_DP} exploits the random stopping technique to convert the \textit{per-iteration} privacy guarantee given in Theorem~\ref{Thm:General_DP} into a privacy guarantee uniform among all players in the algorithm. 
Similar to Theorem~\ref{Thm:General_DP}, the above result only requires $\|\Psi_\sf(w)\|$ be uniformly bounded for all $w\in \W$ and $\sf\in \sF$. This assumption is in fact weaker than the regularity conditions typically assumed in most differentially private iterative algorithms, such as \cite{SGD_Exponential_VS_Gaussian, chaudhuri2011differentially, Chaudhuri_Subsampling, Bassily_ERM, BAssily_NIPS19, Chaudhuri_logistic, DP_OnlineLearning,Thakurta_LASSO,Sarwate_SGD_Update,duchi2013local,Jain_risk_bound, Smith_interaction,Talwar_Privacte_LASSO,Private_ERM_General, kumar2019differentially}. To illustrate this point and better compare our result with the existing differentially private ML algorithms, we instantiate Theorem~\ref{Thm:General_Random_DP} for two popular setups: (1) \textit{one-pass} empirical risk minimization and (2) online gradient descent.

\subsection{Empirical Risk Minimization}\label{Sec:Privacy_Amplification}
Here, we consider an offline learning setting, namely empirical risk minimization (ERM) which is a fundamental problem in machine learning. Given a dataset $\mathbb D = \{x_1, \dots, x_n\}\in \X^{n}$ and a loss function $\ell:\W\times \X\to \R^+$, ERM is formulated as 
$$\inf_{w\in \W} \frac{1}{n}\sum_{t=1}^n\ell(w, x_t).$$
A simple algorithm for approximating the solution of this problem can be implemented with one pass of stochastic gradient descent (SGD) over the data set. This procedure  starts with a random point $W_1$ in $\W$, and then iterates $W_{t+1} = \Pi_{\W}(W_{t}-\eta \nabla\ell(W_{t}, x_{t}))$ for $t\in [n]$. 
This problem can be cast as an online optimization problem by defining $\sf_t(w) = \ell(w, x_t)$ and, thus, the differentially private version of such algorithm can be analyzed using Theorem~\ref{Thm:General_Random_DP}.  Following the DP literature (see, e.g., \cite{Bassily_ERM, Private_ERM_General, Abadi_MomentAccountant, chaudhuri2011differentially,Sarwate_SGD_Update,Bassily_ERM2, BAssily_NIPS19}), we add Gaussian noise to the gradient term, and  thus the magnitude of the noise is assumed to be $\eta_t^2\sigma_t^2$ (as opposed to  $\sigma_t^2$). For brevity, we assume  $\eta_t = \eta$ and $\sigma_t = \sigma$ for all $t\in [n]$. Consequently, we consider the following algorithm, 
which is known as \textit{projected noisy SGD} (PNSGD)
$$W_{t+1} = \Pi_\W\left(W_t-\eta\nabla\ell(W_t, x_t) + \eta \sigma Z_t\right),$$
where $Z_t \sim \N(0, \mathbf{I}_d)$. It can be verified that the above algorithm is an instance of iterative process \eqref{Eq:Iterative_Algorithm2} with update function   $\Psi_{\sf_t}(w) = \Psi^{\mathsf{SGD}}_{x_t}(w)$ where
\begin{equation*}
    \Psi^{\mathsf{SGD}}_{x_t}(w) \coloneqq w - \eta\nabla\ell(w, x_t).
\end{equation*}
In this setup each cost function $\sf_t$ depends only on the data point $x_t$. Thus, the neighboring relationship between two collections of cost functions
$\{\sf_t\}$ and $\{\sf_t'\}$ reduces to that of the datasets $\mathbb D =\{x_1, \dots, x_n\}$ and $\mathbb D' =\{x'_1, \dots, x'_n \}$.
%That is, assuming $x\mapsto \ell(\cdot, x)$ is injective, we have $\{\sf_t\}\stackrel{i}{\sim}\{\sf_t'\}$  if and only if the datasets $\mathbb D \stackrel{i}{\sim}\mathbb D'$, meaning  $x_i\neq x'_i$ and $x_j = x'_j$ for all $j\in [n]\backslash{\{i\}}$. 
Consequently, the definition of DP in Definition~\ref{Def:DP_Online} can be equivalently given in terms of \textit{neighboring datasets} and thus reduces to the original definition of DP in \cite{Dwork_Calibration}.  
%\SA{The map $x\mapsto \ell(\cdot, x)$ must be injective!}
%\FC{[Don't we need to assume some that $\ell(\cdot,x)\neq \ell(\cdot,x')$ if $x\neq x'$?] for the iff condition to hold?]}
% involves pairs of neighboring datasets 
% $\mathbb D = \{x_1, \dots, x_n\}$ and $\mathbb D' = \{x'_1, \dots, x'_n \}$ which differ in only one entry, i.e., there exists an $i\in [n]$ such that $x_i\neq x'_i$ and $x_j = x'_j$ for all $j\in [n]\backslash{\{i\}}$. When two datasets $\mathbb D$ and $\mathbb D'$ are neighboring, we denote it by $\mathbb D\sim\mathbb D'$. 
\begin{algorithm}
\caption{Randomly stopped Projected Noisy SGD (PNSGD)}
\label{alg:PNSGD}
\begin{algorithmic}[1]
\REQUIRE{Dataset $\mathbb D= \{x_1,\ldots,x_n\}$, learning rate $\eta>0$, convex set $\W\subset \R^d$, and noise parameter $\sigma$ }
\STATE{Pick $W_1 \sim \mu_1\in \P(\W)$}
\STATE{Take $T$ uniformly on $[n]$}
\FOR{$t\in\{1,\ldots,T\}$}
\STATE{$W_{t+1} = \Pi_{\mathcal{W}}(W_t - \eta [\nabla_w \ell(W_t,x_{t}) + \sigma Z_{t}])$, \qquad $Z_t\sim \N(0, \mathbf{I}_d)$}
\ENDFOR
\RETURN{$W_{T+1}$} 
\end{algorithmic}
\end{algorithm}

The randomly-stopped PNSGD algorithm is described in Algorithm~\ref{alg:PNSGD}. The privacy guarantee of this algorithm has been recently studied by Feldman et al. \cite{Feldman2018PrivacyAB} under the name of \textit{privacy amplification by iteration}. However, the notion of privacy used in their work is R\'enyi differential privacy. In the following corollary, whose proof is given in Appendix~\ref{Appendix:Cor_PNSGD}, we instantiate Theorem~\ref{Thm:General_Random_DP} to derive the privacy guarantee of this algorithm directly in terms of DP.
%\footnote{\textit{Added in print}: After the first draft of this work, Sordello et al.\ \cite{PNSGD_Shuffled} proposed another approach for homogenizing the privacy guarantee given in Theorem~\ref{Thm:General_DP}, specifically for PNSGD.  Instead of randomizing the stopping time, they considered shuffling the dataset $\mathbb D$ before initializing the algorithm. \SA{What do you think of bringing up this paper?} \MD{I think it is appropriate, but we have to clearly say that they build upon our work! \SA{Convert the footnote to a remark and mention that it showcases the versatility of our approach.}}} 

\textit{Added in print}: After the first draft of this work, Sordello et al.\ \cite{PNSGD_Shuffled} proposed another approach for homogenizing the privacy guarantee given in Theorem~\ref{Thm:General_DP}, specialized for PNSGD.  Instead of randomizing the stopping time, they considered shuffling the dataset $\mathbb D$ before initializing the algorithm and then applied our framework (in particular, Jensen's inequality and \eqref{Eq:EGamma_PNSGD}).  This indeed showcases the versatility of the contraction-based framework presented in our work. 

\begin{corollary}\label{cor:PNSGD}
Let $\W\subset \R^d$ be a convex set and $\{\ell(\cdot, x)\}_{x\in \X}$ be a family of convex $L$-Lipschitz functions over $\W$. Then the randomly-stopped PNSGD algorithm  is $(\eps, \delta)$ with $\eps\geq 0$ and 
 \begin{equation}\label{Eq:Delta_PNSGD}
     \delta = \frac{1}{n}\theta_{e^\eps}\Big(\frac{2L}{\sigma}\Big)\left[1-\theta_{e^\eps}\Big(\frac{\textnormal{dia}(\W)+2\eta L}{\eta\sigma}\Big)\right]^{-1}.
 \end{equation}
\end{corollary}

If we further assume that $w\mapsto \ell(w, x)$ is\footnote{A function $f:\W\to \R$ is $\beta$-smooth if $\|\nabla f(w_1)-\nabla f(w_2)\|\leq \beta\|w_1-w_2\|$ for all $w_1, w_2\in \W$.} $\beta$-smooth for any $x\in \X$, then a standard calculation in convex optimization shows that $w\mapsto \Psi^{\mathsf{SGD}}_{x}(w)$ is $1$-Lipschitz for $\eta\leq \frac{2}{\beta}$ (see Appendix~\ref{Appendix_Contractivity} for a detailed proof). A similar argument as in the proof of Corollary~\ref{cor:PNSGD} reveals that with this extra assumption \eqref{Eq:Delta_PNSGD} can be improved as (see Appendix~\ref{Appendix_Contractivity}) 
\begin{equation}\label{Eq:Delta_PNSGD2}
    \delta = \frac{1}{n}\theta_{e^\eps}\Big(\frac{2L}{\sigma}\Big)\left[1-\theta_{e^\eps}\Big(\frac{\textnormal{dia}(\W)}{\eta\sigma}\Big)\right]^{-1},
\end{equation}
for $\eta\leq \frac{2}{\beta}$. This enables us to formally compare our result with  
\cite{Feldman2018PrivacyAB}. To do so, we need the following definition. Given $\alpha>1$, a mechanism $\M$ is called $(\alpha, \zeta)$-R\'enyi differentially-private (RDP) if $$\sup_{\mathbb D\sim \mathbb D'}D_\alpha(\mu_{n+1}\|\mu'_{n+1})\leq \zeta,$$ where $D_\alpha(\cdot\|\cdot)$ denotes the R\'enyi divergence of order $\alpha$ and $\mu_{n+1}$ and $\mu'_{n+1}$ are the distributions of $W_{n+1}$ and $W'_{n+1}$ the outputs of $\M$ when running on neighboring $\mathbb D$ and $\mathbb D'$ (denoted by $\mathbb D\sim \mathbb D'$), respectively.

\begin{theorem}[{{\cite[Theorem 26]{Feldman2018PrivacyAB}}}]\label{Thm:Feldman26}
Let $\W\subset \R^d$ be a convex set and $\{\ell(\cdot, x)\}_{x\in \X}$ be a family of convex, $L$-Lipschitz and $\beta$-smooth loss functions over $\W$. Then, for any $\eta\leq \frac{2}{\beta}$ and $\alpha>1$, and $\sigma\geq L\sqrt{2(\alpha-1)\alpha}$, the randomly-stopped PNSGD algorithm is $(\alpha, \zeta)$-RDP for 
$$\zeta = \frac{4\alpha L^2\log n}{n\sigma^2 }.$$
\end{theorem}
While Corollary~\ref{cor:PNSGD} provides the privacy guarantee for any $\sigma>0$, this theorem is restricted to $\sigma$ greater than $L\sqrt{2(\alpha-1)\alpha}$. This discrepancy stems from the fact that, unlike $\sE_\gamma$-divergence,  $(\mu\|\nu)\to D_\alpha(\mu\|\nu)$ is not convex for $\alpha>1$. To get around this issue, Feldman et al. \cite[Lemma 25]{Feldman2018PrivacyAB}  presented a ``weak'' form of joint convexity for the R\'enyi divergence. 

In order to compare Corollary~\ref{cor:PNSGD} with Theorem~\ref{Thm:Feldman26}, we need to convert $(\alpha, \zeta)$-RDP guarantee to $(\eps, \delta)$-DP. To do so, we invoke two existing RDP-to-DP conversion formulae, making both analytical and numerical comparison possible. 
\begin{lemma}[{{\cite[Proposition 3]{RenyiDP}}}] \label{Lemma:RDP_DP}If a mechanism $\M$ is $(\alpha, \zeta)$-RDP for $\alpha>1$, then it is $(\eps, e^{-(\alpha-1)(\eps-\zeta)})$-DP.
\end{lemma}
Due to the efficiency of RDP (especially in the context of composition), this lemma has been extensively used in many recent private ML algorithms, see e.g., \cite{Abadi_MomentAccountant, Balle_AnalyticMomentAccountant, Duchi_Federatedprotection, McMahan_GeneralApproach} and has been implemented in Google's open-source TensorFlow Privacy\footnote{
%\FC{[I would add this as a reference, and note here that this is true at the moment of submission.]}
See the function ''$\text{compute}\_\text{eps}$'' in the analysis directory of TensorFlow Privacy in \cite{TensorFlow_Privacy_Analysis}. At the time of submission of this paper, Lemma~\ref{Lemma:RDP_DP} is particularly implemented in the line 216 of \cite{TensorFlow_Privacy_Analysis}.} \cite{TensorFlow_Privacy}. 
%The way it is used in practice is as follows: 
%Once a mechanism is shown to be $(\alpha, \zeta)$-RDP for \textit{all} $\alpha>1$ and some $\zeta = \zeta(\alpha)$, Lemma~\ref{Lemma:RDP_DP} implies that it is $(\eps, \inf_{\alpha>1}e^{-(\alpha-1)(\eps-\zeta)})$-DP for any $\eps\geq 0$. 
In light of this result, Theorem~\ref{Thm:Feldman26} implies that the randomly stopped PNSGD is $(\eps, \hat\delta)$-DP for any $\eps\geq 0$ and 
\begin{equation}\label{Eq:Delta_FMTT}
    \hat\delta \coloneqq \inf_{\alpha\in (1, \alpha^*]} e^{-(\alpha-1)(\eps-\zeta)},
\end{equation}
where $\alpha^*\coloneqq \frac{1}{2}\left[1+\sqrt{1+\frac{2\sigma^2}{L^2}}\right]$ and $\zeta = \frac{4\alpha L^2\log n}{n\sigma^2 }$. The restriction on the range of $\alpha$ was caused by the constraint on $\sigma$ in Theorem~\ref{Thm:Feldman26}. Since $\zeta$ is linear in $\alpha$, the minimization in \eqref{Eq:Delta_FMTT} can be solved analytically:
\begin{equation}\label{Eq:Delta_FMTT2}
   \hat\delta = e^{-(\alpha^{\mathsf{opt}}-1)(\eps-\rho\alpha^{\mathsf{opt}})}, 
\end{equation}
where $\rho\coloneqq \frac{4L^2\log n}{n\sigma^2}$ and 
$$\alpha^{\mathsf{opt}}\coloneqq 
\begin{cases}
\alpha^*,&\text{if}~ \eps^2\geq \rho^2\Big(1+\frac{2\sigma^2}{L^2}\Big),\\
% \alpha^*,&\text{if}~ \eps^2\geq \frac{16L^2(L^2+2\sigma^2)\log^2n}{n^2\sigma^4},\\
\frac{1}{2}+\frac{\eps}{2\rho},& \text{otherwise}.
\end{cases}
$$
Hence, for sufficiently large $\eps$ (i.e., $\eps\geq O(\frac{\log n}{n})$),   $\hat\delta$ behaves approximately like $e^{-O(\eps)}$ while our privacy analysis yields $\delta$ in  \eqref{Eq:Delta_PNSGD2} that decays as $e^{-O(\eps^2)}$. 
\begin{figure*}[t]
\centering
	\begin{subfigure}[t]{0.3\textwidth}
	\includegraphics[scale=0.15]{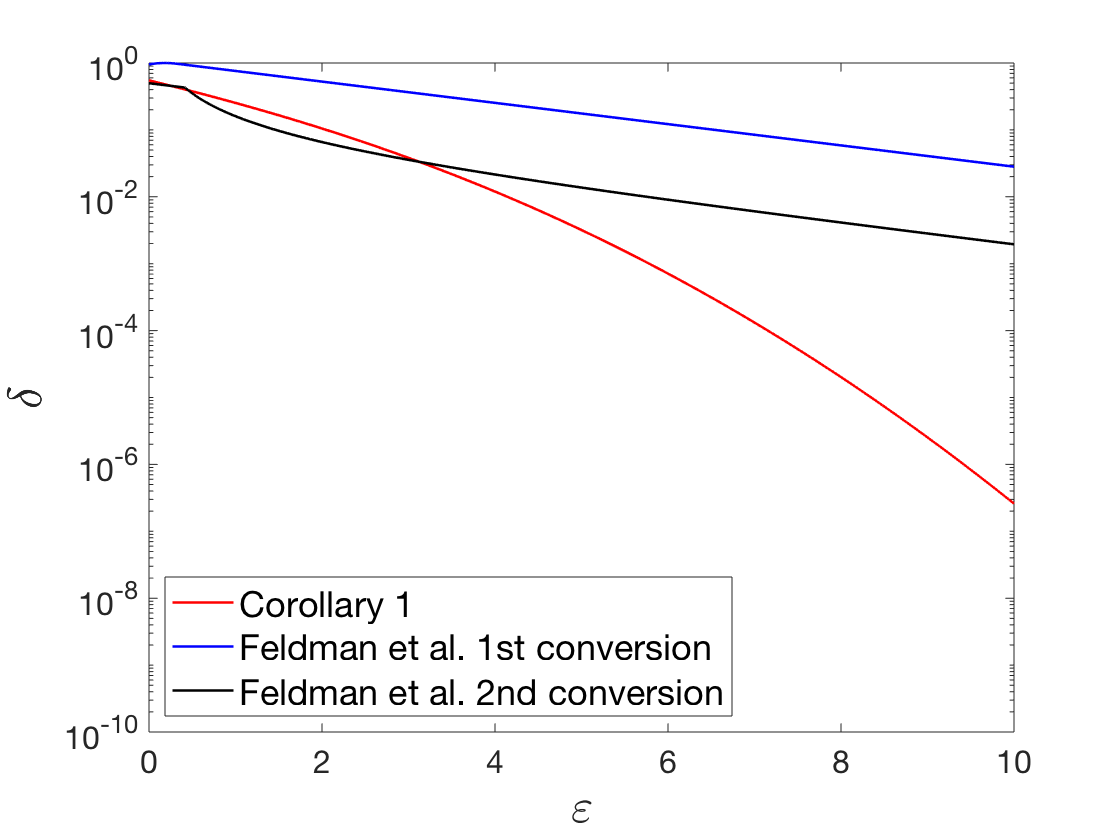}
	\caption{$\eta = 0.2, \sigma = 1$}
	\end{subfigure}
	~~~
	\begin{subfigure}[t]{0.3\textwidth}
	\includegraphics[scale=0.15]{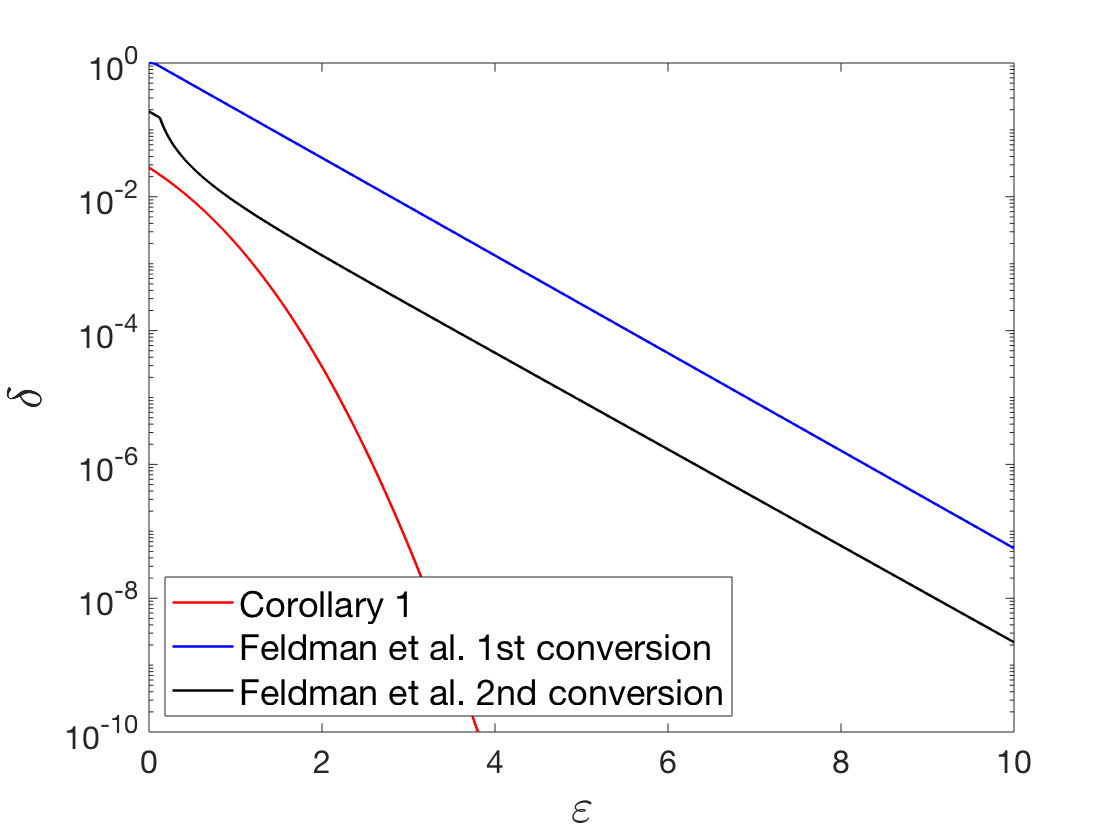}
	\caption{$\eta = 0.1, \sigma = 3$}
	\end{subfigure}
	~~~
	\begin{subfigure}[t]{0.3\textwidth}
	\includegraphics[scale=0.15]{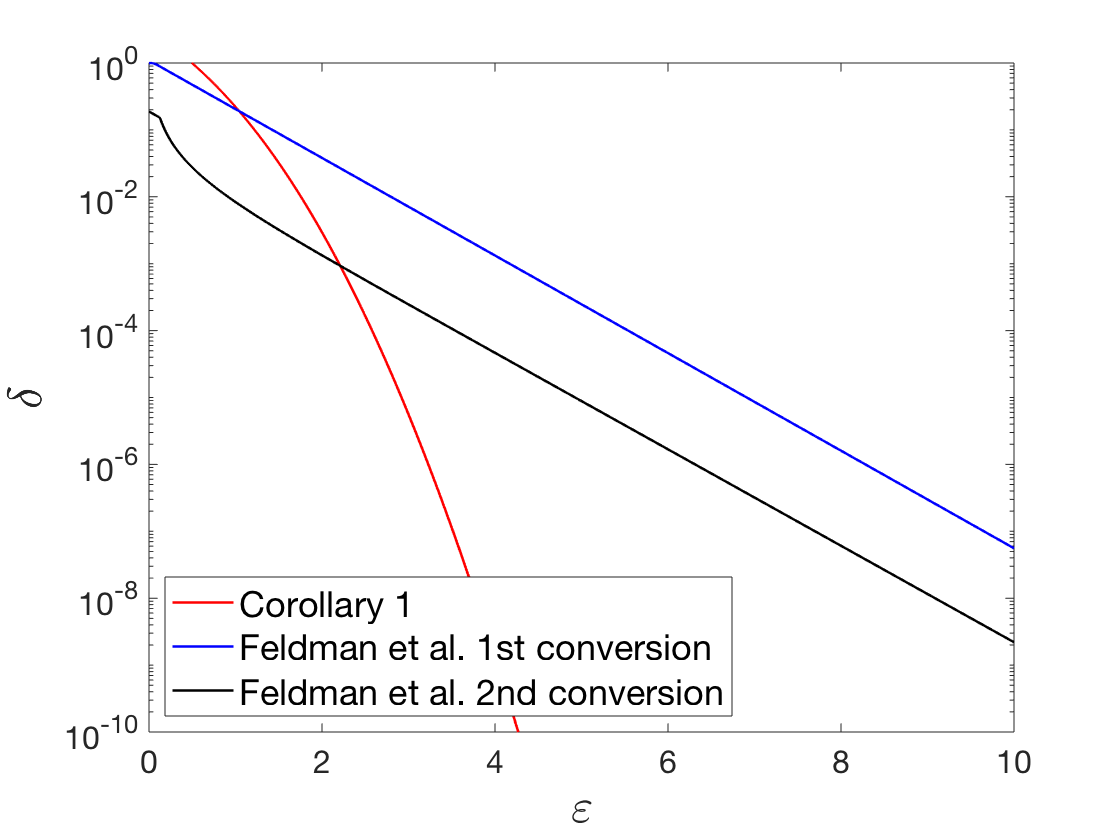}
	\caption{$\eta = 0.05, \sigma = 3$}
	\end{subfigure}
	\caption{The privacy parameters of PNSGD Algorithm \ref{alg:PNSGD}  obtained from Corollary~\ref{cor:PNSGD} and converting \cite[Theorem 26]{Feldman2018PrivacyAB} according to \eqref{Eq:Delta_FMTT2} and \eqref{Eq:Delta_ALCKS}, respectively.  
 	Here, we vary the learning rate $\eta$ and $\sigma$. Other parameters are as follows: $L = 1, \beta = 1$, and $n = 100$. }
	\label{fig:PNSGD}
\end{figure*}

Despite its prevalence in practice, Lemma~\ref{Lemma:RDP_DP} is not tight in general. This issue has recently prompted Asoodeh et al. \cite{asoodeh2020better} to derive the \textit{optimal} translation of $(\alpha, \zeta)$-RDP into $(\eps, \delta)$-DP. Invoking  \cite[Lemma 2]{asoodeh2020better}, we deduce that the randomly stopped PNSGD is $(\eps, \check\delta)$-DP for $\eps\geq 0$ and 
\begin{equation}\label{Eq:Delta_ALCKS}
    \check\delta \coloneqq \inf_{\alpha\in (1, \alpha^*]} \min\Big\{\kappa e^{-(\alpha-1)(\eps-\zeta)}, \frac{e^{(\alpha-1)\zeta}-1}{\alpha\big(e^{(\alpha-1)\eps}-1\big)}\Big\},
\end{equation}
where $\kappa\coloneqq \frac{1}{\alpha}\left(1-\frac{1}{\alpha}\right)^{\alpha-1}$ and $\zeta = \frac{4\alpha L^2\log n}{n\sigma^2 }$. 
Although this result is tighter than Lemma~\ref{Lemma:RDP_DP}, it is not possible to analytically express $\check\delta$. 
In Fig. \ref{fig:PNSGD}, we compare the privacy parameters of the randomly-stopped PNSGD given in Corollary~\ref{cor:PNSGD} with  $\hat\delta$ and $\check\delta$ given in  \eqref{Eq:Delta_FMTT2} and \eqref{Eq:Delta_ALCKS}, respectively. As illustrated in this figure, Corollary~\ref{cor:PNSGD} results in  significantly smaller values of $\delta$, compared to the technique developed in \cite{Feldman2018PrivacyAB}, for sufficiently large $\eps$ across different values of parameters $\eta$ and $\sigma$. The privacy improvement of Corollary \ref{cor:PNSGD} over  \cite{Feldman2018PrivacyAB} is  more pronounced for smaller values of $\frac{\eta}{\sigma}$. In particular, for the reasonable range\footnote{Abadi et al. \cite{Abadi_MomentAccountant} empirically observed that the accuracy of PNSGD algorithm is stable for a learning rate in
the range of $(0.01, 0.07)$ and peaks around $0.052$.} of $\eta\in (0.05, 0.1)$ our technique results in tighter privacy parameters than \cite{Feldman2018PrivacyAB} for all $\eps\geq 2$ as long as $\sigma\geq 3$. It is also worth emphasizing that, unlike \cite{Feldman2018PrivacyAB}, the privacy parameter given in Corollary~\ref{cor:PNSGD} heavily depends on the learning rate $\eta$. Intuitively, the larger the value of $\eta$, the stronger the privacy guarantee of the PNSGD algorithm.  Corollary~\ref{cor:PNSGD} formalizes this intuition by establishing a precise rate at which $\delta$ decreases as $\eta$ increases for any given $\eps$. %This therefore enables us to characterize a more accurate tradeoff between privacy and accuracy achievable by PNSGD algorithm than what would be possible by \cite[Theorem 26]{Feldman2018PrivacyAB}.

\subsection{Online Gradient Descent Algorithm}\label{Sec:OnlinePrivacy}

In this section, we apply Theorem~\ref{Thm:General_Random_DP} to study the privacy guarantee of the online gradient descent (OGD) algorithm \cite{Online_SGD}. Then, by drawing on  standard results from online convex optimization, we expound the trade-off between privacy and utility achievable by this algorithm. 

%To this end, we recall standard results in online convex optimization and instantiate the privacy results in Section~\ref{Sec:Privacy_Iterative}.
The OGD algorithm is an instance of the iterative process \eqref{Eq:Iterative_Algorithm2} with update function
\begin{equation*}
    \Psi^{\mathsf{OGD}}_{\sf_t}(w) \coloneqq w - \eta_{t}\nabla \sf_t(w).
\end{equation*}
Similar to Section~\ref{Sec:Privacy_Amplification}, we add Gaussian noise to the gradient term
and thus the noisy OGD algorithm iterates as 
\begin{equation}
\label{eq:UpdataOGD}
    W_{t+1} = \Pi_{\W}(W_{t} - \eta_{t}\nabla\sf_{t}(W_{t}) + \eta_{t}\sigma_{t} Z_{t}),
\end{equation}
where $Z_t\sim \N(0, \mathbf{I}_d)$.
% As before, we consider a randomly-stopped version of this algorithm by replacing $\Psi_{\sf_t}$ and $\sigma_t$ in Algorithm~\ref{alg:Randomly_Stopped} with $\Psi^{\mathsf{OGD}}_{\sf_t}$ and , respectively.
A standard quantity for measuring the utility of typical online algorithms is regret described in Section~\ref{Sec:OnlinePrivacy}. To take into account our assumptions that the stopping time is random and \textit{only} $W_{n+1}$ is available for analysis, we choose to measure utility in terms of  \textit{stochastic regret} \cite{Online_One_Shot_Frank_Wolfe}. 
%it is implicitly assumed in this definition that the algorithm stops at time $n$ and \textit{all} $W_1, \dots W_{n+1}$ are available for analysis. However, these assumption are not applicable in our setting: the algorithm stops at a random time $T\in [n]$  and only $W_{n+1}$ is available for analysis. We therefore need to consider a variant of regret that takes these requirements into consideration. 
Recall that $\mathsf{F}$ denotes the collection of all possible real-valued cost functions over $\W$.
Here, we assume that $\sF$ consists of all convex functions over $\W$ with uniformly bounded gradients and  
let $\mu$ be a distribution over $\sF$. We further assume that the cost functions $\{\sf_t\}$ are independently sampled from $\sF$ according to $\mu$. 
The \textit{stochastic regret} for the randomly stopped OGD algorithm is given by 
$$SR(n)\coloneqq \E[\sf(W_T)] - \inf_{w\in \W}\sf(w),$$
where the expectation is taken with respect to the randomness in stopping time $T$ and $\sf(w) = \E_\mu[\sf_t(w)]$. 
%Note that $n SR(n)$ can be thought of as the \textit{average} regret of an algorithm when applied several times. 
It is worth noting that stochastic regret is sometimes referred to as \textit{expected excess population loss} in the offline setting, see e.g., \cite{BAssily_NIPS19, Feldman_LinearTime,Feldman_generalization,Bassily_Stability}. 
%The utility of the randomly stopped OGD algorithm is stated next.   
\begin{proposition}
\label{Prop:Utility}
Assume that $\mathcal{W}\subset\mathbb{R}^d$ is a closed convex set, $\sF$ is a family of convex functions over $\W$ with uniformly bounded gradients, $\{\sf_t\}$ are $n$ independent samples from $\sF$, and $P_Z = \N(0, \mathbf{I}_d)$. If $\displaystyle \eta_{t} = \frac{\textnormal{dia}(\W)}{M\sqrt{t}}$ with $\displaystyle M = \sup_{\sf_1\in\sF} \sup_{w\in\W} \lVert \nabla \sf_{1}(w) \rVert$, then the randomly stopped OGD algorithm satisfies
\begin{equation*}
SR(n) \leq \frac{3M\textnormal{dia}(\W)}{2\sqrt{n}} + \frac{d}{2n} \sum_{t=1}^{n} \eta_{t}\sigma_{t}^{2}.
\end{equation*}
\begin{equation*}
    % \E[\sf(W_{T})] - \inf_{w\in\W} \sf(w) \leq \frac{3M\textnormal{dia}(\W)}{2\sqrt{n}} + \frac{d}{2n} \sum_{t=1}^{n} \eta_{t}\sigma_{t}^{2},
\end{equation*}
%where $\sf(w) \coloneqq \E_\mu[\sf_{t}(w)]$.
\end{proposition}

This proposition follows from standard results in online convex optimization, see, e.g., \cite{Hazan_Online}. For the reader's convenience, we provide a full proof in Appendix~\ref{Appendix:ProofPropUtility}.
Combined with Theorem~\ref{Thm:General_Random_DP}, this  proposition can be applied to capture the balance between privacy and utility of OGD algorithm. 
In particular, they elucidate that if $\eta_t = O(\frac{1}{\sqrt{t}})$, then a necessary condition for non-trivial privacy and utility is $\sigma_t\to \infty$ and $\frac{\sigma_t}{\sqrt{t}}\to 0$, i.e., $\{\sigma_t\}$ must be an increasing sequence but with a rate slower than $\sqrt{t}$.

It is worth noting that Feldman et al. \cite{Feldman_LinearTime} recently proposed an offline algorithm, termed \textit{Snowball-SGD}, which is shown to be order optimal in both the expected excess population loss and also privacy guarantee. However, this algorithm differs from ours most notably in the batch size. While the batch size in our analysis is assumed to be one and the stopping time is random, in Snowball-SGD the batch size gradually increases from one to $\lceil\sqrt{d}\rceil$ as the algorithm progresses.

	\small
	\bibliographystyle{IEEEtran}
	\bibliography{reference}
	\normalsize

	\appendix
	\subsection{Proof of Theorem~\ref{Thm:Contraction_EGamma}}\label{Appendix:Proof_THM_Contraction}
	
	Observe that the case $\gamma = 1$ corresponds to Dobrushin's formula \eqref{eq:Dobrushin}, so we assume that $\gamma > 1$.
	
	We begin by showing that, for any two probability measures $\mu,\nu\in\P(\D)$,
	\begin{equation}
		\label{eq:ProofDobrushinFirstPart}
		\sE_\gamma(\mu\sK\|\nu\sK) \leq \sE_\gamma(\mu\|\nu) \sup_{x_1,x_2\in\D} \sE_\gamma(\sK(\cdot \vert x_1)\|\sK(\cdot \vert x_2)).
	\end{equation}
	Let $\phi \coloneqq \mu - \gamma\nu$ and $(\phi^{+},\phi^{-})$ be its Hahn-Jordan decomposition. By the definition of $\sE_\gamma$ in \eqref{eq:DefEgamma}, we have that $\sE_\gamma(\mu\|\nu) = \lVert\phi^{+}\rVert$ and
	\begin{align}
		\nonumber \sE_\gamma(\mu\|\nu) &= \frac{1}{2}\lVert\phi^{+}\rVert + \frac{1}{2}\lVert\phi^{-}\rVert + \frac{1}{2}\lVert\phi^{+}\rVert - \frac{1}{2}\lVert\phi^{-}\rVert\\
		\label{eq:EgammaTVRelation} &= \frac{1}{2} \lVert\phi\rVert + \frac{1}{2}(1-\gamma),
	\end{align}
	where the last equality follows from the fact that $\phi^{+}$ and $\phi^{-}$ are positive measures and thus
	\begin{equation}
		\label{eq:ProofDobrushinIdentityPhiPlusMinusPhi}
		\lVert\phi^{+}\rVert - \lVert\phi^{-}\rVert = \phi^{+}(\D) - \phi^{-}(\D) = \phi(\D) = 1 - \gamma.
	\end{equation}
	\textit{Mutatis mutandis}, we have that
	\begin{equation}
		\label{eq:ProofDobrushinEgammaInitial}
		\sE_\gamma(\mu\sK\|\nu\sK) = \lVert(\phi\sK)^{+}\rVert = \frac{1}{2}\lVert\phi\sK\rVert + \frac{1}{2}(1-\gamma).
	\end{equation}
	It follows from the definition of the Hahn-Jordan decomposition that if $\phi^{+} \equiv 0$, then $(\phi\sK)^{+} \equiv 0$. In this case, \eqref{eq:ProofDobrushinFirstPart} holds trivially as
	\begin{equation*}
		\sE_\gamma(\mu\sK\|\nu\sK) = \lVert(\phi\sK)^{+}\rVert = 0 = \lVert\phi^{+}\rVert = \sE_\gamma(\mu\|\nu).
	\end{equation*}
	Now assume that $\phi^{+}$ is not the trivial measure. By \eqref{eq:ProofDobrushinIdentityPhiPlusMinusPhi},
	\begin{equation*}
		\phi^{+}(\D) - \phi^{-}(\D) = 1 - \gamma < 0,
	\end{equation*}
	which implies that $\phi^{-}(\D) > 0$, i.e., $\phi^{-}$ is not the trivial measure. Hence, both $\phi^{+}/\lVert\phi^{+}\rVert$ and $\phi^{-}/\lVert\phi^{-}\rVert$ are well-defined probability measures. As a result, we have that
	\begin{align}
		\nonumber \lVert\phi\sK\rVert &= \left\lVert\int \sK(\cdot \vert x_1) \text{d}\phi^{+}(x_1) - \int \sK(\cdot \vert x_2) \text{d}\phi^{-}(x_2)\right\rVert\\
		\nonumber &= \bigg\lVert\int\int \Big[\lVert\phi^{+}\rVert \sK(\cdot \vert x_1) - \lVert\phi^{-}\rVert \sK(\cdot \vert x_2)\Big] \frac{\text{d}\phi^{+}(x_1)}{\lVert\phi^{+}\rVert} \frac{\text{d}\phi^{-}(x_2)}{\lVert\phi^{-}\rVert}\bigg\rVert\\
		\label{eq:ProofDobrushinNormPhiK} &\leq \sup_{x_1,x_2\in\D} \left\lVert\lVert\phi^{+}\rVert \sK(\cdot \vert x_1) - \lVert\phi^{-}\rVert \sK(\cdot \vert x_2)\right\rVert.
	\end{align}
	For ease of notation, let
	\begin{equation*}
		\psi(x_{1},x_{2}) \coloneqq \left\lVert\lVert\phi^{+}\rVert \sK(\cdot \vert x_1) - \lVert\phi^{-}\rVert \sK(\cdot \vert x_2)\right\rVert.
	\end{equation*}
	By adding and subtracting the term $\gamma\lVert\phi^{+}\rVert\sK(\cdot \vert x_2)$, the triangle inequality implies that
	\begin{equation*}
		\psi(x_{1},x_{2}) \leq \lVert\phi^{+}\rVert \lVert\sK(\cdot \vert x_1) - \gamma\sK(\cdot \vert x_2)\rVert + \lVert\phi^{-}\rVert - \gamma\lVert\phi^{+}\rVert,
	\end{equation*}
	where we used the inequality $\lVert\phi^{-}\rVert - \gamma\lVert\phi^{+}\rVert \geq 0$. In addition, a direct computation shows that
	\begin{align*}
		\psi(x_{1},x_{2}) &\leq \lVert\phi^{+}\rVert \left(\lVert\sK(\cdot \vert x_1) - \gamma\sK(\cdot \vert x_2)\rVert+ 1-\gamma\right)- (\lVert\phi^{+}\rVert - \lVert\phi^{-}\rVert).
	\end{align*}
	Therefore, \eqref{eq:EgammaTVRelation} and \eqref{eq:ProofDobrushinIdentityPhiPlusMinusPhi} imply that
	\begin{align*}
		\psi(x_{1},x_{2}) \leq 2 \lVert\phi^{+}\rVert \sE_{\gamma}(\sK(\cdot \vert x_1) \Vert \sK(\cdot \vert x_2)) - (1-\gamma).
	\end{align*}
	As a result, \eqref{eq:ProofDobrushinNormPhiK} becomes
	\begin{equation*}
		\lVert\phi\sK\rVert \leq 2\lVert\phi^{+}\rVert \sup_{x_1,x_2\in\D} \sE_{\gamma}(\sK(\cdot \vert x_1) \Vert \sK(\cdot \vert x_2)) - (1-\gamma).
	\end{equation*}
	By \eqref{eq:ProofDobrushinEgammaInitial} and the fact that $\sE_\gamma(\mu\|\nu) = \lVert\phi^{+}\rVert$, the previous inequality becomes
	\begin{equation*}
		\sE_\gamma(\mu\sK\|\nu\sK) \leq \sE_\gamma(\mu\|\nu) \sup_{x_1,x_2\in\D} \sE_\gamma(\sK(\cdot \vert x_1)\|\sK(\cdot \vert x_2))
	\end{equation*}
	or, equivalently,
	\begin{equation}
		\label{eq:ProofDobrushinFirstPartFinal}
		\eta_\gamma(\sK) \leq \sup_{x_1,x_2\in\D} \sE_\gamma(\sK(\cdot \vert x_1)\|\sK(\cdot \vert x_2)).
	\end{equation}
	
	Now we show that the reverse inequality in \eqref{eq:ProofDobrushinFirstPartFinal} holds, and hence the desired equality. For $x\in\D$, let $\delta_{x}$ be the Dirac mass at $x$. Observe that for $x_1,x_2\in\D$ such that $x_1 \neq x_2$, we have that
	\begin{equation*}
		\sE_\gamma(\delta_{x_1}\|\delta_{x_2}) = (\delta_{x_1} - \gamma\delta_{x_2})^{+}(\D) = 1.
	\end{equation*}
	Since $\delta_{x}\sK = \sK(\cdot \vert x)$ for every $x\in\D$, the previous equality implies that
	\begin{equation*}
		\frac{\sE_\gamma(\delta_{x_1}\sK\|\delta_{x_2}\sK)}{\sE_\gamma(\delta_{x_1}\|\delta_{x_2})} = \sE_\gamma(\sK(\cdot \vert x_1)\|\sK(\cdot \vert x_2)).
	\end{equation*}
	Therefore, by the definition of $\eta_\gamma(\sK)$ in \eqref{Eq:SDPI_f_Div},
	\begin{align*}
		\eta_\gamma(\sK) &\geq \sup_{x_1\neq x_2} \frac{\sE_\gamma(\delta_{x_1}\sK\|\delta_{x_2}\sK)}{\sE_\gamma(\delta_{x_1}\|\delta_{x_2})}\\
		&= \sup_{x_1\neq x_2} \sE_\gamma(\sK(\cdot \vert x_1)\|\sK(\cdot \vert x_2)).
	\end{align*}
	Since $\sE_\gamma(\sK(\cdot \vert x)\|\sK(\cdot \vert x)) = 0$ for every $x\in\mathcal{D}$ and $\sE_{\gamma}$ is non-negative, the desired inequality follows. 
	
	Let $\gamma\in(0,1)$. Similar to \eqref{eq:EgammaTVRelation}, it can be shown that, for any probability measures $\mu,\nu\in\mathcal{P}(\mathcal{D})$,
	\begin{equation*}
		\sE_{\gamma}(\mu \Vert \nu) = \frac{1}{2} \lVert\mu-\gamma\nu\rVert - \frac{1}{2}(1-\gamma).
	\end{equation*}
	A straightforward manipulation leads to
	\begin{equation*}
		\sE_{\gamma}(\mu \Vert \nu) = \gamma \left(\frac{1}{2} \lVert\nu-(1/\gamma)\mu\rVert + \frac{1}{2}(1-1/\gamma)\right).
	\end{equation*}
	Since $1/\gamma > 1$, \eqref{eq:EgammaTVRelation} implies that
	\begin{equation}
		\label{eq:ProofDobrushinReciprocityEgamma}
		\sE_{\gamma}(\mu \Vert \nu) = \gamma \sE_{1/\gamma}(\nu \Vert \mu).
	\end{equation}
	In particular, $\sE_{\gamma}(\mu \Vert \nu) = 0$ if and only if $\sE_{1/\gamma}(\nu \Vert \mu) =0$. Therefore, \eqref{eq:ProofDobrushinReciprocityEgamma} and the definition of $\eta_\gamma(\sK)$ in \eqref{Eq:SDPI_f_Div} imply
	\begin{align*}
		\eta_{\gamma}(\sK) &= \sup_{\substack{\mu,\nu\in\mathcal{P}(\mathcal{D}):\\\sE_{\gamma}(\mu\Vert\nu)\neq0}} \frac{\sE_{\gamma}(\mu\sK\Vert\nu\sK)}{\sE_{\gamma}(\mu\Vert\nu)}\\
		&= \sup_{\substack{\mu,\nu\in\mathcal{P}(\mathcal{D}):\\\sE_{1/\gamma}(\nu\Vert\mu)\neq0}} \frac{\sE_{1/\gamma}(\nu\sK\Vert\mu\sK)}{\sE_{1/\gamma}(\nu\Vert\mu)}\\
		&= \eta_{1/\gamma}(\sK),
	\end{align*}
	as we wanted to show.
	%%%%%%%

	\subsection{Proof of Proposition~\ref{Proposition:etaPAGK}}
	\label{Appendix:Prop_etaPAGK}

	By Theorem~\ref{Thm:Contraction_EGamma}, we have that
	\begin{align*}
		\eta_{\gamma}(\sK) &=  \sup_{x_1,x_2\in\mathcal{D}} \sE_{\gamma}(\sK(\cdot \vert x_1)\|\sK(\cdot \vert x_2))\\
		&= \sup_{x_1,x_2\in\mathcal{D}} \sE_{\gamma}(\mathcal{N}(x_1,\sigma^2\mathbf{I}_d)\|\mathcal{N}(x_2,\sigma^2\mathbf{I}_d)),
	\end{align*}
	It could be verified that $r\mapsto\theta_\gamma(r)$ is increasing. Hence, by \eqref{theta_Gaussian}, we conclude that
	\begin{align*}
		\eta_{\gamma}(\sK) &= \sup_{x_1,x_2\in\mathcal{D}} \theta_\gamma\left(\frac{\|x_2-x_1\|}{\sigma}\right)\\
		&= \theta_\gamma\left(\frac{\textnormal{dia}(\mathcal{D})}{\sigma}\right),
	\end{align*}
	as desired.
	%%%%%%%

	% \subsection{Proof of Theorem~\ref{Thm:Informal_SDPI}}
	% \label{Appendix:DPIIf}
	% Recall that, by definition,
	% \begin{equation*}
		%     \mu_{n+1} = \mu_{1} \sK_{\sf_1} \cdots \sK_{\sf_n} \quad \text{and} \quad \mu_{n+1} = \mu_{1} \sK_{\sf'_1} \cdots \sK_{\sf'_n}.
		% \end{equation*}
	% Since $\{\sf_t\}\stackrel{i}{\sim} \{\sf'_t\}$, we have $\sK_{\sf_t} = \sK_{\sf'_t}$ for all $t\in [n]\backslash{\{i\}}$ and $\sK_{\sf_i}\neq \sK_{\sf'_i}$. In particular,
	% \begin{align*}
		%     \mu_{n+1} &= \mu_{i} \sK_{\sf_i} \sK_{\sf_{i+1}} \cdots \sK_{\sf_n}\\
		%     \mu'_{n+1} &= \mu_{i} \sK_{\sf'_i} \sK_{\sf_{i+1}} \cdots \sK_{\sf_n}.
		% \end{align*}
	% By repeatedly invoking \eqref{Eq:SDPI_f_Div}, we obtain that
	% \begin{align*}
		%     D_f(\mu_{n+1}\|\mu'_{n+1}) & = D_f(\mu_{i} \sK_{\sf_i} \cdots \sK_{\sf_n} \| \mu_{i} \sK_{\sf'_i} \cdots \sK_{\sf_n})\\
		%     & \leq D_f(\mu_{i}\sK_{\sf_i}\|\mu_{i}\sK_{\sf'_i})\prod_{t=i+1}^n\eta_f(\sK_{\sf_t}),
		% \end{align*}
	% as we wanted to prove.
	%%%%%%%

	\subsection{Proof of Lemma~\ref{Lem:LB_Minimax}}
	\label{Appendix:ProofLemmaLB_Minimax}
	
	Recall that $X_{1},\ldots,X_{n}$ are i.i.d.\ random variables and $\sK_i$ is the conditional distribution of $Z_{i}$ given $(X_{i},Z^{i-1})$. Let $P_{Z_{1},\ldots,Z_{n}}$ and $Q_{Z_{1},\ldots,Z_{n}}$ be the distribution of $Z^n$ when $X^{n} \sim P_0^{\otimes n}$ and $X^{n} \sim P_1^{\otimes n}$, respectively. By Pinsker's inequality, we have that
	\begin{align}
		\nonumber  \tv^2(P_{Z^{n}},Q_{Z^{n}})& \leq \frac{1}{2} D_\kl(P_{Z^{n}} \Vert Q_{Z^{n}})\\
		\label{eq:ProofPrivateLeCamTVKL} & = \frac{1}{2} \sum_{i=1}^{n} D_\kl(P_{Z_{i} \vert Z^{i-1}} \Vert Q_{Z_{i} \vert Z^{i-1}} \vert P_{Z^{i-1}}),
	\end{align}
	where the last equality follows from the chain rule for KL divergence. Observe that, for all $z^{i}\in\mathcal{Z}^{i}$,
	\begin{align*}
		P_{Z_{i} \vert Z^{i-1}} (z_{i} \vert z^{i-1}) &= \int_{\mathcal{X}} \sK_{i}(z_{i} \vert x_{i},z^{i-1}) \text{d}P_{0}(x_{i})\\
		&= (P_{0} \otimes \delta_{z^{i-1}})\sK_{i}.
	\end{align*}
	\emph{Mutatis mutandis}, we can show that
	\begin{equation*}
		Q_{Z_{i} \vert Z^{i-1}} (z_{i} \vert z^{i-1}) = (P_{1} \otimes \delta_{z^{i-1}})\sK_{i}.
	\end{equation*}
	Since $\sK_{i}\in\Q_{\eps,\delta}$, \eqref{eq:LDPImpliesContractionDf} implies that, for all $z^{i}\in\mathcal{Z}^{i}$,
	\begin{align*}
		D_{\kl}((P_{0} \otimes \delta_{z^{i-1}})\sK_{i} \Vert (P_{1} \otimes \delta_{z^{i-1}})\sK_{i})&\leq \varphi(\eps,\delta) D_{\kl}(P_{0} \otimes \delta_{z^{i-1}} \Vert P_{1} \otimes \delta_{z^{i-1}})\\
		& = \varphi(\eps,\delta) D_{\kl}(P_{0} \Vert P_{1}).
	\end{align*}
	Therefore, \eqref{eq:ProofPrivateLeCamTVKL} becomes
	\begin{equation*}
		\tv^2(P_{Z^{n}},Q_{Z^{n}}) \leq \frac{n\varphi(\eps,\delta)}{2} D_{\kl}(P_{0} \Vert P_{1}).
	\end{equation*}
	Plugging the previous inequality in \eqref{LeCam_TV}, we obtain the desired result.
	%\MD{Shahab, I fixed the previous proof. The conditioning was not quite right (that's where the missing arguments were lost). If you have time, please take a quick look.}
	\subsection{Proof of Corollary~\ref{corollary:LB_Pk}}
	\label{Appendix:ProofCorollaryLB_Pk}
	
	Fix $\omega\in (0,1]$ and consider two distributions $P_0$ and $P_1$ on $\{-\omega^{-\frac{1}{k}}, 0, \omega^{-\frac{1}{k}}\}$ defined as
	$$P_0(-\omega^{-\frac{1}{k}}) = \omega, \qquad P_0(0) = 1-\omega,$$
	and 
	$$P_1(\omega^{-\frac{1}{k}}) = \omega, \qquad P_1(0) = 1-\omega.$$
	It can be verified that both $P_0$ and $P_1$ belong to $\P_k$.  Note that $\ell_2^2(\theta(P_0), \theta(P_1)) = 2\omega^{\frac{2(k-1)}{k}}$.
	Let $M_0^n = P^{\otimes n}_0\sK^n$ and $M_1^n=P^{\otimes n}_1\sK^n$ be the corresponding output distributions of the mechanism $\sK^n=\sK_1\dots\sK_n$, the composition of mechanisms $\sK_i$. Le Cam's bound for $\ell_2^2$-metric yields   
	\begin{align}
		\mathcal R_n(\P_k, \ell_2^2, \eps, \delta)&\geq \omega^{\frac{2(k-1)}{k}}(1-\tv(M^n_0, M^n_1)\nonumber\\
		&\geq \omega^{\frac{2(k-1)}{k}}\left(1-H(M^n_0, M^n_1)\right),\label{Hell1}
	\end{align}
	where the last inequality follows from the fact $\tv(P,Q)\leq H(P,Q)$ for $H(P, Q)$ being the Hellinger distance. Notice that $M_0^n = \prod_{i=1}^n(P_0\sK_i)$ and $M_1^n = \prod_{i=1}^n(P_1\sK_i)$ where each $\sK_i$ for $i\in [n]$ is $(\eps, \delta)$-LDP. 
	It is well known that 
	$$H^2\left(\prod_{i=1}^nP_i, \prod_{i=1}^nQ_i\right) = 2-2\prod_{i=1}^n\left(1-\frac{1}{2}H^2(P_i, Q_i)\right).$$
	Thus, 
	\begin{align}
		H^2(M_0^n, M_1^n)&= 2-2\prod_{i=1}^n\left(1-\frac{1}{2}H^2(P_0\sK_i, P_1\sK_i)\right)\nonumber\\
		&\leq 2-2\prod_{i=1}^n\left(1-\frac{\varphi(\eps, \delta)}{2}H^2(P_0, P_1)\right)\nonumber\\
		&= 2-2\left(1-\frac{\varphi(\eps, \delta)}{2}H^2(P_0, P_1)\right)^n\nonumber\\
		&= 2-2\left(1-\omega \varphi(\eps, \delta)\right)^n. \label{Hell2}
	\end{align}
	Hence, we obtain 
	\eqn{TV_UB_EXample}{\tv(M_0^n, M_1^n)\leq \sqrt{2-2\left(1-\omega \varphi(\eps, \delta)\right)^n}.}
	Plugging \eqref{Hell2} into \eqref{Hell1}, we obtain 
	\begin{align*}
		& \mathcal R_n(\P_k, \ell_2^2, \eps, \delta)\geq \omega^{\frac{2(k-1)}{k}}\left[1-\sqrt{2}\sqrt{1-(1-\omega\varphi(\eps,\delta))^n}\right].
	\end{align*}
	Now, choose
	$\omega = \min\left\{1, \frac{1}{\varphi(\eps, \delta)}\left[1-\left(\frac{7}{8}\right)^{\frac{1}{\sqrt{n}}}\right]\right\}$. Notice that we assume $\delta>0$ and hence $\varphi(\eps, \delta)>0$ regardless of $\eps$. Plugging this choice of $\omega$ into the above bound, we obtain 
	\begin{align*}
		\mathcal R_n(\P_k, \ell_2^2, \eps, \delta)&\gtrsim (\varphi(\eps, \delta))^{-\frac{2(k-1)}{k}}\left[1-\left(\frac{7}{8}\right)^{\frac{1}{\sqrt{n}}}\right]^{\frac{2(k-1)}{k}}\\
		&  \gtrsim (\varphi(\eps, \delta))^{-\frac{2(k-1)}{k}} n^{-\frac{k-1}{k}}.
	\end{align*}

	\subsection{Proof of Lemma~\ref{Lem:LB_Minimax_Alternative}}
	\label{Appendix:ProofLemmaLB_Minimax_Alternative}
	
	Recall that $X_{1},\ldots,X_{n}$ are i.i.d.\ random variables and $\sK_i$ is the conditional distribution of $Z_{i}$ given $X_{i}$. Note that the distribution of $Z^{n}$ is $P_{0}^{\otimes n} K^{n}$ or $P_{1}^{\otimes n} K^{n}$ whenever $X^{n} \sim P_{0}^{\otimes n}$ or $X^{n} \sim P_{1}^{\otimes n}$, respectively. By the definition of contraction coefficient, we have that
	\begin{align*}
		\tv(P_{0}^{\otimes n}K^{n},P_{1}^{\otimes n} K^{n}) &\leq \eta_{\tv}(K^{n}) \tv(P_{0}^{\otimes n},P_{1}^{\otimes n})\\
		&\leq \eta_{\tv}(K^{n}) \sqrt{\frac{n}{2} D_{\kl}(P_{0} \Vert P_{1})},
	\end{align*}
	where the last inequality is due to Pinsker's inequality and the tensorization of KL-divergence. Since in the non-interactive setting we have that $K^{n} = K^{\otimes n}$, Lemma~\ref{lemma:UB_Eta_Kn_general} implies that
	\begin{align*}
		\eta_{\tv}(\sK^{n}) &\leq \max_{i\in [n]}\left[1-\left(\frac{1-\eta_{e^\eps}(\sK_{i})}{e^\eps}\right)^n\right]\\
		&\leq 1 - e^{-n\eps}(1-\delta)^{n},
	\end{align*}
	where the last inequality follows from Theorem~\ref{thm:LDP_Contraction} and the fact that $K_{i}\in\mathcal{Q}_{\eps,\delta}$ for every $i\in[n]$.
	
	%\MD{This proof is new, a quick check won't hurt.}
	%%%%%%%

	\subsection{Proof of Theorem~\ref{Thm:Bayesian_LB}}
	
	\label{Appendix:ProofTheoremBayesian_LB}
	Let $\hat \Theta = \Psi(Z^n)$ be an estimate of $\Theta$ for some $\Psi$ and $p_\zeta\coloneqq P_{\Theta\hat \Theta}(\ell(\Theta, \hat \Theta)\leq \zeta)$ and $q_\zeta\coloneqq (P_{\Theta}P_{\hat \Theta})(\ell(\Theta, \hat \Theta)\leq \zeta)$, i.e., $p_\zeta$ and $q_\zeta$ correspond to the probability of the event $\{\ell(\Theta, \hat\Theta)\leq \zeta\}$ under the joint and product distributions, respectively. By definition, we have for any $\gamma\geq 1$
	\al{I_\gamma(\Theta; \hat \Theta) &= 
		\sE_\gamma(P_{\Theta\hat\Theta}\| P_\Theta P_{\hat\Theta}) \\ 
		&=\sup_{A\subset \T\times \T} \left[P_{\Theta\hat \Theta}(A) - \gamma (P_\Theta P_{\hat \Theta})(A)\right]\\
		&\geq p_\zeta - \gamma q_\zeta\\
		&\geq p_\zeta-\gamma\L(\zeta),}
	where the last inequality follows from the inequality $q_\zeta\leq \L(\zeta)$. Indeed, observe that
	\al{q_\zeta &= \int_{\T}\int_{\T}1_{\{\ell(\theta, \hat\theta)\leq\zeta\}}P_{\Theta}(\text{d}\theta) P_{\hat\Theta}(\text{d}\hat\theta)\\
		&\leq \sup_{t\in \T}\int_{\T}1_{\{\ell(\theta, t)\leq\zeta\}}P_{\Theta}(\text{d}\theta)\\
		& = \L(\zeta).}
	Recalling that $\Pr(\ell(\Theta, \hat \Theta)>\zeta) = 1-p_\zeta$, the above thus implies 
	\eqn{}{\Pr(\ell(\Theta, \hat \Theta)>\zeta)\geq 1-I_\gamma(\Theta; \hat \Theta)-\gamma\L(\zeta).}
	Since $\E[\ell(\Theta, \hat \Theta)]\geq \zeta\Pr(\ell(\Theta, \hat \Theta)\geq \zeta)$ by Markov's inequality, we can write by setting $\gamma = e^\eps$
	\al{R_n^{\mathsf{Bayes}}(P_\Theta, \ell, \eps, \delta)&\geq \zeta \left[1-I_{e^\eps}(\Theta; \hat \Theta)-e^\eps\L(\zeta)\right]\\
		&\geq \zeta \left[1-I_{e^\eps}(\Theta; Z^n)-e^\eps\L(\zeta)\right],}
	where the second inequality comes from the data processing inequality for $I_\gamma$. To further lower bound the right-hand side, we write
	\begin{align*}
		I_{e^\eps}(\Theta; Z^n)& = \int_{\T} \sE_{e^\eps}(P_{Z^n|\Theta=\theta}\|P_{Z^n})P_{\Theta}(\text{d}\theta)\\
		&  \leq \eta_{e^\eps}(P_{Z^n|X^n})\int_{\T} \sE_{e^\eps}(P_{X^n|\Theta=\theta}\|P_{X^n})P_{\Theta}(\text{d}\theta)\\
		& = \eta_{e^\eps}(P_{Z^n|X^n}) I_{e^\eps}(\Theta; X^n),
	\end{align*}
	where the inequality follows from the definition of contraction coefficient.
	%= I_{e^\eps}(\Theta; X^n) \left[1-(1-\delta)^n\right].} 
When $n=1$, we have $\eta_{e^\eps}(\sK)\leq \delta$ as $\sK=P_{Z|X}$ is assumed to be $(\eps, \delta)$-DP.  
%$I_{e^\eps}(\Theta; Z)\leq \eta_{e^\eps}(\sK)I_{e^\eps}(\Theta; X)\leq \delta I_{e^\eps}(\Theta; X)$ where the Markov kernel $\sK$ is specified by $P_{Z|X}$. 
For $n>1$, we invoke \eqref{eq:LDPImpliesContractionDf_n} to obtain $\eta_{e^\eps}(P_{Z^n|X^n}) \leq \varphi_n(\eps, \delta)$.
%%%%%%%

\subsection{Proof of Corollary~\ref{Cor:BHT}}
\label{Appendix:ProofCorollaryBHT}

Let $\beta_n(\alpha)\coloneqq \beta^{\infty, 1}_n(\alpha)$ be the non-private trade-off between type I and type II error probabilities (i.e., $Z^n = X^n$). According to Chernoff-Stein lemma (see, e.g., \cite[Theorem 11.8.3]{cover2012elements}), we have
\eqn{Chernoff}{\lim_{n\to \infty}\frac{1}{n}\log\beta_n(\alpha) = -D_\kl(P_0\|P_1).}
Assume now that, $Z^n$ is the output of  $\sK^{\otimes n}$ for an $(\eps, \delta)$-LDP mechanism $\sK$.  
%observes $Z^n$ and wishes  to distinguish between $H_0$ and $H_1$. That is, individuals pass their data $X_i$, $i\in [n]$ through a (non-interactive) mechanism $\sK\in \Q_{\eps, \delta}$ and generate $Z_i$.
According to \eqref{Chernoff}, we obtain that
\eqn{Chernoff2}{\lim_{n\to \infty}\frac{1}{n}\log\beta^{\eps, \delta}_n(\alpha) = -\sup_{\sK\in \Q_{\eps, \delta}}D_\kl(P_0\sK\|P_1\sK).}
By \eqref{eq:LDPImpliesContractionDf}, we obtain the desired result. 
%\MD{I changed an old reference for \eqref{eq:LDPImpliesContractionDf}. Shahab, does it make sense?}
%%%%%%%

\subsection{Proof of Theorem~\ref{Thm:General_DP}}\label{Appendix:Thm:General_DP}

Let $\mu_1$ be the initial distribution of the iterative process and  $\{\sf_t\} \stackrel{i}{\sim} \{\sf'_t\}$ be a pair of neighboring collections of cost functions. In light of Definition~\ref{Def:DP_Online} and inequality~\eqref{Eq:EGamma_PNSGD}, we have that
\begin{equation}\label{proof_Thm4_1}
	\delta = \sE_{e^\eps}(\mu_{i}\sK_{\sf_i}\|\mu_{i}\sK_{\sf'_i})\prod_{t=i+1}^{n} \eta_{e^\eps}(\sK_{\sf_t}).
\end{equation}
%\MD{I changed a missing reference for Definition~\ref{Def:DP_Online}. Shahab, does it make sense?}

We begin by bounding  $\eta_{\gamma}(\sK_{\sf_t})$. Each kernel $\sK_{\sf_t}$ can be decomposed as
\begin{equation*}
	\sK_{\sf_t} = \Pi_{\W}\circ\sK_{t} \circ \Psi_{\sf_t},
\end{equation*}
where $\sK_t$ is a $\Psi_{\sf_t}(\W)$-constrained additive Gaussian kernel with noise magnitude given by $\sigma_t^2$, as defined in Section~\ref{Section:ContractionEgamma}.
By Theorem~\ref{Thm:Contraction_EGamma}, we have 
\begin{align}
	\eta_{e^\eps}(\sK_{\sf_t}) &= \sup_{w_1,w_2\in\mathcal{W}} \sE_{e^\eps}(\sK_t(\Psi_{\sf_t}(w_1)) \| \sK_t(\Psi_{\sf_t}(w_2)))\nonumber\\
	&= \sup_{w_1,w_2\in\Psi_{\sf_t}(\mathcal{W})} \sE_{e^\eps}(\sK_t(w_1) \| \sK_t(w_2))\nonumber\\
	&= \eta_{{e^\eps}}(\sK_t)\nonumber\\
	&= \theta_{{e^\eps}}\left(\frac{\textnormal{dia}(\Psi_{\sf_{t}}(\mathcal{W})) }{\sigma_t}\right), \label{proof_Thm4_2}
\end{align}
where the last equality comes from Proposition~\ref{Proposition:etaPAGK}.

Next, we look at the first term in the RHS of \eqref{proof_Thm4_1}. By Jensen's inequality, we can write 
\begin{equation*}
	\sE_{e^\eps}(\mu_{i}\sK_{\sf_i} \| \mu_{i} \sK_{\sf_i'}) \leq \int_{\mathcal{W}} \sE_{e^\eps}(\sK_{\sf_i}(w) \| \sK_{\sf_i'}(w)) \text{d} \mu_{i}(w).
\end{equation*}
Observe that, for every $w\in\mathcal{W}$, \begin{equation*}
	\sK_{\sf_i}(w) \sim \Pi_\mathcal{W}(\Psi_{\sf_i}(w) + \sigma_i Z),
\end{equation*}
where $Z \sim \mathcal{N}(0,\mathbf{I}_d)$. A similar relation holds for $\sK_{\sf'_i}$.
% \begin{equation*}
	%     \sK_{\sf'_i}(w) \sim \Pi_\mathcal{W}(\Psi_{\sf'_i}(w) + \sigma_i Z).
	% \end{equation*}
By the data processing inequality, we obtain that
\begin{align*}
	& \sE_{e^\eps}(\sK_{\sf_i}(w) \| \sK_{\sf_i'}(w))  \leq \sE_{e^\eps}(\Psi_{\sf_i}(w) + \sigma_i Z \| \Psi_{\sf'_i}(w) + \sigma_i Z).
\end{align*}
Therefore, Lemma~\ref{Lemma:EgammaGaussian} implies that 
\begin{align*}
	& \sE_{e^\eps}(\mu_{i}\sK_{\sf_i} \| \mu_{i} \sK_{\sf_i'})\leq \int_{\W}\theta_{e^\eps}\left(\frac{\|\Psi_{\sf_i}(y) - \Psi_{\sf'_i}(y)\|}{\sigma_i}\right) \text{d} \mu_{i}(y).
\end{align*}
Since $r\mapsto \theta_\gamma(r)$ is increasing, we conclude that
\begin{equation}
	\label{eq:ThmGaussianMechanismPsi}
	\sE_{e^\eps}(\mu_{i}\sK_{\sf_i} \| \mu_{i} \sK_{\sf_i'}) \leq \theta_{e^\eps}\left(\frac{\psi}{\sigma_i}\right), 
\end{equation}
By plugging \eqref{proof_Thm4_2} and \eqref{eq:ThmGaussianMechanismPsi} into \eqref{proof_Thm4_1}, we obtain the desired result.
%%%%%%%

\subsection{Proof of Theorem~\ref{Thm:General_Random_DP}}
\label{Appendox:Thm_General_RandomDP}

Assume that  $\{\sf_t\}$ and $\{\sf'_t\}$ are collections of cost functions such that $\{\sf_t\}\stackrel{i}{\sim}\{\sf_t'\}$ for some $i\in [n]$. Let $T$ be a uniform random variable over $[n]$. If $\mu_{T+1}$ and $\mu'_{T+1}$ are the distributions of the output of Algorithm~\ref{alg:Randomly_Stopped} applied to $\{\sf_t\}$ and $\{\sf'_t\}$, respectively, then
\begin{align*}
	\mu_{T+1} &= \frac{1}{n} \sum_{t=1}^n \mu_1 \sK_{\sf_1}\dots \sK_{\sf_t},\\
	\mu'_{T+1} &= \frac{1}{n}\sum_{t=1}^{n} \mu_{1}\sK_{\sf'_1}\dots \sK_{\sf'_t}.
\end{align*}
The convexity\footnote{For any convex function $f$ on $\R_+$, its perspective, i.e.,  $(p, q)\mapsto qf\big(\frac{p}{q}\big)$, 
	is convex on $\R_+^2$. Since $D_f(\mu\|\nu) = \E_\nu\Big[f\big(\frac{\text{d}\mu}{\text{d}\nu}\big)\Big]$, it follows that $(\mu, \nu)\mapsto D_f(\mu\|\nu)$ is convex.} of $(\mu, \nu)\mapsto \sE_\gamma(\mu\|\nu)$  and Jensen's inequality imply that
\begin{align}
	\sE_{e^\eps}(\mu_{T+1}\|\mu'_{T+1})\leq  \frac{1}{n} \sum_{t = 1}^n \sE_{e^\eps}( \mu_1 \sK_{\sf_1}\dots \sK_{\sf_t} \| \mu_1 \sK_{\sf'_1}\dots \sK_{\sf'_t}).\label{Proof_RandoM_Jensen}
\end{align}
Recall that $\sf_j = \sf_j'$ for all $j\neq i$. In particular, we have that $\mu_1 \sK_{\sf_1}\dots \sK_{\sf_t} = \mu_1 \sK_{\sf'_1}\dots \sK_{\sf'_t}$ for all $t<i$ and hence
\begin{align}
	\sE_{e^\eps}(\mu_{T+1}\|\mu'_{T+1})\leq  \frac{1}{n} \sum_{t = i}^n   \sE_{e^\eps}(\mu_i \sK_{\sf_i}\dots \sK_{\sf_t} \| \mu_i \sK_{\sf'_i}\dots \sK_{\sf'_t}),\label{Proof_THm_Random1}
\end{align}
where $\mu_{i} = \mu_1 \sK_{\sf_1}\dots \sK_{\sf_{i-1}}$. As in the proof of Theorem~\ref{Thm:General_DP}, each summand can be bounded via inequality~\eqref{Eq:EGamma_PNSGD} to obtain
\begin{align*}
	\sE_{e^\eps}(\mu_{T+1}\|\mu'_{T+1})\leq  \frac{1}{n} \sum_{t = i}^n   \sE_{e^\eps}(\mu_{i}\sK_{\sf_i}\|\mu_{i}\sK_{\sf'_i})\prod_{j=i+1}^t \eta_{e^\eps}(\sK_{\sf_j}).
\end{align*}
Furthermore, \eqref{proof_Thm4_2} and \eqref{eq:ThmGaussianMechanismPsi} imply that
\begin{align*}
	\sE_{e^\eps}(\mu_{T+1}\|\mu'_{T+1})\leq \frac{1}{n} \sum_{t=i}^{n} \theta_{e^\eps}\left(\frac{\psi}{\sigma_{i}}\right) \prod_{j=i+1}^{t} \theta_{{e^\eps}}\left(\frac{\textnormal{dia}(\Psi_{\sf_{j}}(\mathcal{W})) }{\sigma_j}\right),
\end{align*}
from which, and Definition~\ref{Def:DP_Online}, we obtain \eqref{eq:PrivacyRandomlyStopper1}. By further exploiting the monotonicity of $r\mapsto\theta_{\gamma}(r)$, we obtain that
\begin{equation*}
	\delta \leq  \frac{1}{n} \sum_{t = i}^n   \theta_{e^\eps}\Big(\frac{\psi}{\sigma}\Big)\prod_{j=i+1}^t \theta_{e^\eps}\Big(\frac{D}{\sigma}\Big).
\end{equation*}
A straightforward manipulation leads to
\begin{align*}
	\delta &\leq \frac{1}{n} \theta_{e^\eps}\Big(\frac{\psi}{\sigma}\Big)\sum_{t = 0}^{n-i}    \left[\theta_{e^\eps}\Big(\frac{D}{\sigma}\Big)\right]^{t}\\
	&\leq \frac{1}{n} \theta_{e^\eps}\Big(\frac{\psi}{\sigma}\Big) \left[1-\theta_{e^\eps}\left(\frac{D}{\sigma}\right)\right]^{-1},
\end{align*}
as claimed in \eqref{eq:PrivacyRandomlyStopper2}.

\subsection{Proof of Corollary~\ref{cor:PNSGD}}
\label{Appendix:Cor_PNSGD}
Define $\Psi^{\mathsf{SGD}}_{x}(w) = w - \eta\nabla\ell(w, x)$ for $x\in \X$. The PNSGD algorithm can then be expressed as
$$W_{t+1} = \Pi_\W\left(\Psi^{\mathsf{SGD}}_{x_t}(W_t) + \eta\sigma Z_t\right).$$
Applying Theorem~\ref{Thm:General_Random_DP}, we obtain that this algorithm is $(\eps, \delta)$-DP for $\eps\geq 0$ and 
\begin{equation}\label{Eq:Thm5inSGD}
	\delta = \frac{1}{n}\theta_{e^\eps}\Big(\frac{\psi}{\eta\sigma}\Big)\Big[1-\theta_{e^\eps}\Big(\frac{D}{\eta\sigma}\Big)\Big]^{-1},
\end{equation} where 
$$\psi = \sup_{x_1,x_2\in \X}\sup_{w\in \W}\|\Psi^{\mathsf{SGD}}_{x_1}(w) - \Psi^{\mathsf{SGD}}_{x_2}(w)\|,$$
and $D = \max_{t\in [n]}\mathsf{dia}\big(\Psi^{\mathsf{SGD}}_{x_t}(\W))$. Since $\ell(\cdot,x)$ is $L$-Lipschitz for all $x\in\mathcal{X}$, the triangle inequality implies
\begin{align*}
	\|\Psi^{\mathsf{SGD}}_{x_1}(w) - \Psi^{\mathsf{SGD}}_{x_2}(w)\| & = \eta\|\nabla\ell(w, x_1) - \nabla\ell(w, x_2)\|\\
	%&\leq \eta\left(\|\nabla\ell(w, x_1)\| + \|\nabla\ell(w, x_2)\|\right)\nonumber\\
	& \leq 2\eta L, 
\end{align*}
and thus 
\begin{equation}\label{Proof_Cor_SGD1}
	\psi\leq 2\eta L.
\end{equation}
We can also write for each $t\in [n]$
\begin{align}
	\mathsf{dia}\big(\Psi^{\mathsf{SGD}}_{x_t}(\W))& = \sup_{w_1, w_2\in \W} \|\Psi^{\mathsf{SGD}}_{x_t}(w_1) - \Psi^{\mathsf{SGD}}_{x_t}(w_2)\|\nonumber\\
	& = \sup_{w_1, w_2\in \W} \|w_1 - \eta\nabla\ell(w_1, x_t) - w_2 + \eta\nabla\ell(w_2, x_t)\|\nonumber\\
	& \leq \|w_1 - w_2\| + \eta\|\nabla\ell(w_1, x_t) -\nabla\ell(w_2, x_t)\|\nonumber\\
	& \leq \textnormal{dia}(\W) + 2\eta L, \label{Proof_Cor_SGD2} 
\end{align}
implying 
\begin{equation}\label{Proof_Cor_SGD3} 
	D\leq \textnormal{dia}(\W)+2\eta L.
\end{equation}
Plugging \eqref{Proof_Cor_SGD1} and \eqref{Proof_Cor_SGD3} into \eqref{Eq:Thm5inSGD}, we obtain the desired result. 
%%%%%%%

\subsection{Missing proof of Section~\ref{Sec:Privacy_Amplification}}
\label{Appendix_Contractivity}

We start showing that if $f:\mathcal{W}\to\mathbb{R}$ is convex and $\beta$-smooth, then 
$G(w)\coloneqq w-\eta\nabla f(w)$ is $1$-Lipschitz for $\eta\leq \frac{2}{\beta}$. To do so, observe that for any $w_1, w_2\in \W$,
\begin{align}
	\|G(w_1)-G(w_2)\|^2& = \|w_1-w_2 + \eta\nabla f(w_2) - \eta\nabla f(w_1)\|^2 \nonumber\\
	\nonumber & =  \|w_1-w_2\|^2 + \eta^2\|\nabla f(w_1)-\nabla f(w_2)\|^2- 2\eta\langle\nabla f(w_1)-\nabla f(w_2), w_1-w_2\rangle\nonumber\\
	&\leq \|w_1-w_2\|^2 + \eta \Big(\eta-\frac{2}{\beta}\Big) \|\nabla f(w_1)-\nabla f(w_2)\|^2 \label{Proof_AP_Contrac1}\\
	\nonumber & \leq \|w_1-w_2\|^2,
\end{align}
where the inequality in \eqref{Proof_AP_Contrac1} follows from Lemma~3.11 in \cite{Convex_Bubeck}, which states that $\displaystyle \langle\nabla f(w_1)-\nabla f(w_2), w_1-w_2\rangle\geq \frac{1}{\beta}\|\nabla f(w_1)-\nabla f(w_2)\|^2$.

Replacing $f(w)$ with $\ell(w, x)$, we obtain that the function $w\mapsto \Psi^{\mathsf{SGD}}_x(w)$ is $1$-Lipschitz for all $x\in \X$.  To obtain \eqref{Eq:Delta_PNSGD2}, we modify \eqref{Proof_Cor_SGD2}  in the proof of Corollary~\ref{cor:PNSGD} as follows
\begin{align}
	\nonumber D &= \max_{t\in [n]}\mathsf{dia}\big(\Psi^{\mathsf{SGD}}_{x_t}(\W))\\
	& = \sup_{w_1, w_2\in \W} \|\Psi^{\mathsf{SGD}}_{x_t}(w_1) - \Psi^{\mathsf{SGD}}_{x_t}(w_2)\|\\
	& \leq \textnormal{dia}(\W).  \label{Proof_Cor_SGD4} 
\end{align}
Plugging this and \eqref{Proof_Cor_SGD1} into \eqref{Eq:Thm5inSGD}, we obtain \eqref{Eq:Delta_PNSGD2}.
%%%%%%%

\subsection{Proof of Proposition~\ref{Prop:Utility}}
\label{Appendix:ProofPropUtility}

For ease of notation, let $\displaystyle w^{\ast} \in \argmin_{w\in\W} \sf(w)$ and
\begin{equation*}
	\Delta \coloneqq \E[\sf(W_{T})] - \inf_{w\in\W} \sf(w).
\end{equation*}
Since $T$ is uniformly distributed on $[n]$, we have that
\begin{equation*}
	\E\left[\sf(W_{T}) \mid W^{n} \right] = \frac{1}{n} \sum_{t=1}^{n} \E[\sf(W_{t}) \mid W^{n}],
\end{equation*}
where $W^{n} = (W_{1},\ldots,W_{n})$. Therefore,
\begin{equation*}
	\Delta = \E\left[\frac{1}{n} \sum_{t=1}^{n} \big\{\sf(W_{t}) - \sf(w^{\ast})\big\}\right].
\end{equation*}
Since $\sF$ is a family of convex functions, the function $\sf(w) = \E[\sf_{t}(w)]$ is convex. In particular, it follows that
\begin{equation}
	\label{eq:UtilityProofRed1}
	\Delta \leq \E\left[\frac{1}{n} \sum_{t=1}^{n} \langle \nabla\sf(W_{t}), W_{t}-w^{\ast}\rangle\right].
\end{equation}
Define
\begin{equation*}
	G_{t} \coloneqq \nabla\sf_{t}(W_{t}) - \sigma_{t} Z_{t}.
\end{equation*}
Recall that $\sf_{t}$ and $Z_{t}$ are independent of $W_{t}$. Hence, by differentiation under the integral sign,
\begin{equation*}
	\E\left[G_{t} \mid W_{t}\right] = \nabla\sf(W_{t}).
\end{equation*}
As a consequence, we obtain that
\begin{align*}
	\E[\langle G_{t}, W_{t}-w^{\ast}\rangle] &= \E\left[\langle \E[G_{t} \mid W_{t}], W_{t}-w^{\ast}\rangle\right]\\
	&= \E\left[\langle \nabla\sf(W_{t}), W_{t}-w^{\ast}\rangle\right].
\end{align*}
Therefore, \eqref{eq:UtilityProofRed1} becomes
\begin{equation}
	\label{eq:UtilityProofRed2}
	\Delta \leq \frac{1}{n} \sum_{t=1}^{n} \E[\langle G_{t}, W_{t}-w^{\ast}\rangle].
\end{equation}

Recall that, for every $w\in\R^{d}$ and every $w_{0}\in\W$,
\begin{equation*}
	\lVert \Pi_{\W}(w)-w_{0} \rVert \leq \lVert w-w_{0} \rVert.
\end{equation*}
Thus, the update rule in \eqref{eq:UpdataOGD} and the definition of $G_{t}$ imply that
\begin{align*}
	\lVert W_{t+1} - w^{\ast} \rVert^{2} &\leq \lVert W_{t} - \eta_{t}G_{t} - w^{\ast} \rVert^{2}\\
	&= \lVert W_{t}-w^{\ast} \rVert^{2} + \eta_{t}^2 \lVert G_{t} \rVert^{2}- 2 \eta_{t} \langle G_{t}, W_{t}-w^{\ast} \rangle.
\end{align*}
A direct manipulation shows that $\E[\langle G_{t}, W_{t}-w^{\ast} \rangle]$ is bounded above by
\begin{equation*}
	\E\left[\frac{\lVert W_{t}-w^{\ast} \rVert^{2} - \lVert W_{t+1}-w^{\ast} \rVert^{2}}{2\eta_{t}}\right] + \frac{\eta_{t}(M^{2}+\sigma_{t}^{2}d)}{2},
\end{equation*}
where the last inequality uses the fact that
\begin{equation*}
	\E[\lVert G_{t} \rVert^{2}] \leq M^{2} + \sigma_{t}^{2}d.
\end{equation*}
Therefore, \eqref{eq:UtilityProofRed2} becomes
\begin{align*}
	\Delta &\leq \frac{1}{n} \sum_{t=1}^{n} \E\left[\frac{\lVert W_{t}-w^{\ast} \rVert^{2} - \lVert W_{t+1}-w^{\ast} \rVert^{2}}{2\eta_{t}}\right]+ \frac{1}{n} \sum_{t=1}^{n} \frac{\eta_{t}(M^{2}+\sigma_{t}^{2}d)}{2}\\
	&= \frac{1}{n} \sum_{t=1}^{n} \left\{\frac{\E[\lVert W_{t}-w^{\ast} \rVert^{2}]}{2} \left(\frac{1}{\eta_{t}} - \frac{1}{\eta_{t-1}}\right)\right\}+ \frac{1}{n} \sum_{t=1}^{n} \frac{\eta_{t}(M^{2}+\sigma_{t}^{2}d)}{2},
\end{align*}
where for ease of notation we define $1/\eta_{0} = 0$. Since $\displaystyle \eta_{t} = \frac{\textnormal{dia}(\W)}{M\sqrt{t}}$ and $\lVert W_{t}-w^{\ast} \rVert \leq \textnormal{dia}(\W)$ for every $t$,
\begin{equation*}
	\Delta \leq \frac{M\textnormal{dia}(\W)}{2\sqrt{n}} + \frac{1}{n} \sum_{t=1}^{n} \frac{\eta_{t}(M^{2}+\sigma_{t}^{2}d)}{2},
\end{equation*}
Finally, by the inequality $\displaystyle \sum_{t=1}^{n} \frac{1}{\sqrt{t}} \leq 2 \sqrt{n}$, we conclude
\begin{equation*}
	\Delta \leq \frac{M\textnormal{dia}(\W)}{2\sqrt{n}} + \frac{M\textnormal{dia}(\W)}{\sqrt{n}} + \frac{d}{2n} \sum_{t=1}^{n} \eta_{t}\sigma_{t}^{2},
\end{equation*}
as required.

\end{document}